\documentclass[]{aastex7}

\usepackage{CJK}
\usepackage{amsmath}
\usepackage{subcaption}
\usepackage{caption}
\usepackage{xcolor}
\usepackage{multirow}
\defcitealias{Bonomo2017}{B17}
\defcitealias{Fossati2023}{F23}
\defcitealias{2009ApJ...706..785H}{H09}
\defcitealias{2024ApJ...963L...5X}{X24}
\defcitealias{2019MNRAS.488.3067H}{H19}
\defcitealias{2019PASP..131k5003A}{A19}
\defcitealias{2024ApJ...973L..30B}{B24}
\defcitealias{Mancini2018}{M18}
\defcitealias{Casasayas2017}{C17}
\defcitealias{2011ApJ...726...52H}{H11}
\defcitealias{2015MNRAS.450.2279T}{T15}
\defcitealias{2021AJ....161...70P}{P21}
\defcitealias{Barat2025}{B25}
\defcitealias{2019ApJ...885L..12D}{D19}
\defcitealias{2021MNRAS.507.2154V}{V21}
\defcitealias{2019AJ....157...97K}{M19}
\defcitealias{2019NatAs...3.1099G}{G19}
\defcitealias{2025AJ....170...50M}{M25}
\defcitealias{2024Natur.632..752F}{F24}
\defcitealias{2024AJ....168..297T}{T24}
\defcitealias{2024NatAs...8.1008C}{C24}
\defcitealias{2024AJ....168..104S}{S24}
\defcitealias{2022ApJ...940L..35F}{F22}
\defcitealias{2023Natur.623..709B}{B23}
\defcitealias{2024Natur.630..831S}{S24}
\defcitealias{2024Natur.630..836W}{W24}
\defcitealias{2024arXiv240303325B}{BE24}
\defcitealias{2024ApJ...970L..10B}{BT24}
\defcitealias{2021ApJS..255....8R}{R21}
\defcitealias{2014ApJ...785..126K}{K14}
\defcitealias{2022NatAs...6..232S}{S21}
\defcitealias{2026ApJ...996L..28W}{W26}
\defcitealias{2025arXiv251107746Y}{Y25}
\defcitealias{2025AA...703A.264C}{C25}
\defcitealias{2025PNAS..12216193W}{W25}

\received{November 10, 2025}
\revised{March 8, 2026}
\accepted{April 1, 2026}

\submitjournal{ApJ}

\begin{document}
\begin{CJK*}{UTF8}{gbsn}
\title{Unusually Hot Interiors Could Reconcile the Missing Methane Problem for Warm-to-Hot Exoplanets with Hydrogen Atmospheres}

\author[0000-0002-7479-1437, sname=Yu, gname=Xinting]{Xinting Yu (余馨婷)}
\affiliation{Department of Physics and Astronomy, University of Texas at San Antonio\\
1 UTSA Circle\\
San Antonio, TX 78249, USA}
\email{xinting.yu@utsa.edu}

\author[0000-0002-2161-4672, sname=Glein, gname=Christopher]{Christopher R. Glein}
\affiliation{Space Science Division, Southwest Research Institute\\
6220 Culebra Rd, San Antonio, TX 78238, USA}
\email{christopher.glein@swri.org}

\author[0000-0002-5113-8558, sname=Thorngren, gname=Daniel]{Daniel P. Thorngren}
\affiliation{Department of Physics and Astronomy, Johns Hopkins University\\
3400 N. Charles Street, Baltimore, MD 21218, USA}
\email{dpthorngren@jhu.edu}

\author[0009-0008-1902-0788, sname=Murray, gname=David]{David F. Murray}
\affiliation{Department of Physics and Astronomy, University of Texas at San Antonio\\
1 UTSA Circle\\
San Antonio, TX 78249, USA}
\email{yyt458@utsa.edu}

\correspondingauthor{Xinting Yu}
\email{xinting.yu@utsa.edu}

\begin{abstract}
JWST is revolutionizing the field of exoplanet atmospheres by delivering unprecedented spectroscopic constraints on their chemical compositions. It has provided tight constraints on the abundances of dominant carbon- and oxygen-bearing species on numerous warm-to-hot exoplanets with hydrogen-dominated atmospheres. Under thermochemical equilibrium, many of these exoplanets should have abundant methane (CH$_4$); however, CH$_4$ has, so far, only been spotted in a few cases. Here, we present a simple, geochemistry-inspired framework to explore whether elevated intrinsic temperatures (T$\rm_{int}$) can account for the CH$_4$ depletions. Instead of using computationally expensive, forward grid models, our fast analytical framework focuses on two key chemical equilibria: CO-CH$_4$ and CO-CO$_2$, allowing us to quickly constrain the minimum T$\rm_{int}$ that is consistent with JWST-observed abundances of H$_2$O, CO, CH$_4$, and CO$_2$. Applying this framework to 12 warm-to-hot exoplanets, we find that several targets require minimum T$\rm_{int}$ values exceeding those predicted by standard evolution models, while others remain consistent with lower T$\rm_{int}$ solutions or exhibit degeneracies with other solutions. Our sample enables an initial exploration of population-level trends: while many exoplanets broadly follow an empirical T$\rm_{eq}$-T$\rm_{int}$ relation derived from the hot Jupiter mass-radius population, a subset of targets lie well above this trend. The apparent need for hotter interiors suggests that while Ohmic dissipation is probably an important heat source among the general population, additional heating processes, such as tidal heating, may also play important roles for some planets. Our results demonstrate the diagnostic power of atmospheric chemistry as a complementary probe of exoplanet interiors.

\end{abstract}

\keywords{\uat{Exoplanets}{498} --- \uat{Exoplanet atmospheres}{487} --- \uat{Planetary interior}{1248} --- \uat{Exoplanet atmospheric composition}{2021} --- \uat{James Webb Space Telescope}{2291} --- \uat{Exoplanet atmospheric evolution}{2308}}

\section{Introduction}

In only three years since its commissioning, JWST has been delivering data that are revolutionizing the field of exoplanet atmospheres \citep{2024RvMG...90..411K}. Because of their large atmospheric scale heights, warm-to-hot Neptune- and Jupiter-sized planets represent some of the best targets for JWST transmission and emission spectroscopy. To date, atmospheric spectra have been acquired for over a dozen such exoplanets, revealing a rich inventory of carbon-, oxygen-, and sulfur-bearing molecules. However, these observations do not always align with current expectations. Similar to what Hubble Space Telescope (HST) data hinted \citep{2018ApJ...858L...6K}, methane (CH$_4$) remains an enigmatic molecule in exoplanet atmospheres. Even though JWST can now observe at wavelengths that allow us to fully distinguish methane, water (H$_2$O), and other carbon- and oxygen-bearing species, many observed planets still show little to no CH$_4$ in their spectra. Thus far, methane has been detected in only a handful of systems, including two temperate sub-Neptunes (K2-18 b,  \citet{2023ApJ...956L..13M} and TOI-270 d, \citet{2024arXiv240303325B, 2024A&A...683L...2H}) and two warm Neptunes (WASP-80 b, \citet{2023Natur.623..709B} and GJ 3470 b, \citet{2024ApJ...970L..10B}). In contrast, our solar system's giant planets all contain abundant methane. Many other similar types of exoplanets do not appear to possess significant amounts of methane in their atmospheres, while other carbon-, oxygen-, and sulfur-bearing species have been detected.

\begin{figure}[ht]
    \centering
    \includegraphics[width=0.7\textwidth]{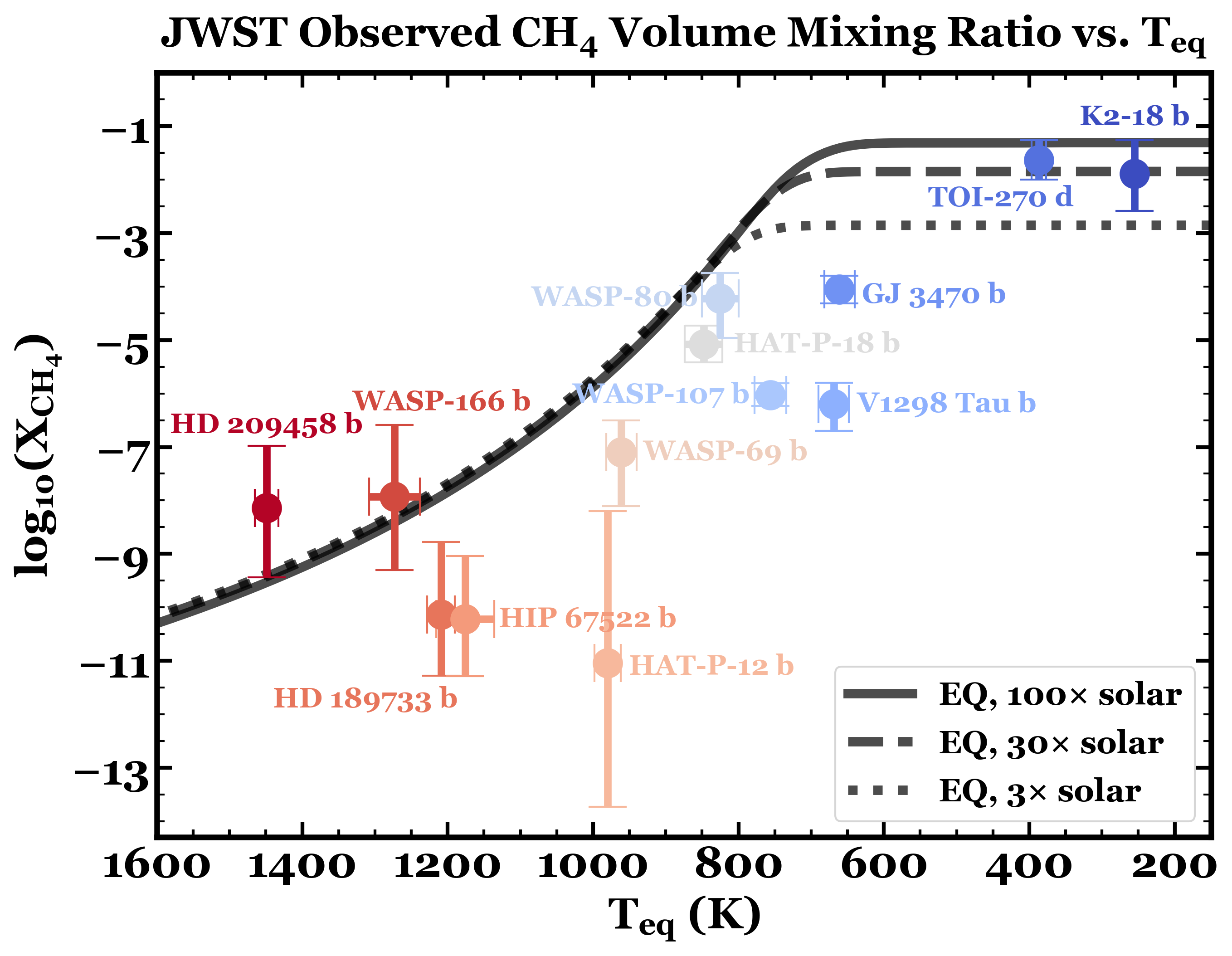}
    \caption{``Missing" methane on volatile-rich exoplanets motivates this study. JWST-measured volume mixing ratios (VMRs) of CH$_4$ are shown as a function of radiative equilibrium temperature for exoplanets with T$\rm_{eq}\leq1600$~K. The solid, dashed, and dotted lines indicate equilibrium CH$_4$ mixing ratios computed with \texttt{FASTCHEM} \citep{2018MNRAS.479..865S, 2022MNRAS.517.4070S, 2024MNRAS.527.7263K} at 0.01~bar (a representative pressure level probed by transmission spectroscopy, e.g., \citet[][]{2024Natur.630..831S}), assuming 3$\times$, 30$\times$, and 100$\times$ protosolar metallicities with the protosolar C/O ratio. Symbol colors correspond to equilibrium temperatures of these planets. Several annotated exoplanets deviate significantly from the expected trend of increasing CH$_4$ abundance with decreasing T$\rm_{eq}$. A complete list of the exoplanets included in this figure is provided in Table~\ref{table:target}.}
    \label{fig:missing_methane}
\end{figure}

In hot exoplanet atmospheres with equilibrium temperatures (T$\rm_{eq}$) from stellar heating that are above 1300~K, methane is typically not expected to be seen, as carbon monoxide (CO) naturally outcompetes CH$_4$ to become the dominant carbon carrier under thermochemical equilibrium, even in the presence of H$_2$. Examples include HD 209458 b (T$\rm_{eq}\sim1450$~K, hereafter, we use zero-albedo $T\rm_{eq}$ for all planets for consistency), which was predicted to have a methane volume mixing ratio (VMR) below 1~ppm because of high temperatures in its upper atmosphere \citep{2011ApJ...737...15M}. Recent JWST observations \citep{2024ApJ...963L...5X, 2025A&A...700A.105B}, indeed, support the theoretical predictions of the lack of methane on this exoplanet. However, in the atmospheres of cooler exoplanets, methane can be the dominant carbon carrier at chemical equilibrium \citep{2002Icar..155..393L}, as it is observed in some exoplanet atmospheres, such as GJ 3470 b (T$\rm_{eq}\sim661$~K) and WASP-80 b (T$\rm_{eq}\sim825$~K) \citep{2023Natur.623..709B, 2024ApJ...970L..10B}. Yet, a growing number of cooler exoplanets with $T\rm_{eq}$ less than 1300~K exhibit a marked depletion of methane despite the detection of other carbon-bearing species; i.e., suggesting that they do not have featureless spectra due to aerosols or the existence of high mean-molecular-weight (MMW) atmospheres. 

In Figure~\ref{fig:missing_methane}, we summarize the JWST-observed CH$_4$ VMRs as a function of equilibrium temperature. It is evident that several exoplanets deviate significantly from (fall below) the expected trend of increasing CH$_4$ abundance with decreasing temperature (see annotated planets in Figure~\ref{fig:missing_methane}). These include HD 189733 b (T$\rm_{eq}=1209$~K) \citep{2024Natur.632..752F,2025AJ....169...38Z}, HIP 67522 b (T$\rm_{eq}=1176$~K) \citep{2024AJ....168..297T},  HAT-P-12 b (T$\rm_{eq}=980$~K) \citep{2025AA...703A.264C}, WASP-107 b (T$\rm_{eq}=756$~K) \citep{2024Natur.630..831S, 2024Natur.630..836W, 2024Natur.625...51D}, and V1298 Tau b (T$\rm_{eq}=669$~K) \citep{Barat2025}. It should be noted that some cooler JWST exoplanet observations were inconclusive because of poor data quality, such as those of GJ 436 b \citep{2025ApJ...982L..39M}; but a lack of methane was hinted by previous Spitzer and HST observations \citep{2010Natur.464.1161S, 2011ApJ...729...41M, 2014Natur.505...66K, 2017AJ....153...86M}. Further observations are necessary to confirm the methane depletion in these systems.

The apparent absence of methane in the atmospheres of these warm-to-hot exoplanets could have several possible causes. These planets could have formed from building blocks with intrinsically low carbon-to-oxygen (C/O) ratios (e.g., by accreting water ice-rich pebbles/planetesimals, \citet{2011ApJ...743L..16O}), which would reduce the overall inventory of carbon-bearing species in their envelopes. However, these planets usually have detections of CO$_2$ and CO, which implies that their C/O ratios are not zero. Photochemistry could also serve to destroy CH$_4$ in the upper atmosphere \citep[e.g.,][]{2015MNRAS.446..345M, 2021ApJ...921...27H}, especially if the photochemically active region penetrates deep into the planetary atmosphere owing to weak vertical mixing or enhanced stellar UV activity. An alternative hypothesis invokes disequilibrium chemistry driven by elevated internal (``geothermal") heating in planetary interiors \citep{2020AJ....160..288F}. The associated internal heat flux of a planet, F$\rm_{int}$, is related to its intrinsic temperature, T$\rm_{int}$, through the Stefan-Boltzmann relation: F$\rm_{int}=\sigma T^4_{int}$, where $\sigma$ is the Stefan-Boltzmann constant. 

It is now well understood that the atmospheric composition observed by JWST often reflects chemical equilibrium at depth, where the vertical transport timescale exceeds the chemical timescale (there, reactions are faster than mixing) \citep{1977Sci...198.1031P, 1996ApJ...472L..37F, 2003IAUS..211..345S, 2011ApJ...737...15M, 2011ApJ...743..191M, 2011ApJ...738...72V, 2012A&A...546A..43V, 2013ApJ...777...34M, 2014ApJ...797...41Z, 2019ApJ...883..194M, 2020AJ....160..288F, 2020AJ....160...63M}. This process of species abundances becoming ``frozen" at a specific deeper part of the atmosphere is called ``quenching," and the point where compositions are frozen is referred to as the ``quench point." For a planet with a hotter interior (i.e., elevated T$\rm_{int}$), the same quench pressure corresponds to a higher temperature, thereby shifting the equilibrium towards CO (or CO$_2$ at high metallicities) at the expense of CH$_4$. The preservation of higher-temperature states can lead to a lower CH$_4$ abundance in the upper atmosphere. Yet, with all species expected to be quenched at some depth in a thick atmosphere, why is CH$_4$ in particular ``missing," while other species appear to be largely consistent with expectations? The underlying reason is that the equilibrium abundance of CH$_4$ is strongly dependent on temperature. For warm-to-hot exoplanets, CH$_4$ equilibrium abundances typically decrease by several orders of magnitude towards the deeper, hotter part of the atmosphere, while the equilibrium abundances of CO and H$_2$O remain almost constant. The equilibrium abundance of CO$_2$ also decreases with depth, but the decrease is not as drastic as that of CH$_4$ \citep[e.g.,][]{2014RSPTA.37230073M}. Consequently, CH$_4$ can be an especially sensitive tracer of interior thermal structure.

\begin{figure}
    \centering
    \includegraphics[width=0.9\textwidth]{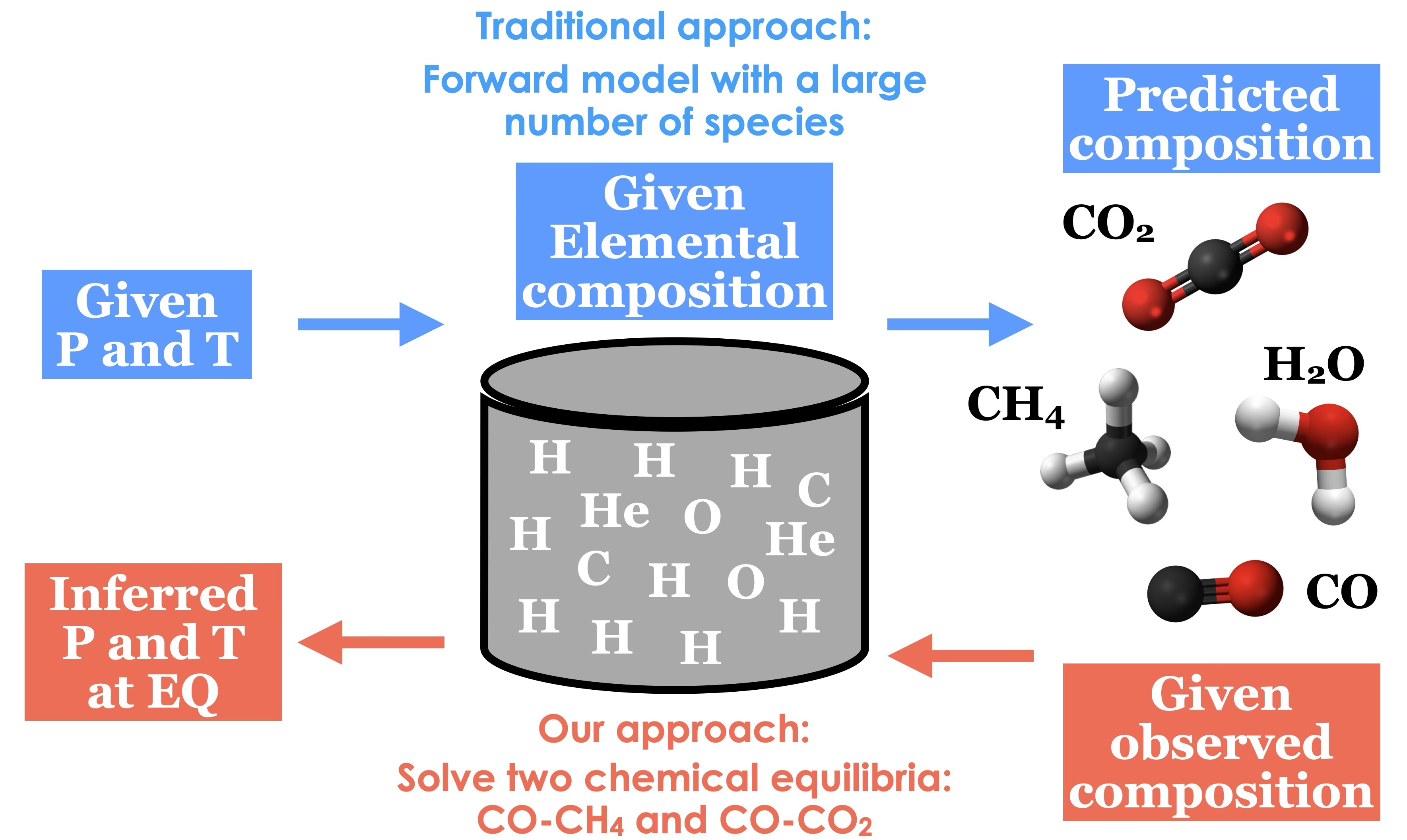}
    \caption{Summary of our method and comparison to the traditional grid modeling approach.}
    \label{fig:method_fig}
\end{figure}

Here, we employ a simple, geochemistry-inspired analytical model to examine whether elevated internal heat within planets can explain the observed ``missing" methane by cooking CH$_4$ to CO/CO$_2$. Compared to previous works, which mostly adopted the single CO-CH$_4$ equilibrium approach \citep[e.g.,][]{2014ApJ...797...41Z, 2020AJ....160..288F}, our new approach advances this line of work in two key ways. First, we employ a two-equilibria approach by incorporating both CO-CH$_4$ and CO-CO$_2$ equilibria simultaneously, which we show is critical to fully capture the quench behavior of all major carbon- and oxygen-bearing species. In addition, we use an inverted approach: instead of prescribing a pressure-temperature (P-T) profile and predicting chemical abundances, we use the observed abundances to infer the P-T conditions consistent with chemical equilibrium (see a schematic of our methodology in Figure~\ref{fig:method_fig} and details in the Methods section). As a result, our simple approach provides a fast, intuitive framework for inferring T$\rm_{int}$ that is compatible with JWST atmospheric observations and can be readily extended to future exoplanet targets with atmospheric compositions of key carbon and oxygen carrier molecules, not only with JWST, but also with ARIEL and other telescopes that operate in the infrared. 

This paper is structured as follows. We describe our analytical model in Section~\ref{sec:methods}. In the Results section, we first benchmark our simple model using WASP-107 b, a well-studied target that has multiple JWST observations and forward grid-model retrievals, including independently derived T$\rm_{int}$ (Section~\ref{sec:benchmark}). We then apply our method to 11 additional warm-to-hot exoplanets with JWST transmission and emission spectroscopy observations (Section~\ref{sec:crowd}). Finally, we compare our chemistry-constrained T$\rm_{int}$ values to those predicted by planet evolution models to assess whether these individual systems exhibit evidence of excess internal heat, and we explore population-level trends for T$\rm_{int}$ (Section~\ref{sec:discussion}). We also discuss the caveats of our model in Section~\ref{sec:discussion} before giving concluding remarks in Section~\ref{sec:conclusion}.

\section{Methods} \label{sec:methods}
\subsection{Target Selection}

In this work, we consider a diverse sample of JWST-observed exoplanet targets with atmospheric spectral features. Our sample spans a wide range of sizes (2-15~$R_\oplus$), masses (4-400~$M_\oplus$), from ``super-Earths" to ``Jupiters," and equilibrium temperatures (350-1300~K). An upper temperature limit of 1300~K is chosen because T$\rm_{int}$ has minimal influence on the observable atmospheric composition above this threshold \citep{2023MNRAS.522.2525A}. We additionally include HD 209458 b (T$\rm_{eq}=1449$~K) to demonstrate this effect. All targets, summarized in Table~\ref{table:target}, have H$_2$-dominated low MMW atmospheres and are observed to have non-featureless spectra (i.e., show evidence of molecules). In addition, all targets are expected to have thick atmospheres that allow thermochemical equilibrium to be reached at depth. Because temperature increases with depth inside planets, thick atmospheres will eventually get hot. Our targets are meant to be wide-ranging but should not be seen as exhaustive, as new data are constantly being released.

We exclude targets with featureless spectra or tenuous atmospheres, such as LHS 1140 b \citep{2024ApJ...968L..22D, 2024ApJ...975..146H}, several sub-Neptunes from JWST's COMPASS program that are observed to have featureless spectra \citep{2024AJ....168...77W, 2025AJ....169...15A, 2025AJ....169..249T, 2024AJ....168..276S}, and the TRAPPIST-1 system exoplanets, which likely do not have thick atmospheres \citep{2023Natur.620..746Z, 2023ApJ...955L..22L, 2023Natur.618...39G, 2025ApJ...990L..52E, 2025ApJ...990L..53G}. We also exclude the sub-Neptune K2-18 b, even though its atmosphere shows clear detections of CO$_2$ and CH$_4$. The atmosphere of K2-18 b is cold enough to trigger condensation, removing water vapor from the gas phase \citep{2021ApJ...921...27H, 2024ApJ...975..146H}. As a result, the observed upper limits on H$_2$O abundance from JWST \citep{2023ApJ...956L..13M, 2025arXiv250712622H} likely reflect the saturated vapor pressure of water rather than the full water abundance from the deeper interior. Our geochemical model requires the deep-atmosphere VMR of H$_2$O, which cannot be directly measured on K2-18 b.

\begin{deluxetable*}{lccc}
\tablecaption{List of the 12 Exoplanet Targets Ranked by Equilibrium Temperature and the Published JWST Data Used in This Work. \label{table:target}}
\tablehead{
\colhead{Target} & 
\colhead{$T_{\mathrm{eq}}$ [K]} & 
\colhead{JWST Retrieval Method} &
\colhead{JWST Data References} 
}
\startdata
HD 209458 b & $1449\pm16$ & PLATON free retrieval for CH$_4$\tablenotemark{a} & \citet{2024ApJ...963L...5X} \\
WASP-166 b & $1273\pm35$ & POSEIDON free retrieval & \citet{2025AJ....170...50M} \\
HD 189733 b & $1209\pm19$ & CHIMERA free retrieval & \citet{2024Natur.632..752F} \\
HIP 67522 b & $1176\pm40$ & CHIMERA grid retrieval & \citet{2024AJ....168..297T} \\
HAT-P-12 b & $980\pm18$ & ARCiS free retrieval & \citet{2025AA...703A.264C} \\
WASP-69 b & $961\pm21$ & PICASO equilibrium grid retrieval & \citet{2024AJ....168..104S} \\
HAT-P-18 b & $848\pm26$ & ATMO free retrieval & \citet{2022ApJ...940L..35F} \\
\multirow{2}{*}{WASP-80 b} & \multirow{2}{*}{$825\pm25$} & AURORA free retrieval & \citet{2023Natur.623..709B} \\
& & CHIMERA free retrieval & \citet{2025PNAS..12216193W} \\
\multirow{2}{*}{WASP-107 b} & \multirow{2}{*}{$756\pm22$} & ATMO and NEMESIS free retrievals & \citet{2024Natur.630..831S}\\
 &  & AURORA and CHIMERA free retrievals & \citet{2024Natur.630..836W}\\
V1298 Tau b & $669\pm21$ & PICASO free retrieval & \citet{Barat2025}\\
GJ 3470 b & $661\pm21$ & AURORA free retrieval & \citet{2024ApJ...970L..10B} \\
TOI-270 d & $387\pm10$  & SCARLET free retrieval & \citet{2024arXiv240303325B} \\
\enddata
\tablenotetext{a}{The VMRs of other species, including H$_2$O, CO$_2$, and CO, were calculated assuming chemical equilibrium with the retrieved temperature, metallicity (logZ), and C/O ratio.}
\tablecomments{Planets are ordered by decreasing equilibrium temperature ($T_{\mathrm{eq}}$), assuming zero albedo.}
\end{deluxetable*}

\subsection{Chemical Constraints on CO-CH$_4$ and CO-CO$_2$ Equilibria and Quenching} \label{sec:chemical}

All of these warm-to-hot exoplanet targets are expected to have thick hydrogen-dominated envelopes that increase in temperature and pressure with depth. Thus, it is reasonable to assume that thermochemical equilibrium will be reached at some depth. As gas parcels are transported upward to cooler, lower-pressure regions, chemical reactions can become kinetically inhibited, causing the abundances of certain species to ``freeze." As a result, at altitudes above the quench point, the VMRs of the quenched species would remain fixed and reflect equilibrium conditions from the deep atmosphere at the quench temperature and pressure. 

We adopt an updated version of the analytical methodology developed by \citet{2025ApJ...985..187G}, which uses two key chemical equilibria, CO-CH$_4$ and CO-CO$_2$, to constrain the quench pressures and temperatures that are consistent with JWST observations of C-H-O species. Our previous study used CO$_2$-CH$_4$ rather than CO-CH$_4$, but that was for a metal-enriched sub-Neptune; CO-CH$_4$ is more generally applicable across a wide range of giant planets. The basic idea of our method is to invert the usual approach: instead of using a P-T profile to predict chemical equilibrium abundances, we use the observed abundances to infer the pressure-temperature conditions where chemical equilibrium was last achieved. With key observational data now available, the community can consider inverse modeling in addition to predictive modeling.

Here, we assume that the abundance of CH$_4$ is quenched at depth and is controlled by the more abundant oxidized carbon species (i.e., CO or CO$_2$). For most of our targets, the observed abundance of CO is higher than that of CO$_2$; thus, CH$_4$ equilibrates with CO at depth. Then, at a lower temperature, the less abundant one of the two carbon oxides (in this case, CO$_2$, \citet{2025RNAAS...9..108W}) would reach equilibrium with CO. We adopt this two-equilibria approach rather than the single CO-CH$_4$ equilibrium approach that is commonly used in the literature \citep[e.g.,][]{2020AJ....160..288F}. The single-equilibrium approach often fails to capture the quench behavior of CO$_2$, an observationally-important carbon species (see Appendix Section~\ref{sec:one_eq}).

\subsubsection{CO-CH$_4$ Equilibrium} 
In a hydrogen-dominated atmosphere, the dominant net reaction that describes chemical equilibrium between CO and CH$_4$ is:
\begin{equation}
\rm{CH}_4 (g) + H_2O (g) \rightleftharpoons CO (g)+ 3H_2 (g).
\label{eq:r1}
\end{equation}
We can write the equilibrium constant of this reaction as:
\begin{equation*}
K_1(T_{\rm CO-CH_4}) = \frac{a^3_{\rm H_2}a_{\rm CO}}{a_{\rm H_2O}a_{\rm CH_4}},
\end{equation*}
where T$_{\rm CO-CH_4}$ is the quench temperature of the CO-CH$_4$ reaction, and $a_i$ is the thermodynamic activity of species $i$, which can be expressed as:
\begin{equation}
a_i = \frac{f_i}{f^0_i}=\frac{\phi_i}{\phi^0_i}\frac{P_i}{P^0_i},
\label{eq:a_i}
\end{equation}
where $f_i$ is the fugacity of species $i$ and $f^0_i$ is the fugacity of species $i$ in the standard state (both have units of pressure). The fugacity of each gas phase species is the product of the partial pressure of the gas and its fugacity coefficient, where $\phi_i$ is the fugacity coefficient for the real gas, and $\phi^0_i$ is the fugacity coefficient for a gas at standard state. By convention, the gas at standard state is an ideal gas at a pressure of 1~bar; we therefore have $\phi^0_i=1$ and $P^0_i=1$ bar. And if we assume that the gases in Reaction~\ref{eq:r1} follow ideal gas behavior, which is typical for the pressures and temperatures at the quench point \citep{LEMMON-RP10}, then $\phi_i=1$. Thus, we can simplify Equation~\ref{eq:a_i} to:
\begin{equation*}
a_i = P_i = X_iP,
\end{equation*}
where $X_i$ is the mole fraction (or volume mixing ratio, VMR) of gas species $i$ and $P$ is the total pressure in bar. Thus, we can rewrite the equilibrium constant as:
\begin{equation}
\rm K_1(T_{CO-CH_4}) = \frac{X^3_{\rm H_2}X_{\rm CO}}{X_{\rm H_2O}X_{\rm CH_4}}P^2.
\label{eq:K1}
\end{equation}

Rearranging Equation~\ref{eq:K1} gives the pressure along the equilibrium curve for a given temperature and quenched chemical composition:
\begin{equation}
P = \sqrt{\frac{\rm K_1(T_{CO-CH_4})X_{\rm H_2O}X_{\rm CH_4}}{X^3_{\rm H_2}X_{\rm CO}}}.
\label{eq:K1_P}
\end{equation}

Thus, if we know the mole fractions of H$_2$O, CH$_4$, CO, and H$_2$ and the equilibrium constant: K$_1(\rm T_{CO-CH_4})$, we can plot the P-T relationship that is consistent with CO-CH$_4$ chemical equilibrium everywhere along the curve. The mole fractions of H$_2$O, CH$_4$, and CO can be directly retrieved from JWST observations. Since the VMR of H$_2$ is not directly constrained by JWST, we therefore assume:
\begin{equation*}
X_{\rm H_2} + X_{\rm He} + X_{\rm H_2O} + X_{\rm CH_4} +  X_{\rm CO_2} + X_{\rm CO} \approx 1,
\end{equation*}
If we also assume the He/H$_2$ ratio to be protosolar; i.e., He/H$_2=0.2$ \citep{2021SSRv..217...44L}, then we can express $X_{\rm H_2}$ as:
\begin{equation}
X_{\rm H_2} = [1 - (X_{\rm H_2O} + X_{\rm CH_4} +  X_{\rm CO_2} + X_{\rm CO})]/1.2.
\label{eq:XH2}
\end{equation}
This approximation is reasonable if nitrogen-, sulfur-, and other unobserved C-H-O-bearing species do not comprise significant fractions of the atmosphere where equilibrium is reached. Current observations suggest that nitrogen is generally underabundant in exoplanet atmospheres, at least in the form of ammonia (NH$_3$) \citep[e.g.,][]{2023ApJ...956L..13M, 2024arXiv240303325B, 2025ApJ...985..187G, 2025arXiv250712622H, 2025arXiv251007367N}. While some exoplanets appear to be enriched in sulfur \citep{2024AJ....168..297T, 2024Natur.630..836W,2024ApJ...970L..10B,2024Natur.632..752F,2024NatAs...8.1008C,2024arXiv240303325B, Barat2025}, we have explicitly tested that including nitrogen- or sulfur-bearing species does not affect the derived mole fraction of H$_2$, since their abundances remain small ($<1\%$) compared to those of dominant carbon- and oxygen-bearing species. Likewise, our results are not sensitive to the assumed He/H$_2$ ratio unless H$_2$ ceases to be the dominant atmospheric constituent \citep[e.g.,][]{2015ApJ...807....8H, 2025arXiv250622537D}. 

For ideal gases, the equilibrium constant itself ($K$) is a function of only temperature. We computed the equilibrium constant for Reaction~\ref{eq:r1} using standard Gibbs free energies of formation from the NIST-JANAF Thermochemical Tables \citep{chase1998nist}, and we fit numerical values to the following equation:
\begin{equation*}
    \rm log{K_1(T_{CO-CH_4})} = -3.43 - \frac{10228}{\rm T~(\rm K)} + 5.27log(T~(K)) - 7.45\times10^{-4}(T~(K)),
\end{equation*}
which reproduces data from 298--3000 K within an absolute error in logK$_1(T)$ of 0.03.

\subsubsection{CO-CO$_2$ Equilibrium} 
The reaction that governs CO-CO$_2$ equilibrium can be written as follows:
\begin{equation}
    \rm CO_2 (g) + H_2 (g) \rightleftharpoons CO (g)+ H_2O (g).
\label{eq:r2}
\end{equation}
The equilibrium constant of this reaction can be expressed as (also assuming ideal gas behavior):
\begin{equation}
    \rm K_2(T_{CO-CO_2}) = \frac{\rm X_{CO}X_{H_2O}}{\rm X_{CO_2}X_{H_2}},
    \label{eq:K2}
\end{equation}
where T$_{\rm CO-CO_2}$ is the quench temperature of the CO-CO$_2$ reaction. 

Unlike CO-CH$_4$, the law of mass action in this case is independent of pressure because there are equal numbers of gaseous molecules on both sides of Reaction~\ref{eq:r2}. We fit this equilibrium constant using a similar approach as for Reaction~\ref{eq:r1} and obtained:
\begin{equation*}
    \rm log{K_2(T_{CO-CO_2})} = 5.90 - \frac{2335}{\rm T~(\rm K)} - 1.25log(T~(K)) + 2.73\times10^{-5}(T~(K)),
\end{equation*}
which reproduces data from 298--3000 K within an absolute error in log${K_2(T)}$ of 0.01. 

\subsubsection{Combining the Two Equilibria with Quench Temperature Relationship} 
\citet{2025ApJ...985..187G} presented a useful, simple empirical scaling relationship that was derived from Earth geothermal gas data \citep{1987ApGC....2..143G} to convert quench temperatures between CO$_2$-CH$_4$ equilibrium and CO-CO$_2$ equilibrium:
\begin{equation}
    \rm T_{CO-CO_2} (K) \approx 0.8T_{CO_2-CH_4} (K).
    \label{eq:quenchtemp}
\end{equation}
This approximation was tested against the analytical expressions of \citet{2014ApJ...797...41Z} and the modeling data of \citet{2021ApJ...914...38Y} and works well. The CO-CH$_4$ timescale should be similar to that of the CO$_2$-CH$_4$ reaction because the interconversion between CO and CO$_2$ is expected to remain rapid at temperatures where CH$_4$ becomes quenched. As a result, CO and CO$_2$ are likely to stay near equilibrium at the CH$_4$ quench level, implying that $\rm T_{CO_2-CH_4} \approx \rm T_{CO-CH_4}$. The timescale approach of \citet{2014ApJ...797...41Z} also explicitly relates the CO-CH$_4$ and CO-CO$_2$ reactions. Thus, here we can assume the quench temperatures of CO-CH$_4$ and CO-CO$_2$ follow the same relationship as for CO$_2$-CH$_4$ and CO-CO$_2$, an approximation that reproduces the full kinetic-transport quench temperatures to within 100~K in the \citet{2014ApJ...797...41Z} framework in \citet{2025ApJ...985..187G}:
\begin{equation}
    \rm T_{CO-CO_2} (K) \approx 0.8T_{CO-CH_4} (K).
    \label{eq:quenchtemp2}
\end{equation}
This relationship allows us to relate the quench points of the two reactions.

Combining the equilibrium constant of the CO-CO$_2$ reaction, K$_2(\rm T_{CO-CO_2})$ (Equation~\ref{eq:K2}) with the mole fraction of hydrogen (Equation~\ref{eq:XH2}) allows us to derive an independent constraint on the mole fraction of CO$_2$ using the mole fractions of CH$_4$, CO, and H$_2$O. This leads to a quadratic expression for X$_{\rm CO_2}$:
\begin{equation}
    \rm X^2_{CO_2} - (1-X_{H_2O} - X_{CO} - X_{CH_4})X_{CO_2} + \frac{1.2X_{CO}X_{H_2O}}{K_2(T_{CO-CO_2})} = 0,
    \label{eq:CO_2}
\end{equation}
To make it more compact, we can define $\rm b = (1-X_{H_2O} - X_{CO} - X_{CH_4})$ and $\rm c = 1.2X_{CO}X_{H_2O}/K_2(T_{CO-CO_2})$. There are two solutions to Equation~\ref{eq:CO_2}:
\begin{equation*}
    \rm X_{CO_2} = \frac{b \pm \sqrt{b^2 - 4c}}{2}.
\end{equation*}
Here, we take only the negative-sign root as the physically relevant solution:
\begin{equation}
    \rm X_{CO_2} = \frac{b - \sqrt{b^2 - 4c}}{2}.
    \label{eq:CO_2-2}
\end{equation}
The positive root produces $\rm X_{CO_2}>0.5$, which would imply CO$_2$ dominating the atmospheric composition and is therefore incompatible with the H$_2$-dominated atmosphere considered in this work and with elemental abundance constraints.

\subsubsection{Implementing the Framework and Interpretation} 

With the above Equations~\ref{eq:K1}, \ref{eq:XH2}, \ref{eq:quenchtemp2}, and \ref{eq:CO_2-2}, we can define P-T relationships that are consistent with observational data for the chemical equilibria of CO-CH$_4$ and CO-CO$_2$. The computational framework flows as follows: 

\begin{enumerate}
\item Iterate a temperature grid that spans a relevant temperature range for the target planet, and convert the CO-CH$_4$ quench temperature to the corresponding CO-CO$_2$ quench temperature using Equation~\ref{eq:quenchtemp2}.
\item Iterate H$_2$O, CH$_4$, and CO mole fractions within the retrieved 1$\sigma$ mole fractions using JWST observational data, and calculate the mole fraction of CO$_2$ using the quadratic solution of Equation~\ref{eq:CO_2-2}. Only keep combinations of H$_2$O-CO-CH$_4$ mole fractions that lead to CO$_2$ mole fractions within the JWST retrieved 1$\sigma$ CO$_2$ mole fraction.
\item For the JWST retrieval results that provide additional, independent constraints on the ratios of species mole fractions, keep combinations of H$_2$O-CO-CH$_4$-CO$_2$ mole fractions that are within the JWST retrieved 1$\sigma$ ratio range.
\item The remaining combinations of H$_2$O-CO-CH$_4$-CO$_2$ mole fractions are used to calculate the hydrogen mole fraction using Equation~\ref{eq:XH2}.
\item The chemical equilibrium pressure can be calculated using Equation~\ref{eq:K1_P} with the resulting H$_2$O-CO-CH$_4$-H$_2$ combinations.
\end{enumerate}

Using the methodology described above, at each temperature, we have a continuum of P-T relationships that are consistent with the JWST observed atmospheric molecular composition (referred to as ``chemically-derived P-T relationships" or ``chemically-derived P-T region" hereafter), assuming that chemical equilibrium was reached for the CO-CH$_4$ and CO-CO$_2$ reactions at depth. Here, we define the envelope of these relationships in P-T space as the ``max-P" and ``min-P" tracks. These P-T relationships derived from chemistry can then be overplotted on the generated atmospheric pressure-temperature profiles (referred to as ``atmospheric P-T profiles" or ``P-T profiles" hereafter) from Section~\ref{sec:PT} below, which were computed with elemental abundances that are consistent with observations but for a wide range of unknown intrinsic temperature T$\rm_{int}$ values. Two outcomes can be envisioned:
\begin{itemize}
    \item If the chemically-derived P-T relationships intersect an atmospheric P-T profile. The crossing pressure (P$_{\rm cros}$) and temperature (T$_{\rm cros}$) represent conditions where the observed H$_2$O-CO-CH$_4$-CO$_2$ abundances in the upper atmosphere from JWST are consistent with chemical equilibrium being reached at P$_{\rm cros}$ and T$_{\rm cros}$. In other words, the chemical equilibrium abundances of species that reach P$_{\rm cros}$ and T$_{\rm cros}$ are ``frozen" until they are transported up to the observed pressure level. The crossing points are essentially the quench points for the CO-CH$_4$ reaction. CO-CO$_2$ equilibrium is reached at a lower temperature (higher in the atmosphere), following Equation~\ref{eq:quenchtemp2}. Even though vertical mixing is not directly prescribed in our model, the crossing points implicitly encode the range of vertical mixing strengths that are capable of producing the observed atmospheric abundances.    
    \item If the chemically-derived P-T relationships have no intersection with an atmospheric P-T profile. It means that the observed atmospheric composition cannot be reproduced under chemical equilibrium anywhere along the P-T profile. In this case, the P-T profile must be adjusted, typically by increasing the unknown T$\rm_{int}$ value until an intersection occurs. T$\rm_{int}$ is a free parameter in this approach. Note that, because of uncertainties in observational data and internal heating, in general, we should not expect to find a single atmospheric P-T solution. Instead, we seek to determine limiting values and delineate the region of consistency. 
\end{itemize}

A key advantage of our model is that vertical mixing does not need to be prescribed explicitly, unlike in typical kinetic approaches \citep[e.g.,][]{2011ApJ...737...15M, 2011ApJ...738...72V, 2014ApJ...797...41Z}. Instead, the allowed quench temperatures and pressures inferred from the crossing points can be used to determine the range of eddy diffusion coefficients consistent with the observations. In other words, if there is a crossing between the chemically-derived P-T relationships and a P-T profile, there must exist some vertical mixing strength that can explain the observations. If no intersection occurs for a given P-T profile, the observed atmospheric abundances cannot be explained by a deeper equilibrium at that thermal structure, regardless of the vertical mixing strength.

Because T$\rm_{int}$ is the only free parameter for computing the P-T profiles, which cannot be observed directly for transiting exoplanets, this methodology enables us to determine the T$\rm_{int}$ that is required for the exoplanet to reproduce its observed atmospheric composition. In this work, we typically adopt free retrieval results whenever possible (see Table~\ref{table:target} for the JWST retrieval data that are used), rather than grid retrieval, when multiple retrieval results are presented in publications. This decision is made to avoid any undesired (or inadequately explained) imposed physical constraints (which may introduce internal inconsistencies or circular reasoning) in grid retrieval models and allows for more flexibility in the retrieved abundances for each molecule.

\subsection{Climate Model} \label{sec:PT}
We employ an open-source 1D climate model \texttt{PICASO 3.0} \citep{2019ApJ...878...70B,2023ApJ...942...71M} to generate atmospheric P-T profiles for our target exoplanets. We adopt correlated-$k$ coefficients from \citet{Lupu+22} and opacities from \citet{batalha2020resampled}. Input stellar spectra are interpolated from the Phoenix stellar spectrum grid \citep{2013A&A...553A...6H}. The P-T profiles require several input parameters: the incident stellar flux, the bulk elemental composition of the atmosphere, the Bond albedo of the planet, and the internal heat flux of the planet, parameterized as the intrinsic temperature T$\rm_{int}$.

In \texttt{PICASO 3.0}, the Bond albedo is tuned by varying the semi-major axis so that the stellar flux matches the value implied by the desired Bond albedo \citep{2024ApJ...975..146H}. However, the current model does not yet treat atmospheric clouds self-consistently \citep{2023ApJ...942...71M}. The bulk elemental composition is commonly expressed as the metallicity, which assumes that heavy elements are uniformly enhanced relative to protosolar values. This convention is largely a means of accounting for the observed atmospheric composition and does not introduce another variable. The \citet{Lupu+22} database also accounts for cases where bulk carbon and oxygen abundances are non-uniformly enhanced by providing a range of C/O ratios for each metallicity. We constrain the bulk elemental abundances of our targets using atmospheric compositions inferred from JWST observations. Because T$\rm_{int}$ is the only input parameter that cannot be directly observed and is an unknown parameter, we generate a set of P-T profiles spanning a range of plausible T$\rm_{int}$ values.

We compute the elemental C/H and O$_{gas}$/H ratios in the atmosphere of each planet using the retrieved VMRs of the dominant carbon and oxygen carrier species: H$_2$O ($X_{\rm H_2O}$), CO ($X_{\rm CO}$), CH$_4$ ($X_{\rm CH_4}$), and CO$_2$ ($X_{\rm CO_2}$).
\begin{equation}
    \left(\frac{\rm C}{\rm H}\right)_{\rm atm} \approx \frac{X_{\rm CH_4} + X_{\rm CO} + X_{\rm CO_2}}{2X_{\rm H_2} + 2X_{\rm H_2O} + 4X_{\rm CH_4}}
\end{equation}

\begin{equation}
    \left(\frac{\rm O_{gas}}{\rm H}\right)_{\rm atm} \approx \frac{X_{\rm H_2O} + X_{\rm CO} + 2X_{\rm CO_2}}{2X_{\rm H_2} + 2X_{\rm H_2O} + 4X_{\rm CH_4}}.
\end{equation}

The designation O$_{gas}$ is used to distinguish between oxygen atoms in gaseous species compared with oxygen sequestered in solid phases (like silicate minerals in the interior). The VMRs of H$_2$O, CO, CH$_4$, and CO$_2$ can be found in the JWST retrieval data from the observable part of the atmosphere, and the VMR of H$_2$ can be estimated using Equation~\ref{eq:XH2}. These derived elemental ratios can then be compared to the protosolar abundances of carbon and oxygen from \citet{2009LanB...4B..712L} (which is used in \citet{Lupu+22}'s correlated-$k$ coefficients table for various C/O ratios), where (C/H)$_{\rm solar}=2.776\times10^{-4}$, (O$_{\rm gas}$/H)$_{\rm solar}=6.062\times10^{-4}$, and the protosolar C/O ratio is 0.458. For clarity, throughout this work, we use the term ``solar" to mean protosolar abundances (i.e., bulk solar system composition). We iterate over the JWST-observed 1$\sigma$ VMR ranges of H$_2$O, CO, CH$_4$, and CO$_2$ to determine the allowed range of elemental abundances. These are then converted into metallicity and C/O ratios that serve as inputs to generate self-consistent P-T profiles for each exoplanet.

Because of uncertainties in the observed molecular abundances, multiple atmospheric P-T solutions with various T$\rm_{int}$ could reproduce the data. In this work, we define the minimum T$\rm_{int}$ as the value corresponding to the P-T profile that just begins to intersect the chemically-derived P-T relationships. Since our goal is to establish a lower bound on T$\rm_{int}$, we deliberately construct the hottest plausible P-T profiles for a given T$\rm_{int}$. Within the 1$\sigma$ observational uncertainties, this is achieved by adopting the highest metallicity and lowest C/O ratio consistent with the data, as these conditions would lead to hotter P-T profiles \citep[e.g.,][]{2015ApJ...813...47M}. Because metallicity has a stronger influence on P-T structures than the C/O ratio, we first select the maximum allowed metallicity and then recompute the corresponding allowed C/O ratio range at that metallicity before adopting the final C/O ratio. Note that the final adopted C/O ratio may be higher than the minimum value obtained when considering all possible metallicities. This procedure ensures internal consistency. Following similar reasoning, we assume clear atmospheres and a zero Bond albedo for all of our atmospheric models to maximize absorbed stellar irradiation and produce the warmest possible P-T profiles. A discussion of how clouds could alter this interpretation is provided in Section~\ref{sec:cloud}.

\subsection{Forward Model to Test the Simple Model}
To evaluate the robustness of the conclusions drawn from our analytical framework, we also performed forward modeling for a subset of exoplanet targets using an open-source 1D photochemical kinetics code, VULCAN \citep{2017ApJS..228...20T, 2021ApJ...923..264T}. These models provide a self-consistent check that our simplified equilibrium--quench approach captures the key chemical behavior of the observed atmospheres. 

For each planet, we adopt the same elemental abundances as the inputs of the atmospheric P-T profile, which are derived from JWST observations as shown in Section~\ref{sec:PT}. Vertical mixing is parameterized by the eddy diffusion coefficient, K$\rm_{zz}$, following a simple two-layer prescription used in previous studies \citep{2021ApJ...922L..27T, 2024ApJ...975..146H}:
\begin{equation}
\rm K_{zz}(P) =
\begin{cases}
\rm K_{deep}, & P > P_{RCB} \\
\rm K_{deep} \left(\frac{P_{RCB}}{P}\right)^{0.4}, & P \le P_{RCB} ,
\end{cases}
\end{equation}
where K$\rm_{deep}$ is a constant mixing coefficient in the deep, convective region and P$\rm_{RCB}$ is the pressure at the planet's radiative-convective boundary (RCB) obtained from our PICASO climate model. This parameterization captures the expected vertical mixing strength transition from vigorous convection to weaker mixing in the radiative zone of the atmosphere. Note that we did not turn on photochemistry in most of our forward models, so we can isolate the effect of vertical quenching, allowing a direct comparison with our simple equilibrium-quench approach. For a subset of runs that include photochemistry, we adopted the stellar spectra from the MUSCLES treasury survey, selecting spectra with the same, or when unavailable, the closest spectral type to each host star \citep{2016ApJ...820...89F, 2016ApJ...824..101Y, 2023AJ....166...35B}. The influence of photochemistry on our results is further discussed in Section~\ref{sec:photochem}.

\subsection{Thermal Evolution Models}
To obtain theoretically expected T$\rm_{int}$ for comparison with our observationally inferred minimum---or apparent---T$\rm_{int}$ values, we employed structural evolution models in a Bayesian retrieval framework. The forward models solve the equations of hydrostatic equilibrium, mass conservation, and the equations of state in 1D \citep[as in][]{2019ApJ...884L...6T}. For hydrogen and helium, the \citet{2021ApJ...917....4C} equation of state was used at a protosolar H/He ratio; for the metals, we used a 50-50 ice-rock mixture by mass from ANEOS \citep{thompson1990aneos}. We place the first 10 $M_\oplus$ of heavy elements in a core, and mix any remaining metal mass into the envelope. Planets were evolved forward from an initial hot state using the atmosphere grid of \citet{2007ApJ...659.1661F}.

To fit the observations to our model, we employed the Bayesian statistical model of \citet{2019ApJ...884L...6T}. The parameters were mass, the bulk metallicity, the age, and a hot Jupiter heating parameter as fitted in \citet{2018AJ....155..214T}. The model was constrained by the observed mass, radius, and age (Appendix Table~\ref{table:target_property}), as well as the mass-metallicity prior from \citet{2016ApJ...831...64T} and the flux-heating prior from \citet{2018AJ....155..214T}. For planets with an incident stellar flux below $2\times10^8$ erg s$^{-1}$ cm$^{-2}$, inflation is not observed in the population \citep{2011ApJS..197...12D}, so heating was set to zero. Radioactive heating from the core was included following \citet{2023ApJ...945L..36T} but was found to be negligible. After running these models and evaluating convergence using the Gelman-Rubin statistic and corner plots, we obtained the distributions of theoretically expected T$\rm_{int}$ for each exoplanet target in Table~\ref{table:target}.

\section{Results}
\subsection{Benchmarking the Simple Model}  \label{sec:benchmark}

\begin{figure}
    \centering
    \includegraphics[width=0.8\textwidth]{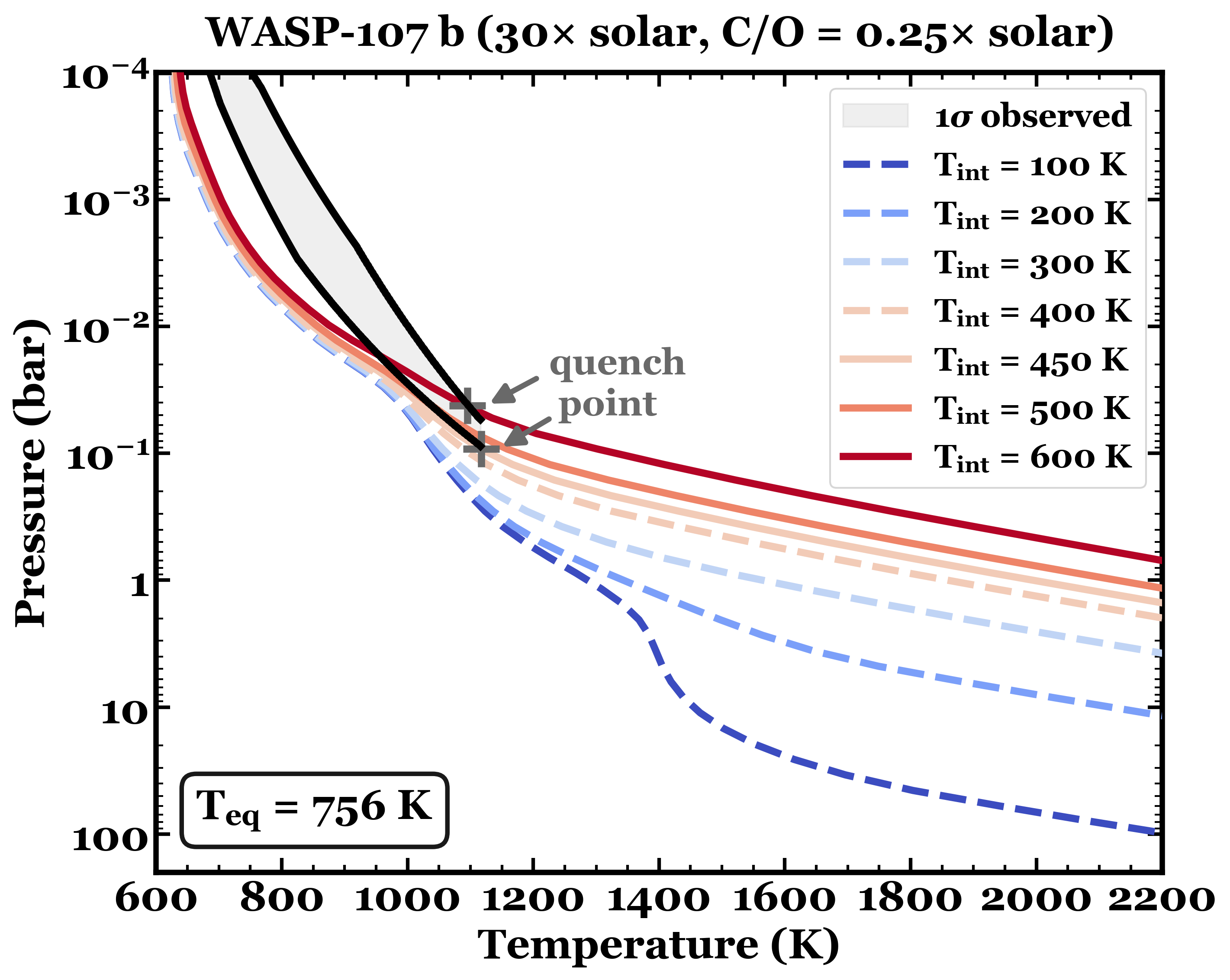}
    \caption{Inferring the intrinsic temperature T$\rm_{int}$ of ``missing methane" exoplanets, with WASP-107 b as an example. Colored curves show atmospheric pressure-temperature (P-T) profiles computed for different T$\rm_{int}$ values, based on 30$\times$ solar metallicity and a C/O ratio of 0.25$\times$ solar. The gray shaded region between the black curves represents the chemically-derived P-T relationships that are consistent with JWST-observed H$_2$O-CO-CH$_4$-CO$_2$ abundances under chemical equilibrium (using the ATMO free retrieval mixing ratios). Solid colored curves represent P-T profiles that intersect the chemically-derived P-T region, while the dashed colored curves represent atmospheric P-T profiles with lower T$\rm_{int}$ values that are incompatible with the measured composition. Quench points (gray plus signs for T$\rm_{int}=450$~K and T$\rm_{int}=600$~K profiles as examples) mark the first intersections between the solid P-T profiles and the chemically-derived P-T region. The minimum T$\rm_{int}$ is defined as the coolest P-T profile that intersects the chemically-derived P-T region (450~K for WASP-107 b). At altitudes above the quench point, species abundances are vertically mixed and remain ``frozen," producing the atmospheric compositions observed in transmission spectroscopy.}
    \label{fig:wasp107b_atmo}
\end{figure}

\begin{figure}
    \centering
    \includegraphics[width=0.8\textwidth]{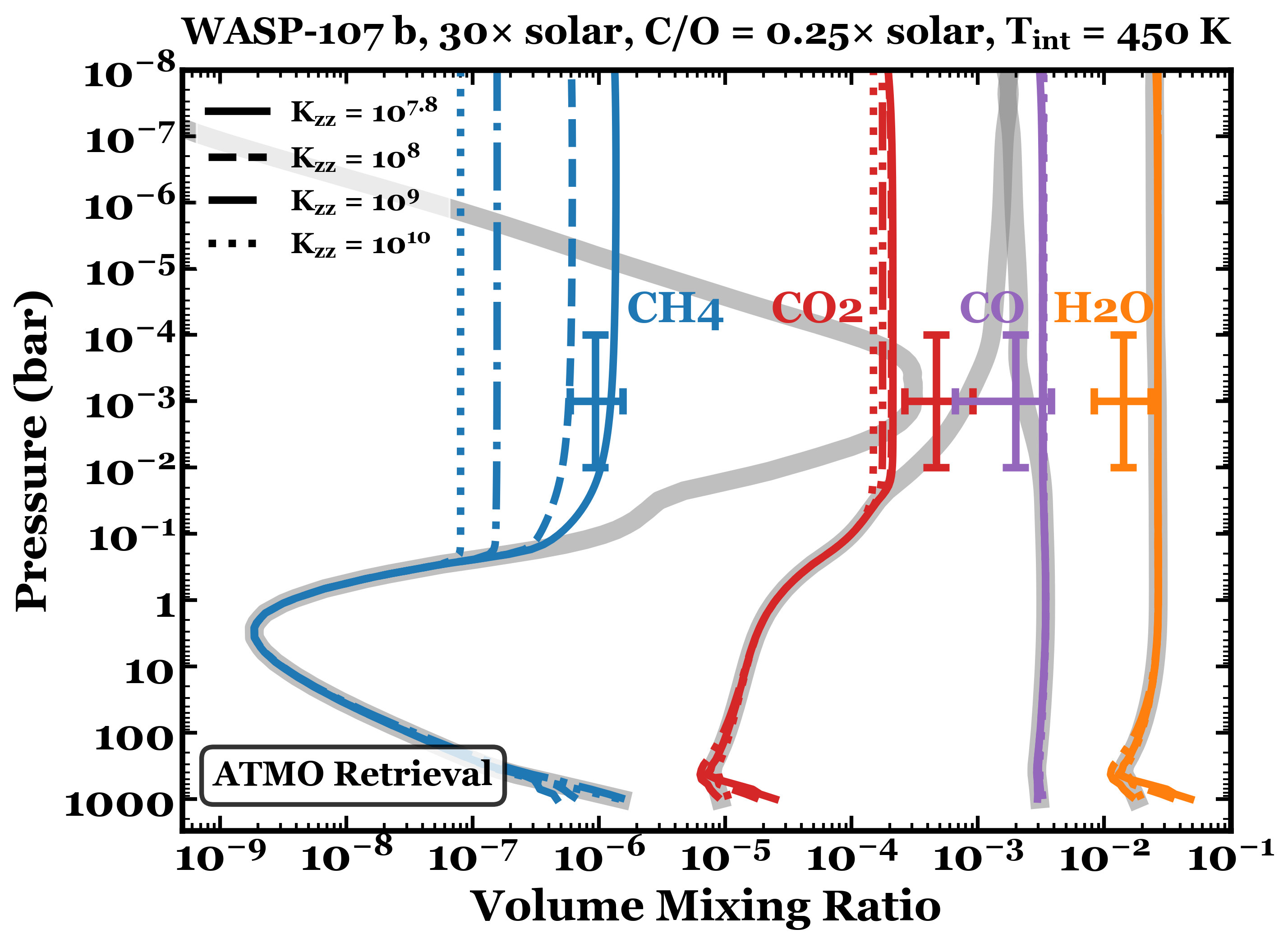}
    \caption{Volume mixing ratios of key species for WASP-107 b as predicted by forward chemical disequilibrium modeling with VULCAN, using different values of K$\rm_{deep}$ for the eddy diffusion coefficient (K$\rm_{zz}$) parameterization. JWST-observed abundances and 1$\sigma$ error bars are shown based on the ATMO free retrieval results from \citet{2024Natur.630..831S}. Thick gray lines represent chemical equilibrium profiles. Solid, dashed, dashed-dotted, and dotted lines correspond to K$\rm_{deep}$ values of 10$^{7.8}$, 10$^{8}$, 10$^{9}$, and 10$^{10}$~cm$^2$~s$^{-1}$, respectively. This plot illustrates how methane goes ``missing" around the 1~mbar level, which can be explained by quenching of deeper CO-CH$_4$ equilibrium.}
    \label{fig:wasp107b_forward}
\end{figure}

To see if our methodology is sensible, we start with WASP-107 b, a super-Neptune (or sub-Saturn) observed by JWST with several instruments, which clearly lacks methane compared to chemical equilibrium (see Figure~\ref{fig:missing_methane}). The T$\rm_{int}$ of WASP-107 b has been rigorously constrained through a grid of forward disequilibrium models \citep{2024Natur.630..831S, 2024Natur.630..836W}. Following the approach in Section~\ref{sec:chemical}, we first calculate the max-P and min-P curves in black in Figure~\ref{fig:wasp107b_atmo}, based on the 1$\sigma$ retrieved mole fraction posteriors of H$_2$O, CO, CH$_4$, and CO$_2$ with the ATMO free retrieval model applied to NIRSpec observations from \citet{2024Natur.630..831S} (2.70-5.16~$\mu$m). The adopted VMRs of the key species are summarized in Appendix Table~\ref{table:vmr}. The max-P and min-P curves define the chemically-derived P-T region. We then calculate the range of atmospheric P-T profiles for WASP-107 b with various T$\rm_{int}$ values. Following the approach in Section~\ref{sec:PT}, we calculate the elemental abundances of WASP-107 b's atmosphere using the 1$\sigma$ mole fraction posteriors of H$_2$O, CO, CH$_4$, and CO$_2$. Carbon is enriched by 2.0-10.2$\times$solar, and oxygen by 9.4-28.8$\times$solar, yielding C/O ratios between 0.08-0.74$\times$solar, based on protosolar abundances from \citet{2009LanB...4B..712L}. 

Given the uncertainty of the retrieved mole fractions, we first select the metallicity that generates the hottest possible P-T profiles to derive the minimum T$\rm_{int}$ that is necessary to be consistent with observations, specifically, 30$\times$ solar metallicity. Using this metallicity, we recompute the allowed C/O ratio range and find that it remains between 0.08-0.74$\times$solar. Therefore, here we adopt a C/O ratio of 0.25, the lowest C/O ratio allowed by the k-coefficient opacity table of \citet{Lupu+22} used in PICASO. Using these elemental abundances, we generate a suite of atmospheric P-T profiles with T$\rm_{int}$ values from 100 to 600 K (in 100 K increments) and include an intermediate T$\rm_{int}$ at 450 K. 

Now, if we examine both the chemically-derived P-T region and the atmospheric P-T profiles with various T$\rm_{int}$ in Figure~\ref{fig:wasp107b_atmo}, we can see that WASP-107 b is clearly out of thermochemical equilibrium in its observable atmosphere (10$^{-4}$--10$^{-2}$~bar), as none of the P-T profiles intersect the chemically-derived P-T region at these pressures. Evidently, pressures and temperatures at the observed level are not high enough for thermochemical equilibrium to be reached among C-H-O species. The atmospheric P-T profiles only begin to intersect the max-P (lower black) curve at $\sim0.1$~bar when T$\rm_{int}\geq450$~K. 

If we only use the CO-CH$_4$ reaction to define the chemically-derived P-T region, it extends across all pressures, as shown in the red region of Appendix Figure~\ref{fig:alternative-COCH4}. In contrast, when both CO-CH$_4$ and CO-CO$_2$ equilibria are included, the chemically-derived P-T region becomes truncated at lower pressures because the CO-CO$_2$ speciation is independent of pressure. The absence of the chemically-derived P-T region at high pressures arises because combinations of CO-CH$_4$-H$_2$O abundances fail to reproduce CO$_2$ VMRs within the observed 1$\sigma$ range. 

To further investigate what inputs control the shapes of the max-P and min-P curves, we perform sensitivity tests by varying the upper and lower limits of each species and evaluating how these perturbations affect the resulting chemically-derived P-T region (Appendix Figure~\ref{fig:sensitivity}). The shape of the max-P curve is primarily determined by the upper limits of CH$_4$ and H$_2$O and the lower limit of CO, which is also evident based on Equation~\ref{eq:K1_P}. On the other hand, the min-P (upper black) curve is mainly set by the lower bounds of CH$_4$ and H$_2$O and the upper limit of CO. The lower limit of CO$_2$ influences both curves, but for reasons not as obvious as for the other species, since we discard combinations of CO-CH$_4$-H$_2$O abundances that yield VMRs of CO$_2$ outside the observed 1$\sigma$ range. 

For WASP-107 b, we can thus establish that the minimum T$\rm_{int}$ has to be 450 K to match the observed composition, which shows remarkable agreement with the $460\pm40$~K value reported in \citet{2024Natur.630..831S} from grid retrievals. However, our crossing pressure (i.e., the quench pressure) of $\sim$0.1~bar is lower than their quench pressure of $\sim1$~bar. By examining closely the forward disequilibrium grid modeling results of \citet{2024Natur.630..831S}, we find that the grid model overpredicts the CO VMR by $\sim$0.4~dex (a factor of $\sim$2.5) compared to the free retrieval result (see their Extended Data Figure 7). If we adopt the grid model CO VMR instead of the free retrieval CO VMR in our model, the crossing pressure can shift to $\sim$1~bar, but then the minimum T$\rm_{int}$ would decrease to 300~K. Yet, there is still evidence of internal heating.

If we assume that the free retrieval results for H$_2$O, CO, CH$_4$, and CO$_2$ are more accurate, then a high T$\rm_{int}$ is needed to explain the observed CH$_4$ depletion. However, the inferred K$\rm_{zz}$ value from \citet{2024Natur.630..831S} may need to be reconsidered, as we found a much shallower quench point, indicating that the previous high K$_{zz}$ solution may be overestimated. In our forward models (Figure~\ref{fig:wasp107b_forward}), we find that both K$\rm_{zz}=10^{7.8}$ and $10^8$~cm$^2$~s$^{-1}$ can reproduce the observations, with quench pressures around 0.1 bar. These K$\rm_{zz}$ are substantially lower than the K$\rm_{zz}=10^{11.6}$~cm$^2$~s$^{-1}$ value from \citet{2024Natur.630..831S}. This difference mainly arises from two factors: 1) the limited K$\rm_{zz}$ range used in their grid, which only explored K$\rm_{zz}>10^9$~cm$^2$~s$^{-1}$ to accommodate strong vertical mixing that is needed to loft aerosol particles \citep{2013A&A...558A..91P} to the observed pressures (as gray clouds are needed to fit the data); 2) the overpredicted CO in their grid compared to the free retrieval results, which drives the chemically-derived P-T region deeper, favoring higher quench pressures and thus a high K$\rm_{zz}$ solution.

The above example for WASP-107 b demonstrates a key utility of our model: it does not require an assumed K$\rm_{zz}$ grid. We simply evaluate whether and where the atmospheric P-T profile intersects the chemically-derived P-T region. Vertical transport strength can only be inferred after finding the quench point, rather than imposed \textit{a priori}. If the P-T profile crosses the chemically-derived P-T region, there must exist some K$\rm_{zz}$ solution that would lead to the observed abundances. In a sense, our approach is equivalent to sampling an effectively continuous and infinite K$\rm_{zz}$ space without introducing bias by selecting a preferred K$\rm_{zz}$ range. This procedure is unlike most grid-based retrievals, which must do so for computational efficiency. The exception would be cases where other disequilibrium processes, such as photochemistry, dramatically alter some of the species abundances in a very sluggish atmosphere (see Section~\ref{sec:photochem}). Our framework can thus complement future grid-based retrievals by identifying the parameter space that needs to be explored and guiding the choice of priors for K$\rm_{zz}$. In this way, our model can first efficiently map the chemically plausible regions of parameter space, while grid-based retrievals can then perform more detailed, physics-based modeling to refine those solutions with higher precision.

\begin{figure}
  \centering
  \begin{subfigure}{0.8\textwidth}
    \centering
    \includegraphics[width=\textwidth]{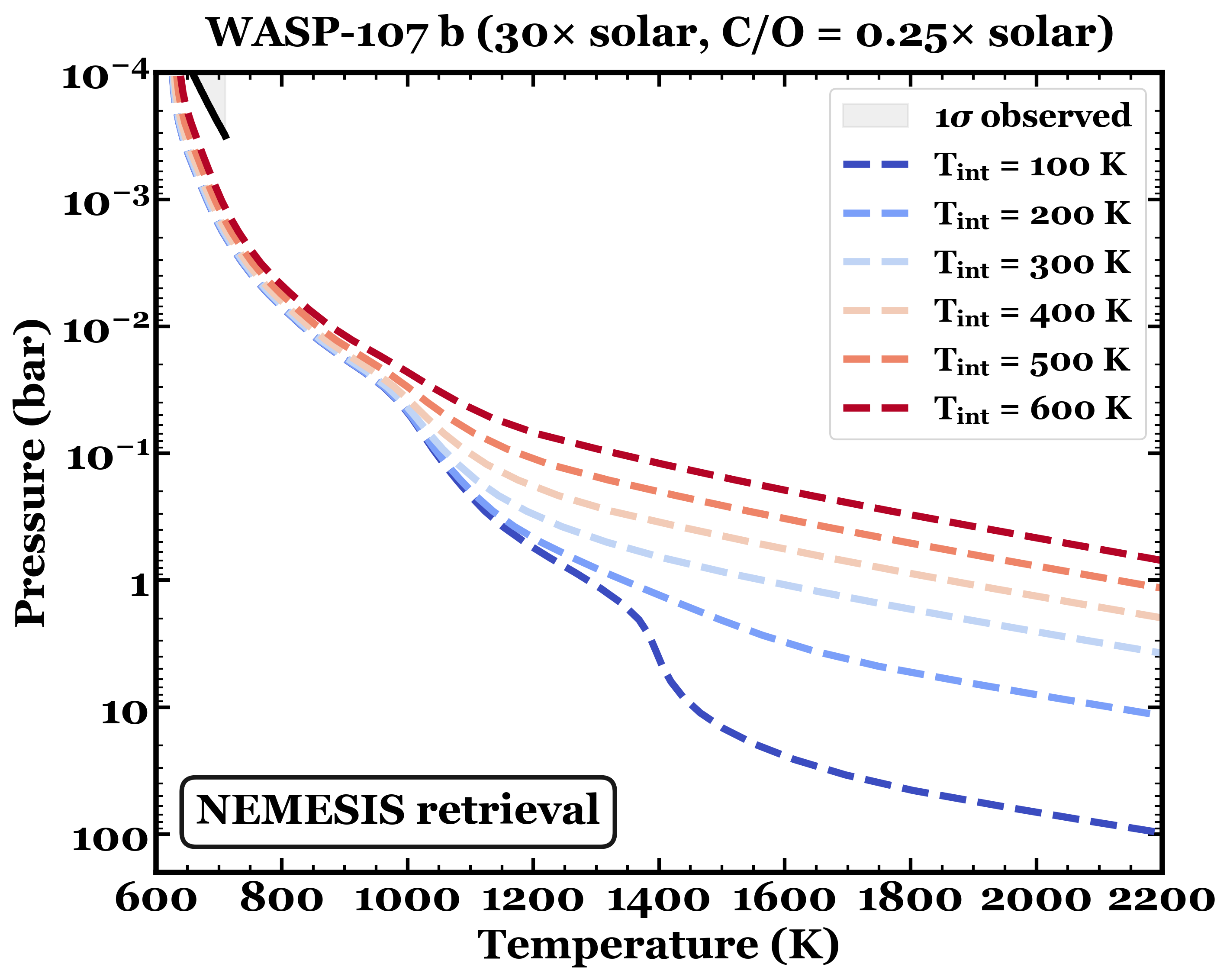}
  \end{subfigure}
  \hfill
  \begin{subfigure}{0.8\textwidth}
    \centering
    \includegraphics[width=\textwidth]{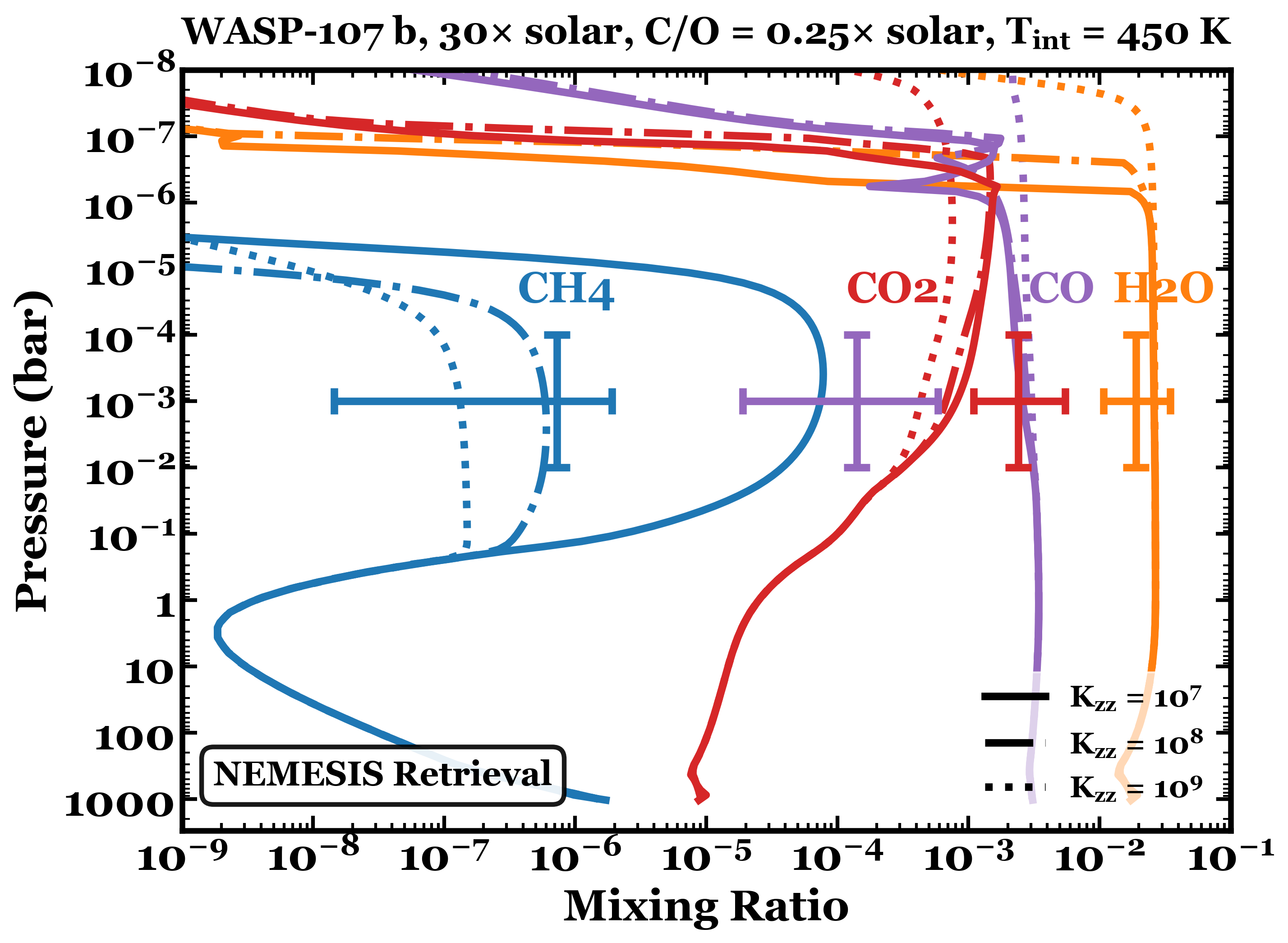}
  \end{subfigure}

\caption{Top panel: Same as Figure~\ref{fig:wasp107b_atmo}, but with elemental abundances of the climate model P-T profiles and the chemically-derived P-T region calculated using the NEMESIS free retrieval results. The gray region at the upper-left is very small here. Because there is no overlap between the colored curves and this region, chemical equilibrium cannot create the observed molecular speciation of C-H-O species. Bottom panel: Same as Figure~\ref{fig:wasp107b_forward}, but with VMRs of key species predicted using VULCAN with photochemistry and compared to JWST-observed abundances (with 1$\sigma$ error bars) from the NEMESIS free retrieval. Solid, dashed-dotted, and dotted lines correspond to K$\rm_{deep}$ values of 10$^{7}$, 10$^{8}$, and 10$^{9}$~cm$^2$~s$^{-1}$, respectively.}
  \label{fig:wasp107b_nemesis}
\end{figure}

Another retrieval model, NEMESIS, has also been used to fit the same JWST transmission spectrum dataset of WASP-107 b \citep{2024Natur.630..831S}. This alternative allows us to test our methodology using different retrieval results. In contrast to the ATMO free retrieval, the chemically-derived P-T region using the NEMESIS free retrieval results is confined to the lowest pressures ($<3\times10^{-4}$~bar) in the atmosphere (Figure~\ref{fig:wasp107b_nemesis} top panel), and it does not intersect with any of the computed P-T profiles regardless of the assumed T$\rm_{int}$. 

This result suggests several possibilities. If the observed abundances reflect quenched abundances of carbon species from the deep interior under thermochemical equilibrium, CO should be at least one or more orders of magnitude more abundant than CO$_2$. Yet, the NEMESIS retrieval yields CO$_2$ with log(VMR)~=~-2.62 exceeding CO with log(VMR)~=~-3.85, a ratio exceeding unity even when their respective 1$\sigma$ uncertainties are considered (i.e., CO$_2-1\sigma> $CO~$+1\sigma$). Such a high CO$_2$/CO ratio is incompatible with equilibrium chemistry in a reducing environment and causes our model to reject all the H$_2$O-CO-CH$_4$ combinations at Step 2 (see Section~\ref{sec:chemical}). In other words, none of the free-retrieved H$_2$O-CO-CH$_4$ abundance combinations can reproduce the retrieved CO$_2$ abundance via Equation~\ref{eq:CO_2-2} under equilibrium assumptions. The implication is that the NEMESIS free retrieval results may be unphysical or have systematically underestimated uncertainties. 

However, quenched abundances of atmospheric species can be further altered before they reach the observable levels in the upper atmosphere. This potential complication brings us to the second and third possibilities. The second possibility involves alternative disequilibrium processes, namely, photochemistry. Even though photochemistry is typically most effective in the microbar to millibar regime, under conditions of strong stellar UV irradiation or sluggish vertical mixing, photochemistry could have the potential to destroy enough CO to produce a CO$_2$/CO ratio above 1 \citep[e.g.,][]{2021ApJ...921...27H}. The third possibility invokes chemical pathways not typically included in current exoplanet atmospheric models, such as heterogeneous or catalytic reactions on aerosols and condensates \citep{2021JGRE..12606655G}, dust-grain catalysis from impact-delivered material \citep{2004Icar..168..475K, 2020PSJ.....1...11Z, 2023PSJ.....4..169W}, exogenic volatile delivery that alters the chemical compositions of atmospheres \citep[e.g.,][]{2008JGRE..11310006H}, or unexpectedly fast gas-phase reaction pathways \citep{2025ApJ...985..187G} occurring above the quench level.

In practice, when the 1$\sigma$ chemically-derived P-T region fails to intersect any atmospheric P-T profiles, this finding should be interpreted as a flag for 1) unphysical retrieval results or retrievals with underestimated errors; 2) strong photochemical modification above the quench point; or 3) missing kinetics, heterogeneous, or exogenic processes in the upper atmosphere. Discriminating between the first two would require photochemistry-transport modeling with slow vertical mixing that leads to CO$_2$/CO$~>1$ while fitting other observed C-H-O species. As an initial test, we performed forward photochemical modeling using VULCAN in an attempt to reproduce the high CO$_2$/CO ratio from the NEMESIS retrieval (see Figure~\ref{fig:wasp107b_nemesis} bottom panel). Even with low K$\rm_{zz}$ values, the CO$_2$/CO ratio in the observable atmosphere never exceeds unity, and further reductions of K$\rm_{zz}$ lead to excessive CH$_4$ relative to the retrieved 1$\sigma$ abundance. Thus, we conclude that the NEMESIS retrieval results are either unphysical or have underestimated uncertainties, or additional, currently unmodeled disequilibrium chemistry would be required to match the data. This exercise highlights the importance of employing multiple retrieval frameworks for the same dataset. If only the NEMESIS retrieval were available, it would have been difficult to identify the high T$\rm_{int}$ solution for WASP-107 b.

\begin{figure}
  \centering
  \begin{subfigure}{.49\textwidth}
    \centering
    \includegraphics[width=\textwidth]{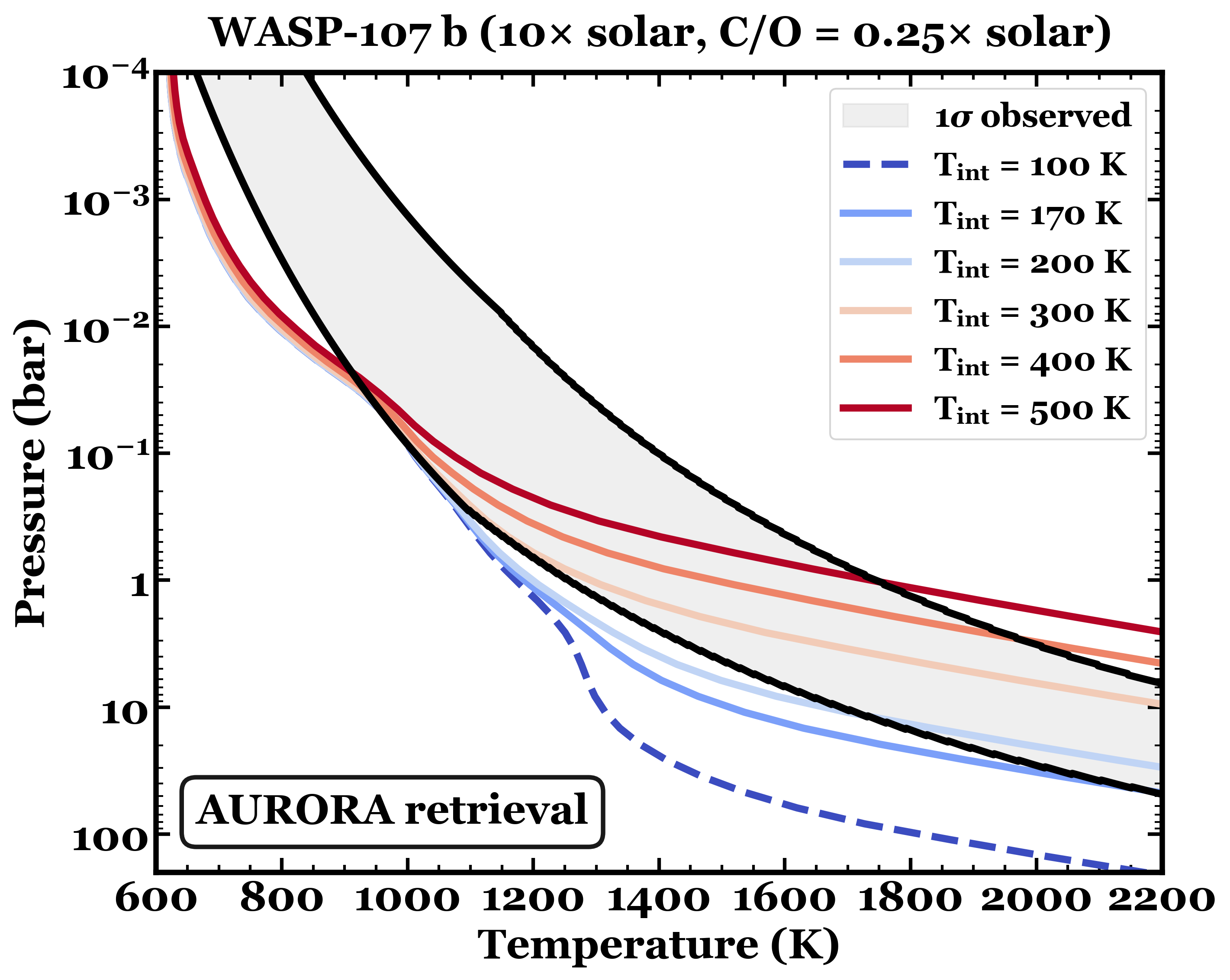}
  \end{subfigure}
  \hfill
  \begin{subfigure}{.49\textwidth}
    \centering
    \includegraphics[width=\textwidth]{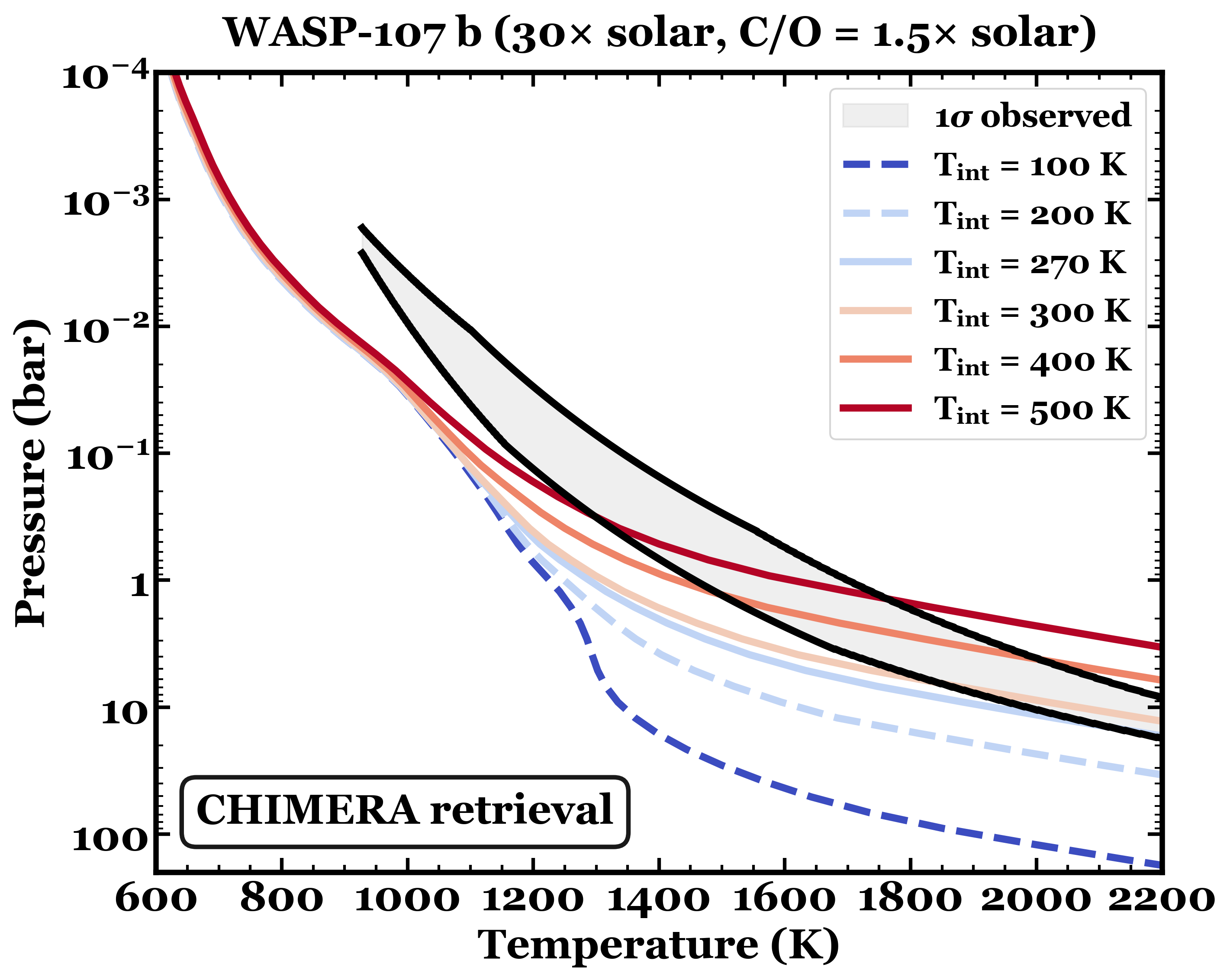}
  \end{subfigure}
  
  \begin{subfigure}{.49\textwidth}
    \centering
\includegraphics[width=\textwidth]{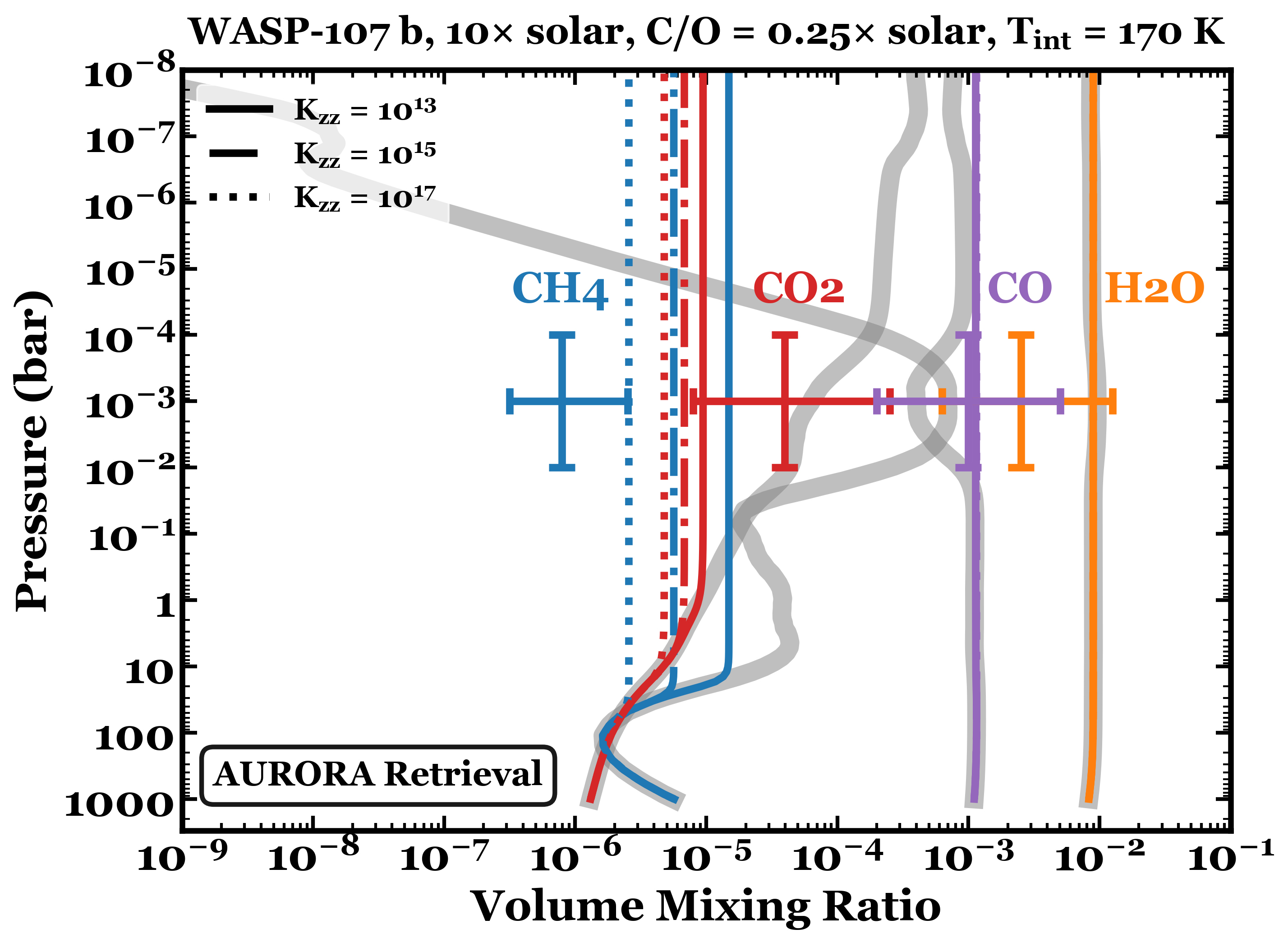}
  \end{subfigure}
  \hfill
  \begin{subfigure}{.49\textwidth}
    \centering
    \includegraphics[width=\textwidth]{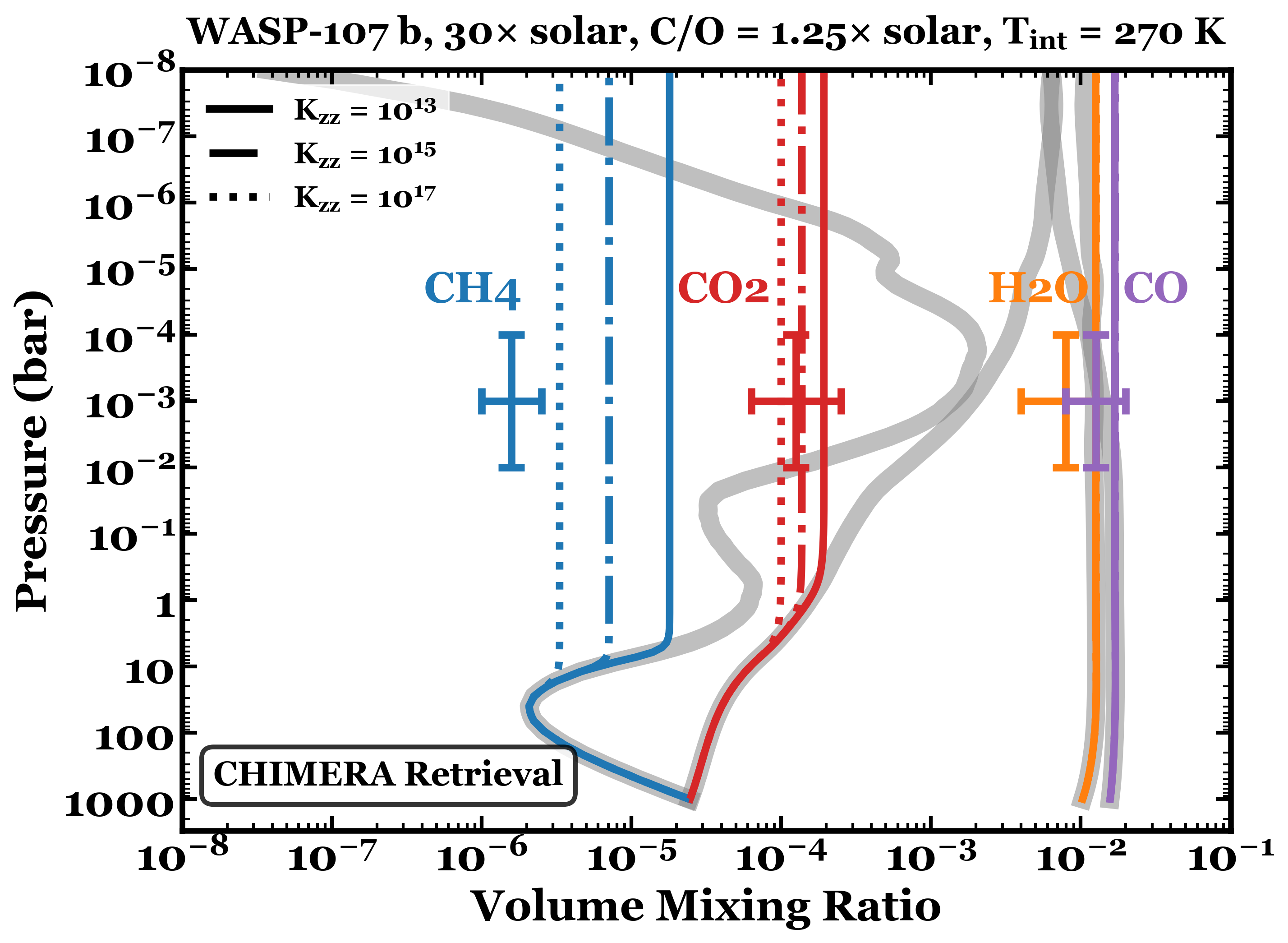}
  \end{subfigure}
\caption{Top panels: Same as Figure~\ref{fig:wasp107b_atmo}, but climate model P-T profiles and chemically-derived P-T regions calculated using the AURORA (left) and CHIMERA (right) free retrieval results from \citet{2024Natur.630..836W}. Bottom panels: Same as Figure~\ref{fig:wasp107b_forward}, but using JWST-observed abundances (with 1$\sigma$ error bars) from the AURORA (left) and CHIMERA (right) free retrievals. Solid, dashed-dotted, and dotted lines correspond to K$\rm_{deep}$ values of 10$^{13}$, 10$^{15}$, and 10$^{17}$~cm$^2$~s$^{-1}$, respectively.}
  \label{fig:WASP107b_aurora}
\end{figure}

In addition to the NIRSpec dataset, WASP-107 b's transmission spectrum has also been observed with the NIRCAM and MIRI instruments on JWST. Two other independent retrieval tools, AURORA and CHIMERA, have been applied to this broader wavelength dataset (2.45--12.2~$\mu$m) \citep{2024Natur.630..836W}. We can therefore test our methodology using different observational data and retrieval methods. The results of our model are shown in the top panels of Figure~\ref{fig:WASP107b_aurora}. The minimum T$\rm_{int}$ required to match the observed H$_2$O-CO-CH$_4$-CO$_2$ abundances drops to 170 K and 270 K for the AURORA and CHIMERA free retrievals, respectively. These values are not only significantly lower than the grid-retrieved T$\rm_{int}$ of 345~K reported in \citet{2024Natur.630..836W}, but they are also much lower than the minimum T$\rm_{int}$ we derived using the ATMO free retrieval results from \citet{2024Natur.630..831S}.

We further test whether these lower T$\rm_{int}$ solutions are physically consistent by performing forward disequilibrium modeling with VULCAN, see the bottom panels of Figure~\ref{fig:WASP107b_aurora}. Using the atmospheric P-T profiles corresponding to the minimum T$\rm_{int}$ for each case, we find that extremely strong vertical mixing is required to reproduce the observed abundances in the upper atmosphere, with K$\rm_{deep}$ exceeding 10$^{17}$~cm$^2$~s$^{-1}$ for both retrievals. Carbon species must be dredged up and quenched from deeper in the atmosphere to compensate for the lower T$\rm_{int}$. The higher T$\rm_{int}$ of 345~K found by \citet{2024Natur.630..836W} is probably caused by the limited K$\rm_{zz}$ range of 10$^{7}$--10$^{9}$~cm$^2$~s$^{-1}$ used in their grid retrieval. Here we show again that our methodology does not impose such constraints and simply allows solutions for any K$\rm_{zz}$ values. 

However, the question remains: is K$\rm_{zz}$ of 10$^{17}$~cm$^2$~s$^{-1}$ physically plausible for WASP-107 b? While most works treat K$\rm_{zz}$ as a free parameter, \citet{2009ApJ...699..564S} provided a scaling relation to estimate its value theoretically:
\begin{equation}
\rm K_{deep} \approx wL,
\end{equation}
where $w$ is the horizontally-averaged global root-mean-square (RMS) vertical velocity and $L$ is the characteristic vertical length scale of the atmosphere. $L$ is typically assumed to be the scale height. Since $w$ cannot exceed the sound speed (C$_s$), an upper limit on K$\rm_{deep}$ is C$_sH$. For WASP-107 b, this upper limit of K$\rm_{deep}$ is on the order of 10$^{13}$~cm$^2$~s$^{-1}$. Therefore, our derived T$\rm_{int}$ values from the AURORA and CHIMERA retrievals should be regarded as very conservative lower bounds rather than unique physical solutions, because our framework does not impose an upper limit on K$\rm_{zz}$. If we instead reject unrealistically high K$\rm_{zz}$ values and adopt physically motivated limits on vertical mixing, then higher T$\rm_{int}$ values are required to match the observed atmospheric composition, bringing the results into closer agreement with grid-based retrieval results. This exercise demonstrates that a wider wavelength range involving multiple instruments may not necessarily give more consistent compositions. While we are still learning how to best reduce, retrieve, and present JWST-observed exoplanet data, these results suggest that, as of now, multiple data reduction pipelines and multiple retrievals may be necessary to decipher the true atmospheric compositions of exoplanets.

\subsection{Looking at the Crowd of Planets}  \label{sec:crowd}

\begin{figure}

  \begin{subfigure}{.49\textwidth}
    \centering
    \includegraphics[width=\textwidth]{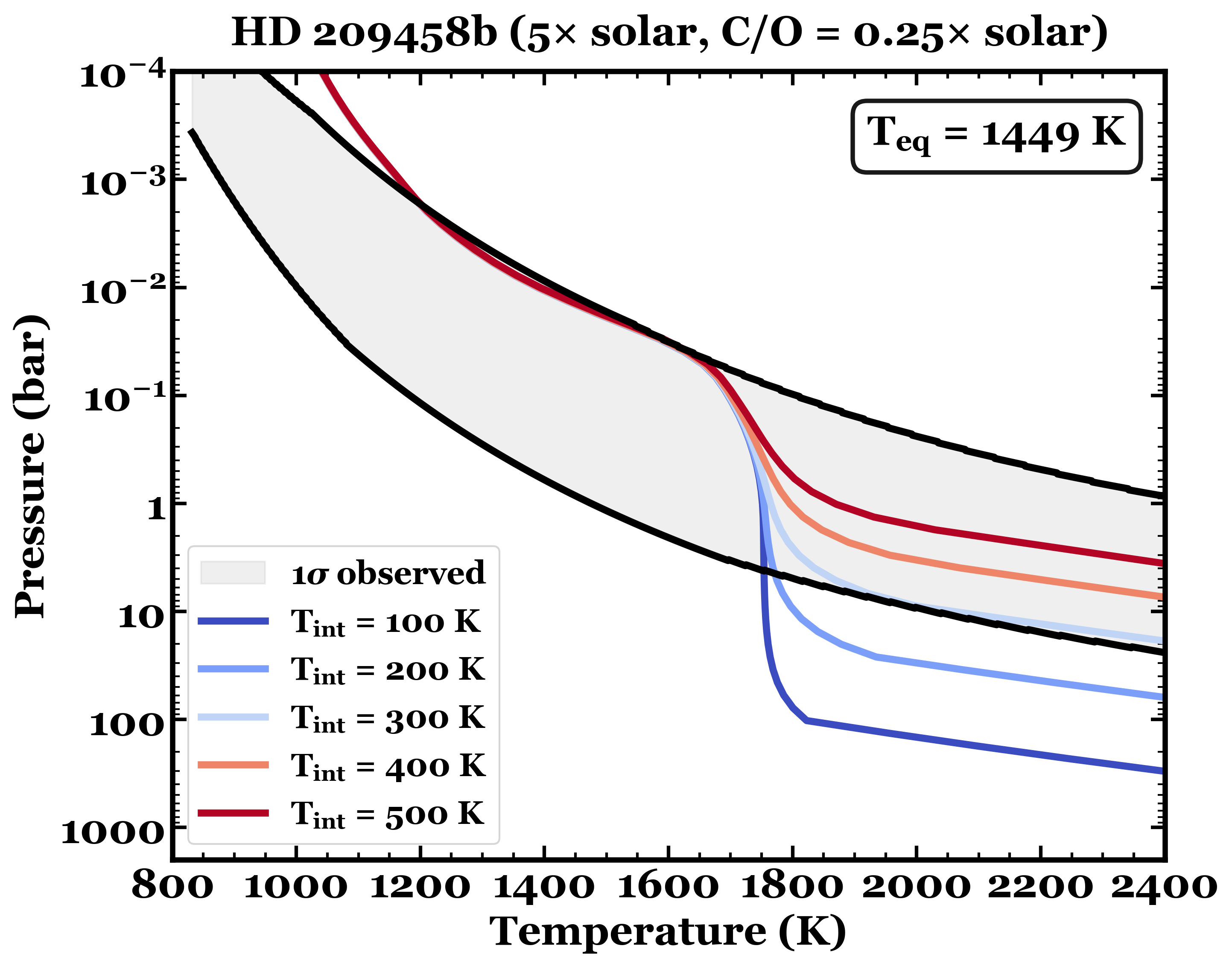}
  \end{subfigure}
  \hfill
  \begin{subfigure}{.49\textwidth}
    \centering
    \includegraphics[width=\textwidth]{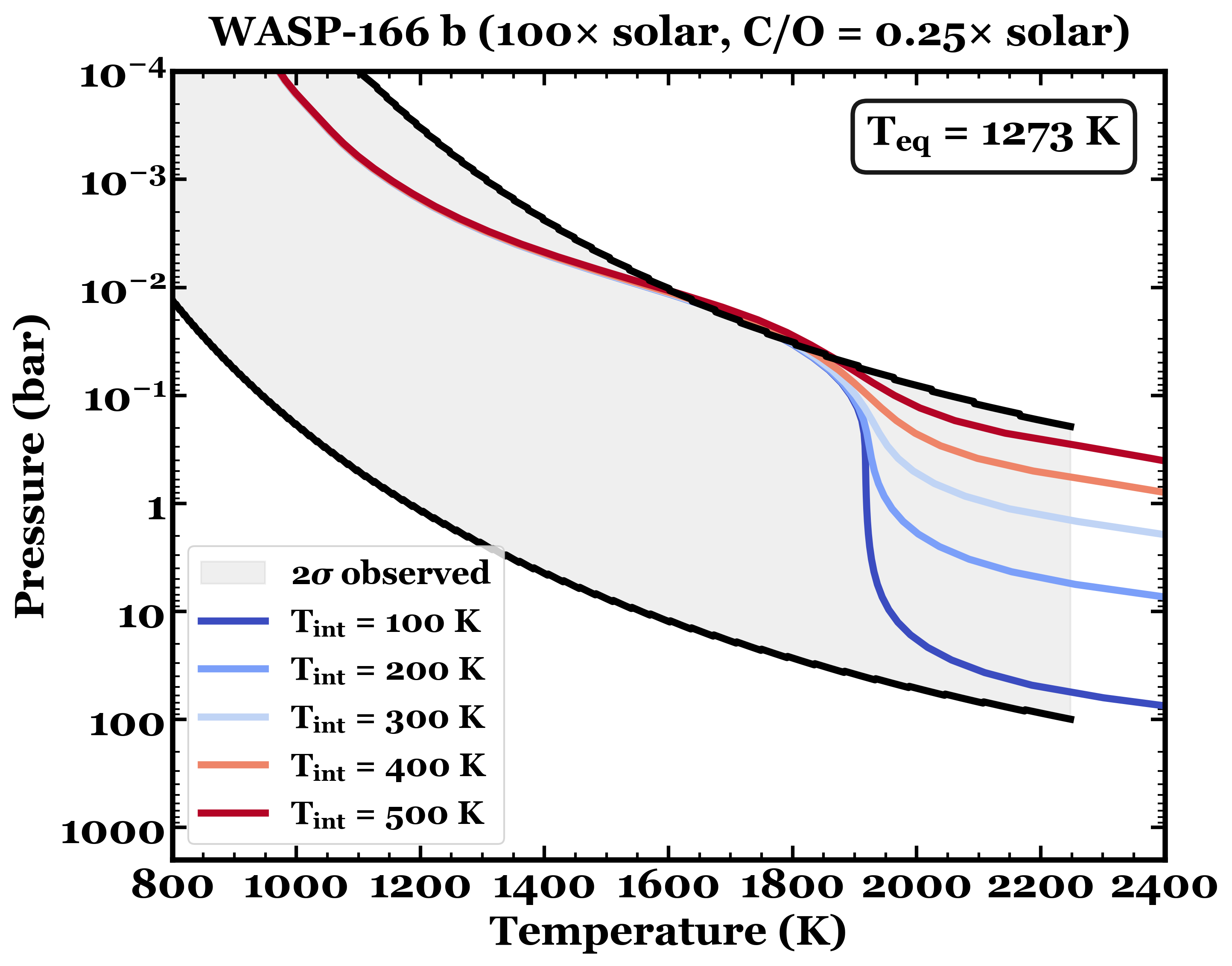}
  \end{subfigure}  
  
  \centering
  \begin{subfigure}{.49\textwidth}
    \centering
\includegraphics[width=\textwidth]{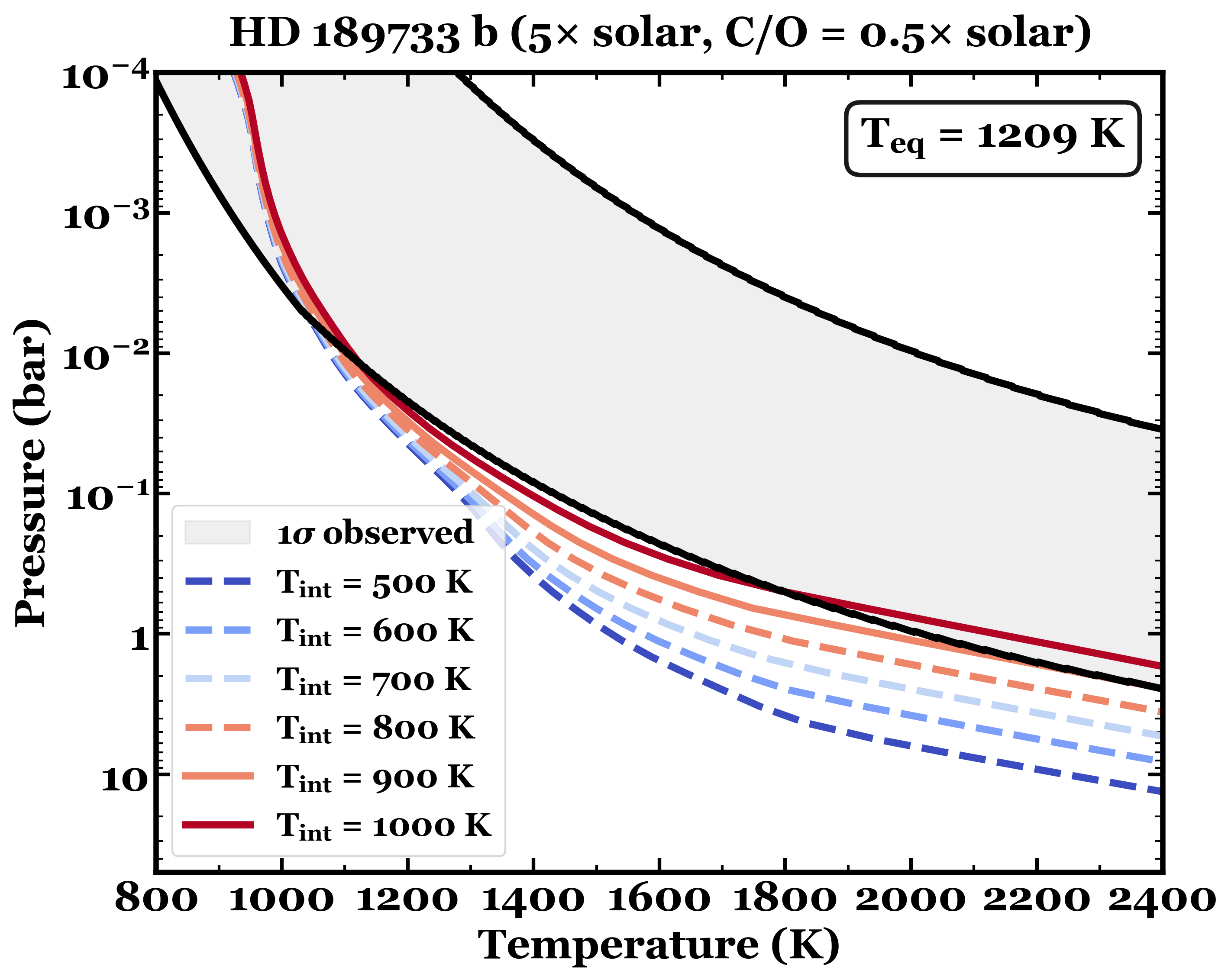}
  \end{subfigure}
  \hfill
  \begin{subfigure}{.49\textwidth}
    \centering
    \includegraphics[width=\textwidth]{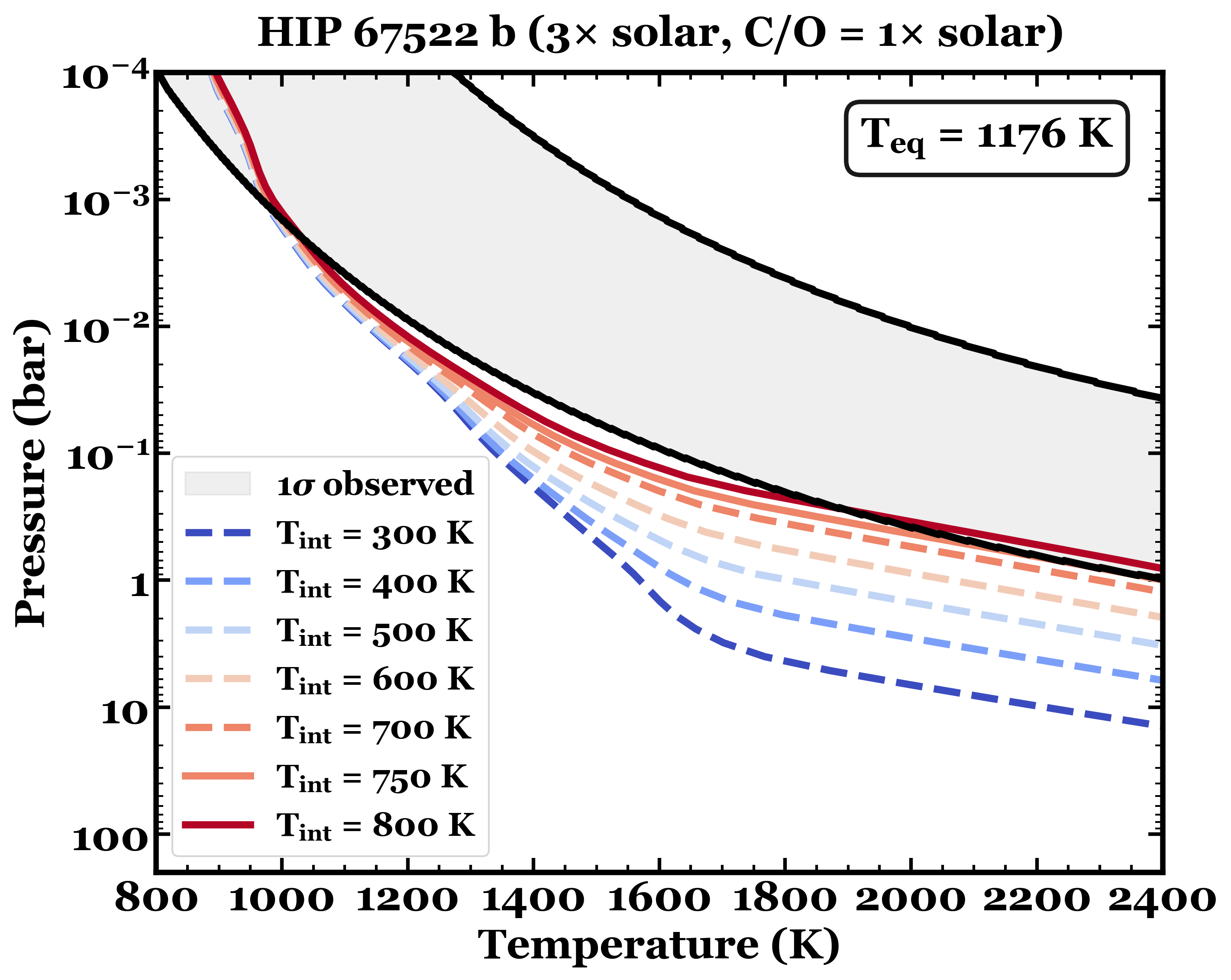}
  \end{subfigure}
  
  \begin{subfigure}{.49\textwidth}
    \centering
    \includegraphics[width=\textwidth]{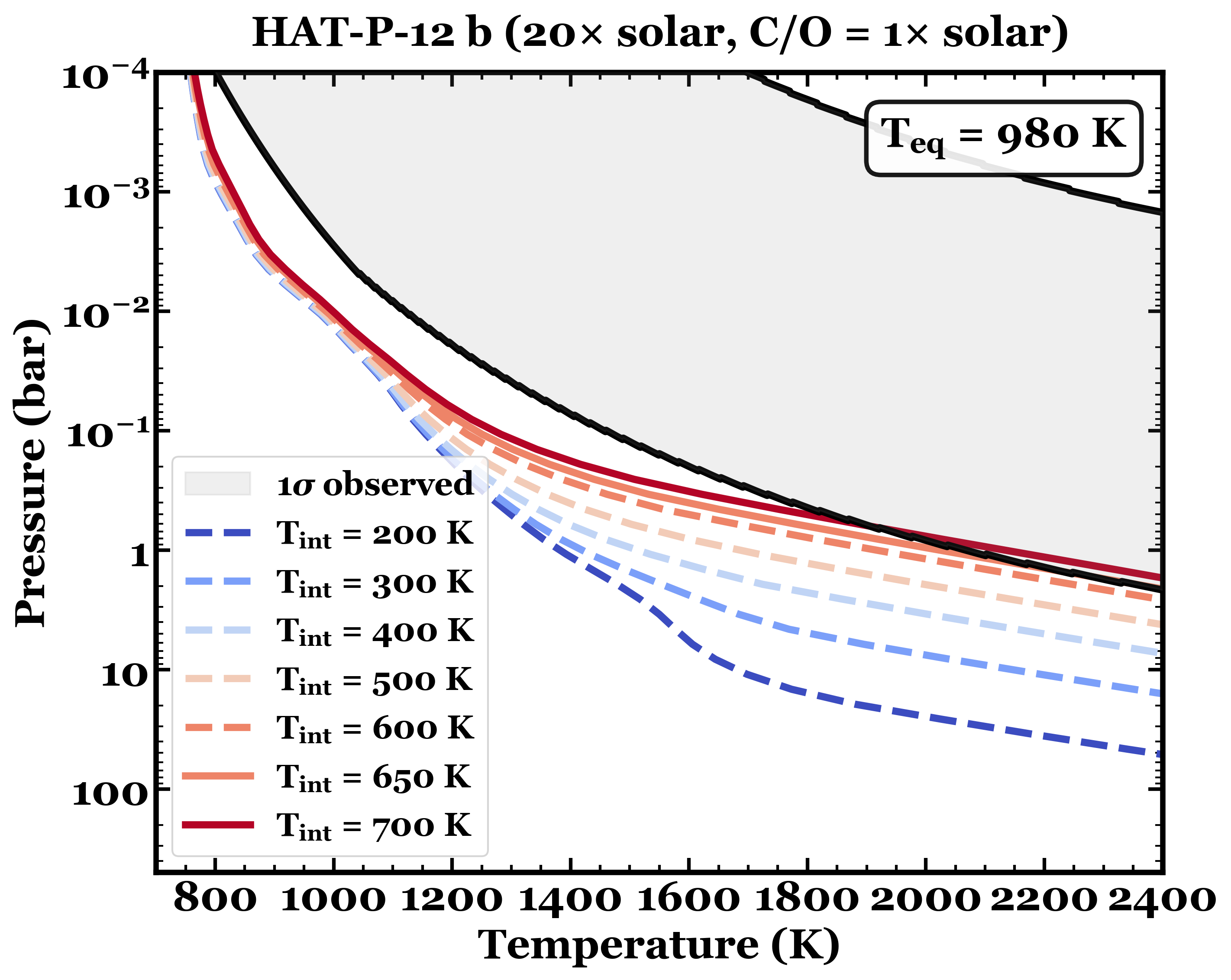}
  \end{subfigure}
  \hfill
  \begin{subfigure}{.49\textwidth}
    \centering
    \includegraphics[width=\textwidth]{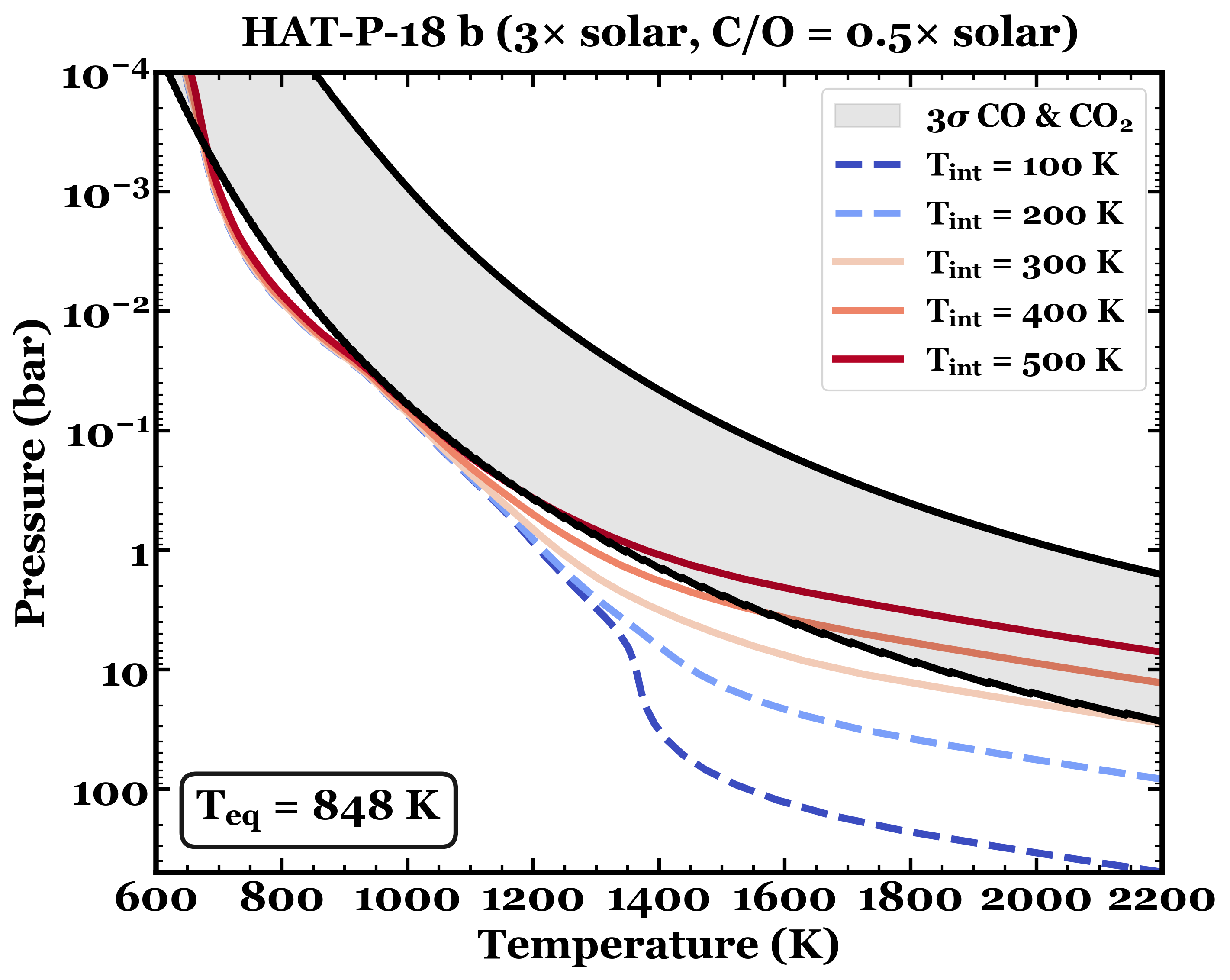}
  \end{subfigure}
  \caption{Results for exoplanets with transmission spectroscopy data. The subplots are of the same form as Figure~\ref{fig:wasp107b_atmo}, but with elemental abundances for the atmospheric P-T profiles and the chemically-derived P-T region calculated using the retrieval results listed in Appendix Table~\ref{table:vmr}.}
  \label{fig:yes_targets}
\end{figure}

Now that we have benchmarked our simple model against the published results of WASP-107 b, we may apply this methodology to the remaining eleven targets. Similar to WASP-107 b, we typically use retrieval abundances while considering their 1$\sigma$ uncertainties to define the chemically-derived P-T region for each exoplanet. The adopted VMRs of the key C-H-O species used in the calculations are summarized in Appendix Table~\ref{table:vmr}. We also summarize calculated and adopted elemental abundances, which are used as inputs for the climate models, in Appendix Table~\ref{table:elementalratio}. The results are shown in Section~\ref{sec:transmission} for exoplanets with transmission spectroscopy observations and in Section~\ref{sec:emission} for exoplanets with emission spectroscopy observations. 

\subsubsection{Targets with transmission spectroscopy data}  \label{sec:transmission}

Transmission spectroscopy probes the limbs of planets, which samples an average of the morning and evening terminators and thus partially integrates over both hemispheres. Although some planets have been shown to exhibit limb asymmetries \citep{2025ApJ...989L..17F}, most targets in our sample do not have known asymmetries. Therefore, for planets with transmission spectroscopy data, we compute P-T profiles assuming globally averaged irradiation, corresponding to uniform redistribution of stellar flux across the entire planetary surface (i.e., r$_\mathrm{st} = 0.5$ in \texttt{PICASO}). 

For the hottest (T$\rm_{eq}>1300$~K) target with transmission spectroscopy data, HD 209458 b, the chemically-derived P-T regions intersect every generated atmospheric P-T profile, regardless of the choice of T$\rm_{int}$. These intersections span all the way from the upper atmosphere ($\sim$10$^{-3}$~bar) down to 1-10 bar. This result indicates that the observed atmospheric compositions are consistent with equilibrium chemistry across a wide range of pressures and temperatures. Thus, similar to conclusions for the retrievals done in \citet{2024ApJ...963L...5X} and \citet{2025arXiv250616232B}, with current observational data, we cannot distinguish between equilibrium chemistry and disequilibrium chemistry driven by transport-induced quenching. The former would produce crossings higher up in the atmosphere (10$^{-3}$-10$^{-1}$~bar), while the latter have crossings deeper in the atmosphere ($>$10$^{-1}$~bar). Currently, both solutions remain viable with the present data. A kinetic argument or a different chemical tracer is needed to break this degeneracy. 

However, in such hot atmospheres, CO is naturally preferred over CH$_4$, and the CH$_4$ abundance would need to be far below current upper limits before any meaningful T$\rm_{int}$ constraints could be obtained. In any event, there is no ``missing" methane for this exoplanet. This finding agrees with forward modeling results from \citet{2023MNRAS.522.2525A}, which show that T$\rm_{int}$ minimally affects the observable atmospheric compositions of hot explanets with T$\rm_{eq}>1300$~K.

For WASP-166 b, we were unable to find a viable chemically-derived P-T solution based on the retrieved 1$\sigma$ atmospheric composition. This inconsistency arose because the POSEIDON retrieval reports high CO$_2$ and low CO abundances, similar to the NEMESIS retrieval for WASP-107 b. The retrieved H$_2$O-CO-CH$_4$ combinations could not reproduce the CO$_2$ abundance within its 1$\sigma$ range, leading our model to reject all possible combinations of these C-H-O species. Therefore, WASP-166 b represents an ideal target for future investigations using multiple independent retrieval pipelines and photochemical-transport modeling. We repeated our calculations with 2$\sigma$ uncertainties, which yields a solution that we present here. Similar to HD 209458 b, WASP-166 b shows a broad crossing region that spans from the observed pressure levels to the deep atmosphere, regardless of the choice of T$\rm_{int}$. Thus, we also cannot distinguish between equilibrium and disequilibrium scenarios given the current observations. 

The next two lower temperature exoplanets around 1200~K, HD 189733 b and HIP 67522 b, are more illuminating. Both are significantly cooler than HD 209458 b and WASP-166 b, but still feature missing methane (see also Figure~\ref{fig:missing_methane}). Unlike the hottest two targets, whose chemically-derived P-T regions form a continuous band from low to high pressures, HD 189733 b and HIP 67522 b show two separate crossing regions between the chemically-derived P-T relationships and the atmospheric P-T profiles. This behavior indicates that the observed abundances could either reflect 1) equilibrium chemistry at the observed pressures or 2) disequilibrium chemistry, with the observed abundances being quenched from deeper atmospheric layers. 

Now, we attempt to determine which of these scenarios is more plausible. For HD 189733 b, the chemically-derived P-T relationships intersect the P-T profiles near 10$^{-2}$ bar and again at a pressure of $\sim$1 bar. The deeper crossing only occurs for P-T profiles with T$\rm_{int}\geq900$~K. The shallower crossing region is consistent with the observed abundances of C-H-O species assuming thermochemical equilibrium, which is the solution that \citet{2024Natur.632..752F} preferred. In their retrieval, the adopted atmospheric P-T profile is nearly isothermal to $\sim$1~bar, so enhanced vertical mixing would actually transport more CH$_4$ upward (since CH$_4$ is the preferred form of carbon over CO at higher pressures under isothermal conditions), which is contrary to the observed depletion of CH$_4$ in the upper atmosphere. In contrast, our models with higher T$_\mathrm{int}$ predict deeper, hotter quench points that naturally suppress CH$_4$. This difference arises because our models use more realistic atmospheric P-T profiles with extended convective zones instead of isothermal assumptions. 

In Figure~\ref{fig:hd189_forward}, it can be deduced that indeed both chemical equilibrium at the observed pressure levels (thick gray lines) or chemical disequilibrium resulting from a hot interior with T$\rm_{int}=900$~K (colored lines) could explain the observed C-H-O species distribution. However, the shallow crossing region is unlikely to represent the quench point since previous works demonstrated that quenching typically occurs at pressures $\gtrsim0.1$~bar, even for low K$_{zz}$ values \citep{2011ApJ...738...72V, 2014ApJ...797...41Z}. The reaction kinetics are inhibited at low gas densities. We therefore adopt the second deeper crossing region as the physically meaningful solution that is consistent with observations. This logic leads to a minimum T$\rm_{int}$ of 900~K for HD 189733 b. One might wonder whether CH$_4$ and CO$_2$ could re-equilibrate in the shallow region after being quenched deeper down; we consider this scenario to be thermodynamically possible but kinetically unlikely, given previous findings such as those of \citet{2011ApJ...738...72V}. HIP 67522 b shows a similar pattern as HD 189733 b, with an initial low-pressure crossing at $\sim10^{-3}$~bar and a second, deeper crossing at $\sim$1~bar. As with HD 189733 b, we disfavor the shallow low-pressure crossing region for the same reason and adopt the deeper crossing solution, which requires a minimum T$\rm_{int}$ of 750~K to explain the observed atmospheric composition and the missing methane on HIP 67522 b. 

\begin{figure}
\ContinuedFloat
  \centering
  \begin{subfigure}{.49\textwidth}
    \centering
    \includegraphics[width=\textwidth]{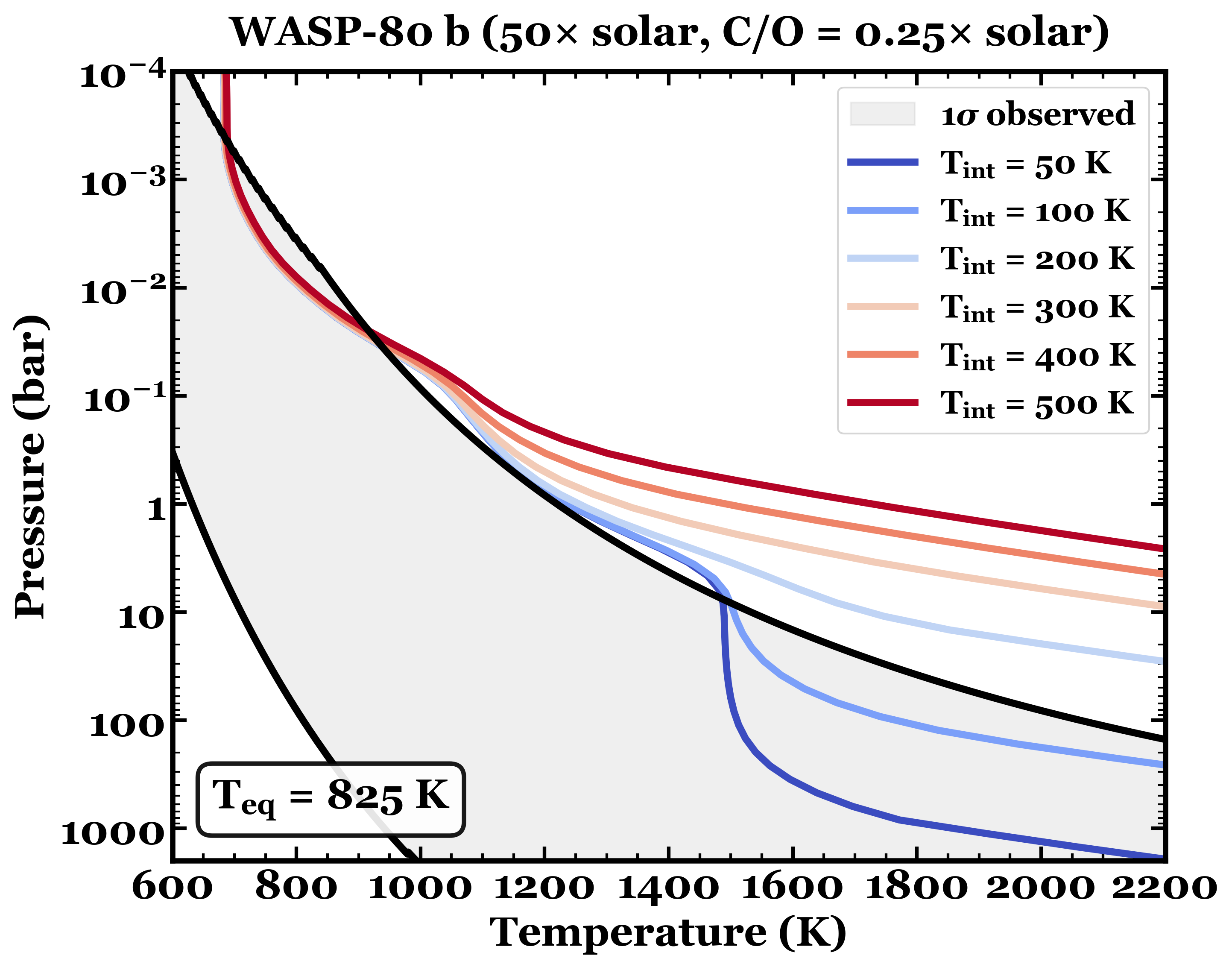}
  \end{subfigure}
  \hfill
  \begin{subfigure}{.49\textwidth}
    \centering
    \includegraphics[width=\textwidth]{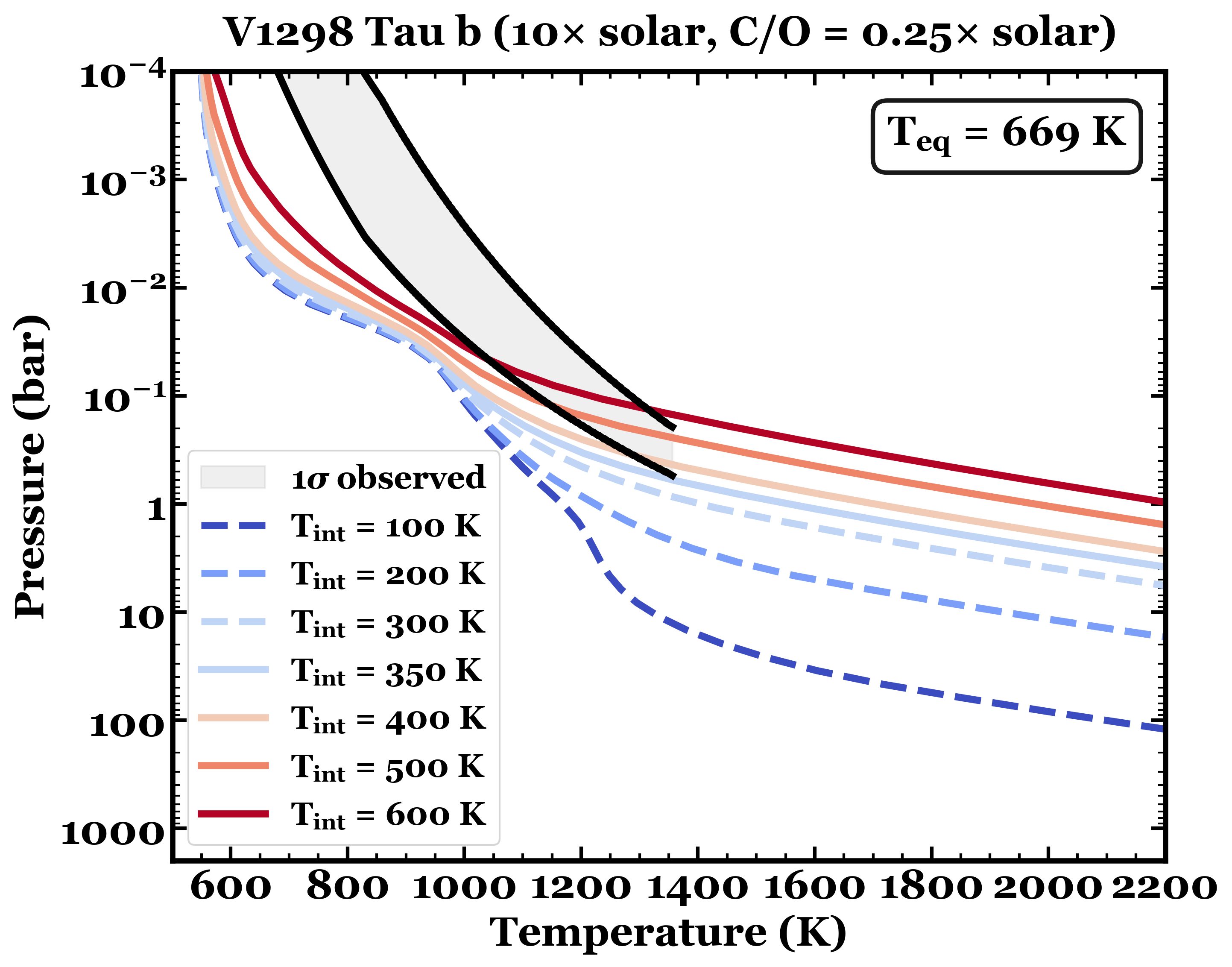}
  \end{subfigure}

    \begin{subfigure}{.49\textwidth}
    \centering
    \includegraphics[width=\textwidth]{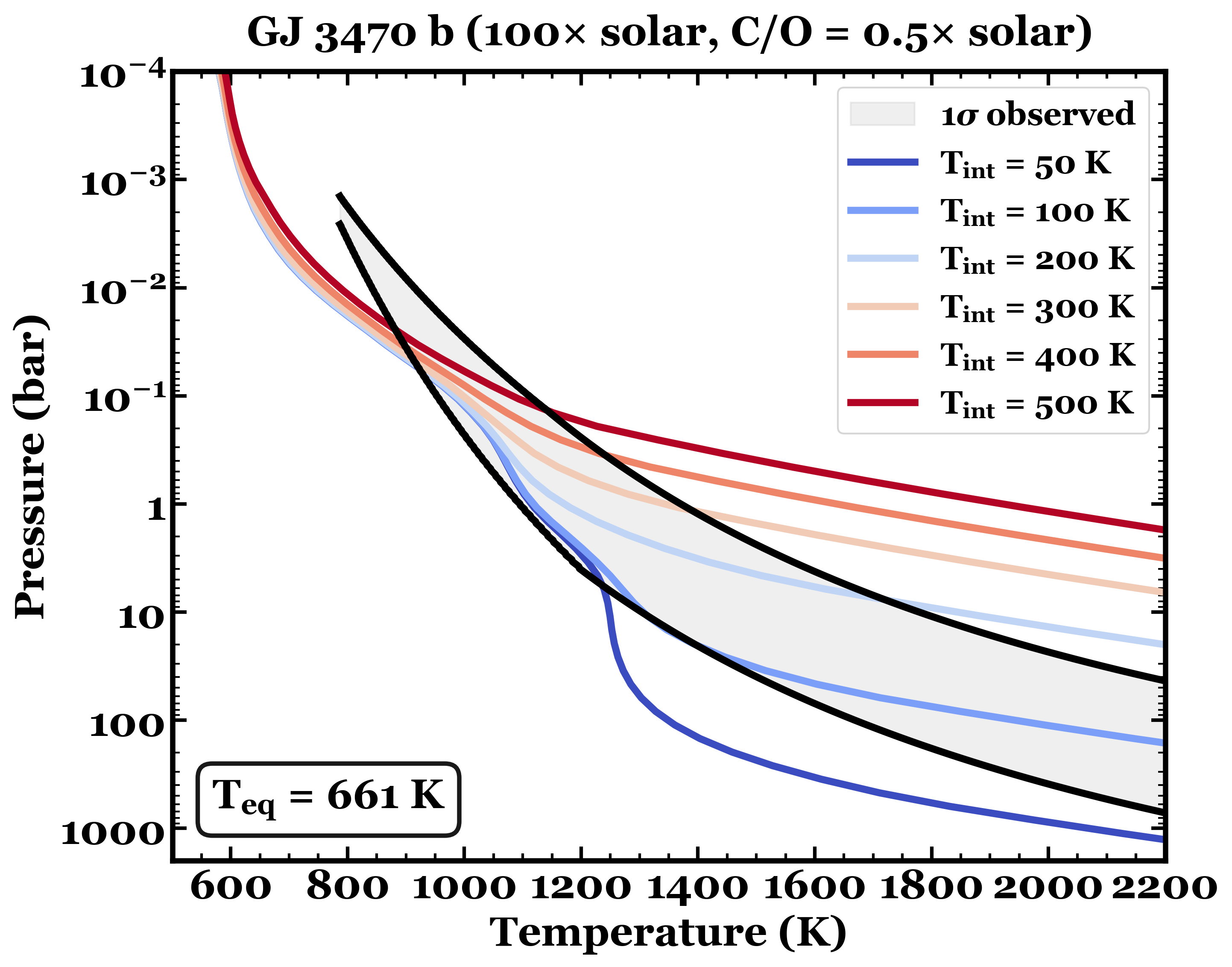}
  \end{subfigure}
  \hfill
  \begin{subfigure}{.49\textwidth}
    \centering
    \includegraphics[width=\textwidth]{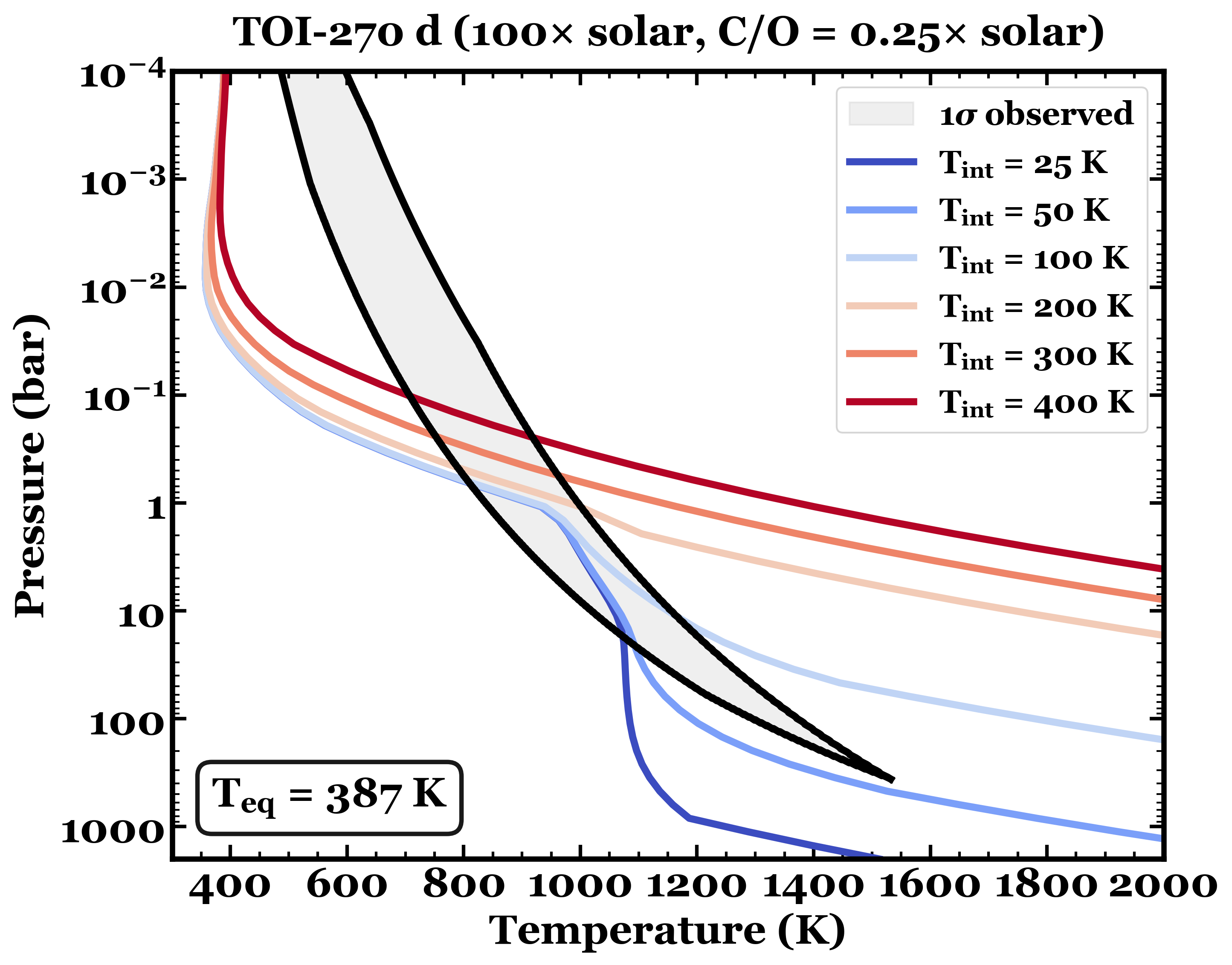}
  \end{subfigure}
  \caption{Figure~\ref{fig:yes_targets} (continued). Results for the remaining exoplanets with transmission spectroscopy data.}
\end{figure}

Moving on to cooler targets with T$\rm_{eq}$ below 1000 K, HAT-P-12 b (T$\rm_{eq}=980$~K) and HAT-P-18 b (T$\rm_{eq}=848$~K), both are ``missing-methane" planets \citep{2022ApJ...940L..35F, 2025AA...703A.264C}. For HAT-P-12 b, we use the joint retrieval results of JWST/NIRSpec and HST/WFC3 \citep{2025AA...703A.264C}. The chemically-derived P-T region and the atmospheric P-T profiles exhibit a crossing region around 1~bar for hotter P-T profiles with T$\rm_{int}\geq650$~K. Similar to the ATMO retrieval of WASP-107 b, this allows us to constrain the minimum T$\rm_{int}$ of HAT-P-12 b to be 650~K. The forward modeling in \citet{2025AA...703A.264C} concluded that T$\rm_{int}>800$~K is necessary to explain the methane depletion on HAT-P-12 b, which is broadly consistent with our derived minimum  T$\rm_{int}$. We also ran forward disequilibrium models to verify this minimum T$\rm_{int}$ solution. The results can be found in the Appendix section (see Figure~\ref{fig:HATP12b}).

For HAT-P-18 b, using the 1$\sigma$ retrieved C-H-O species abundances from \citet{2022ApJ...940L..35F} would only lead to a shallow crossing region at $\sim10^{-4}$~bar (see Figure~\ref{fig:HATP18b} in the Appendix section), which would indicate that the observed abundances are consistent with thermochemical equilibrium at the observed pressure levels. However, given relatively low temperatures in the atmosphere of HAT-P-18 b, the true quench point is likely to occur at higher pressures. Similar to the NEMESIS retrieval for WASP-107 b, this retrieval also produces a CO$_2$/CO ratio greater than unity, leading our model to reject all the H$_2$O-CO-CH$_4$ combinations at higher pressures and temperatures. Thus, similar to WASP-166 b, HAT-P-18 b is also a prime target for reanalysis using multiple retrieval tools and photochemical modeling. Indeed, \citet{2022ApJ...940L..35F} explored the photochemistry solution and showed that inefficient vertical mixing could reproduce the retrieved atmospheric composition. However, the current JWST data for HAT-P-18 b has limited wavelength coverage (NIRISS/SOSS, up to 2.8~$\mu$m), which cannot tightly constrain the abundance of CO$_2$ or CO, both of which have more diagnostic vibrational modes at longer wavelengths. Since our focus is on testing whether elevated T$\rm_{int}$ could account for missing methane, we expanded the uncertainty ranges for CO and CO$_2$ to 2$\sigma$ and 3$\sigma$ and recomputed the chemically-derived P-T region. Both allow deeper crossings than in the 1$\sigma$ case. For the 2$\sigma$ CO/CO$_2$ case, the minimum T$\rm_{int}$ consistent with observations is $\sim$350~K (see Figure~\ref{fig:HATP18b} in the Appendix section), and for the 3$\sigma$ case it drops to $\sim300$~K (see Figure~\ref{fig:yes_targets}). We therefore adopt 300~K as the minimum T$\rm_{int}$ for HAT-P-18 b under current constraints. Future longer-wavelength JWST/NIRSpec observations (GO 5844) that can better constrain the abundances of CO and CO$_2$ would allow us to distinguish whether photochemistry or high internal heat (or both) is responsible for the observed CH$_4$ depletion, and more tightly constrain the minimum T$\rm_{int}$ in the latter case.

\begin{figure}
    \centering
    \includegraphics[width=0.8\textwidth]{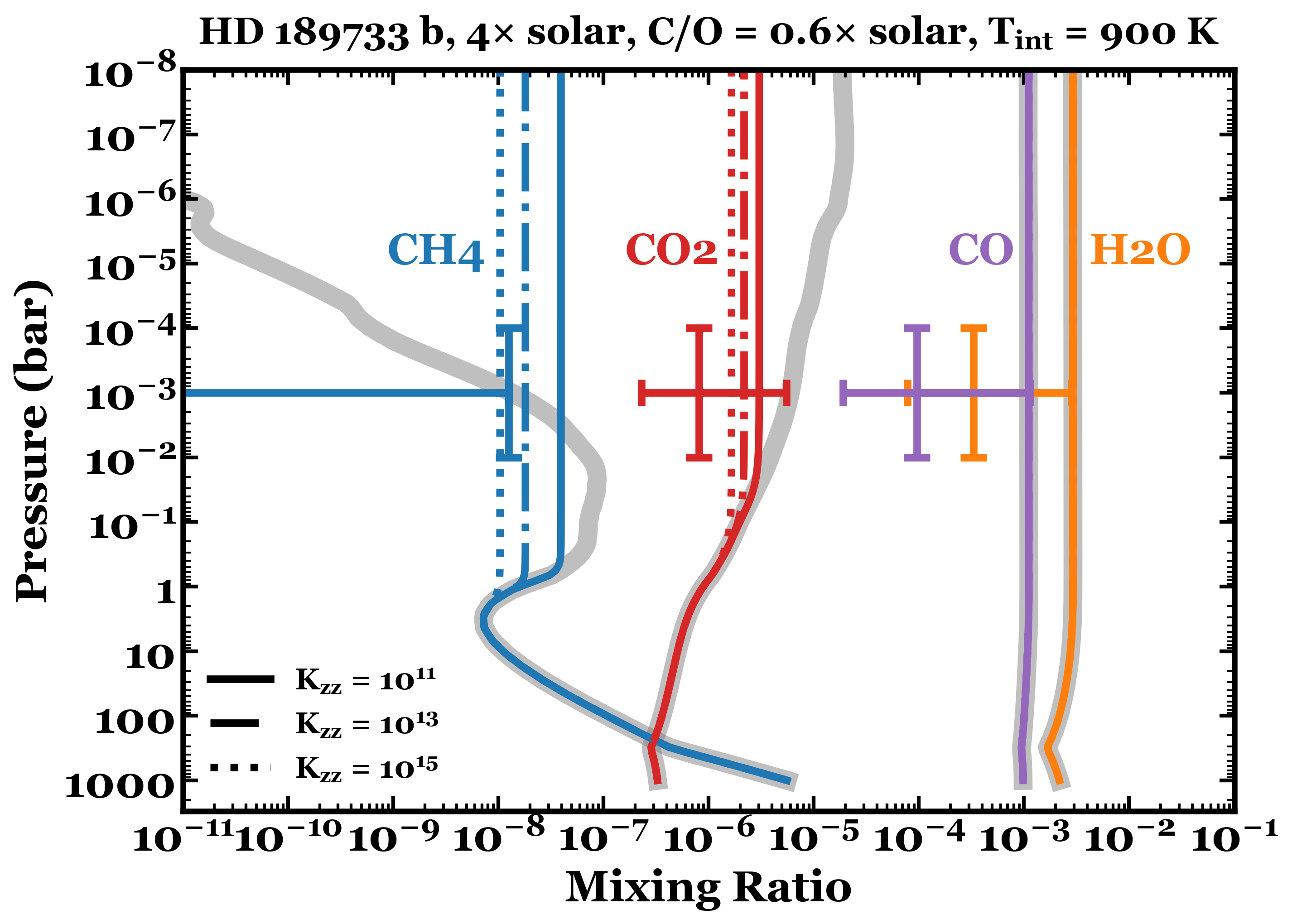}
    \caption{Forward modeling results for HD 189733 b using JWST-observed abundances (with 1$\sigma$ error bars) from \citet{2024Natur.632..752F}. The notations are the same as those in Figure~\ref{fig:wasp107b_forward}. Solid, dashed-dotted, and dotted lines correspond to K$\rm_{deep}$ values of 10$^{11}$, 10$^{13}$, and 10$^{15}$~cm$^2$~s$^{-1}$, respectively.}
    \label{fig:hd189_forward}
\end{figure}

Our next target, WASP-80 b, has a similar equilibrium temperature (T$\rm_{eq}=825$~K) as HAT-P-18 b, but WASP-80 b is one of the first exoplanets confirmed to possess methane in its atmosphere \citep{2023Natur.623..709B, 2025PNAS..12216193W}. Following the same methodology, we find that WASP-80 b's observed atmospheric composition is consistent either with thermochemical equilibrium at the observed pressure levels ($10^{-3.5}-10^{-1.5}$~bar) or with quenching from deeper pressures ($>1$~bar). The atmosphere must not be too hot to keep methane from going missing. Thus, only some of the cooler P-T profiles with T$\rm_{int}\leq200$~K appear to be consistent with the observational data. However, it would be premature to interpret this agreement as an upper limit on T$\rm_{int}$. Because of large 1$\sigma$ uncertainties in the retrieved C-H-O species abundances, a wide range of atmospheric metallicities is possible for this planet. In fact, metallicities as low as 2$\times$ solar are consistent with the data. When we generate P-T profiles for a 2$\times$ solar atmosphere over a range of T$\rm_{int}$ values, we find that, similar to HD 209458 b, we can no longer distinguish equilibrium from disequilibrium scenarios for WASP-80 b. If we assume the disequilibrium scenario is more likely, all T$\rm_{int}$ values for the 2$\times$ solar case would lead to viable quench points. Therefore, given current uncertainty in the measured metallicity of the atmosphere, we adopt a strict minimum T$\rm_{int}$ of 0~K for WASP-80 b, as any T$\rm_{int}$ value could be consistent with the observations. Note that this finding does not mean that WASP-80 b has no internal heat. Rather, it means that there is currently no evidence that heat flow from the interior is affecting the atmospheric composition. Future observations are needed to improve the precision of the mixing ratios of WASP-80 b's atmospheric constituents.

V1298 Tau b is one of the youngest warm exoplanets, with an age of $17\pm2$~Myr, that has been characterized by JWST. Previous grid retrieval using JWST data concluded that a T$\rm_{int}$ of $\sim$500~K is required to explain a lack of methane and the observed abundances of other C-H-O species \citep{Barat2025}. Here, our methodology arrives at a broadly similar conclusion. Like the example of WASP-107 b, the chemically-derived P-T region does not intersect the P-T profiles in the observable atmosphere, indicating that the atmosphere is out of thermochemical equilibrium at those pressures. Intersections would only occur at higher pressures ($\sim10^{-1}$~bar) and only for P-T profiles with elevated T$\rm_{int}$. From these results, we infer a minimum T$\rm_{int}$ of 350~K, lower than the grid-retrieval estimate (500~K). Yet, we find that this minimum T$\rm_{int}$ solution is viable with forward disequilibrium models and present the results in the Appendix section (see Figure~\ref{fig:HATP12b}). This case again highlights the efficiency of our approach, which can recover comparable results without the computational overhead of running extensive forward grid models (in the case of V1298 Tau b, 3960 unique forward models were run in \citet{Barat2025} to determine the best-fit T$\rm_{int}$!).

For our coolest two targets, GJ 3470 b and TOI-270 d, both have methane detected in their JWST transmission spectra \citep{2024ApJ...970L..10B, 2024arXiv240303325B}. Similar to WASP-80 b, the high CH$_4$ abundances observed in their atmospheres push the max-P curves to sufficiently low pressures that they intersect with all P-T profiles, regardless of the T$\rm_{int}$ value. For GJ 3470 b, even though it has a modest eccentricity that could promote some tidal heating, the atmospheric observations alone cannot constrain its T$\rm_{int}$. Our results agree with grid retrievals performed for GJ 3470 b, which recovered a broad allowable T$\rm_{int}$ range of $250\pm200$~K \citep{2024ApJ...970L..10B}. For TOI-270 d, \citet{2025ApJ...985..187G} demonstrated that low-metallicity P-T profiles do not intersect the chemically-derived P-T region. However, at 100$\times$ metallicity, all P-T profiles across the explored T$\rm_{int}$ range produce reasonable crossings. Specific quench conditions can only be derived if we assume a value of T$\rm_{int}$. Similar to GJ 3470 b, we cannot use the C-H-O speciation alone to distinguish which T$\rm_{int}$ values are favored for TOI-270 d. For both planets, we adopt a minimum T$\rm_{int}$ of 0~K. These two planets, along with WASP-80 b, demonstrate that when methane is abundant, the minimum T$\rm_{int}$ effectively drops to zero, making it difficult to place meaningful constraints on the internal heat. Determining an upper limit on T$\rm_{int}$ for these planets is also challenging, as higher T$\rm_{int}$ values can still produce viable crossings at shallower pressures, though not as shallow as the first crossing region of HD 189733 b (i.e., $>10^{-2}$~bar), a behavior that could be explained by slower vertical mixing in the atmosphere.

\subsubsection{Targets with emission spectroscopy data}  \label{sec:emission}

\begin{figure}
  \centering
  \begin{subfigure}{.49\textwidth}
    \centering
    \includegraphics[width=\textwidth]{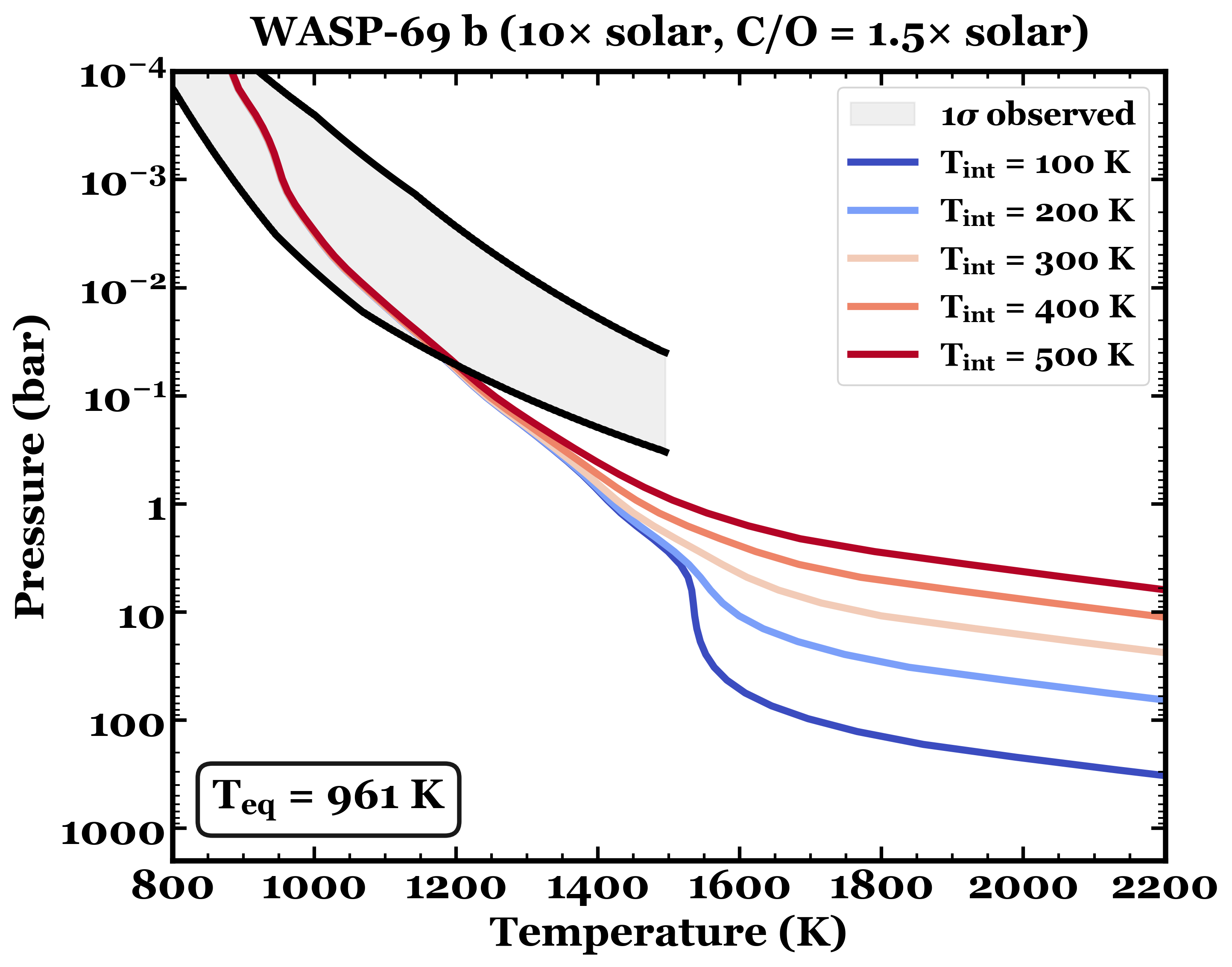}
  \end{subfigure}
  \hfill
  \begin{subfigure}{.49\textwidth}
    \centering
    \includegraphics[width=\textwidth]{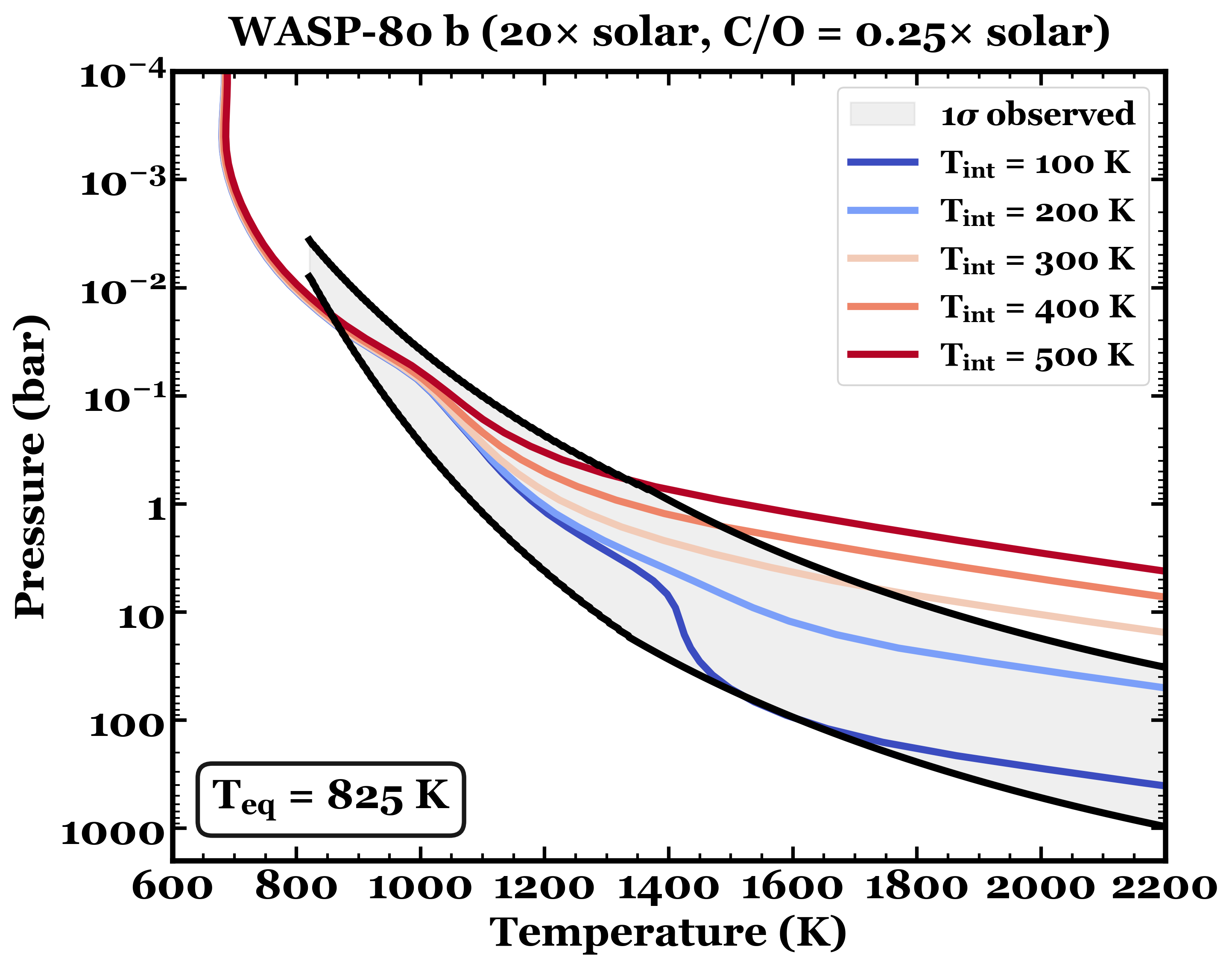}
  \end{subfigure}
  \caption{Results for exoplanets with emission spectroscopy data. The subplots follow the same notations as in Figure~\ref{fig:wasp107b_atmo}.}
  \label{fig:emission}
\end{figure}

We now move on to look at targets with emission spectroscopy measurements: WASP-69 b and WASP-80 b. The atmospheric spectroscopy constraints in these cases are derived from the dayside emission spectra of these planets. Existing transmission and emission observations have shown that hot Jupiter atmospheres are three-dimensional in nature \citep{2025ApJ...989L..17F}. Many of the close-in hot Jupiters are tidally locked to their host stars, producing strong day-night temperature contrasts. For planets with large day-night asymmetries, the emission spectra are controlled primarily by the hotter dayside of the planet. Therefore, for emission-based targets with inefficient heat redistribution, we adopt a stellar heating contribution factor (r$_\mathrm{st}$) of 1, corresponding to no heat redistribution, following \citet{2024AJ....168..104S}. This specification assumes that the dayside atmosphere controls the emission spectrum. If the heat of the planet were evenly distributed, we would assume uniform heat redistribution by using r$_\mathrm{st} = 0.5$ to construct the P-T profiles. The results of our analysis are shown in Figure~\ref{fig:emission}.

WASP-69 b has been inferred to have inefficient day-night heat redistribution, and the temperature on the dayside is likely spatially inhomogeneous. Grid retrievals using a single P-T profile failed to reproduce the observed emission spectra \citep{2024AJ....168..104S}, but using two P-T profiles and combining the two emission spectra led to much better fits of the data. Several additional models were also explored to reproduce the emission data, all assuming chemical equilibrium. Because of the inhomogeneity of the dayside, free retrievals fail to produce reasonable results on molecular abundances (E. Schlawin, private communications). Thus, for our analysis, we adopt the range of mixing ratios that encompasses the full spread of equilibrium models presented in \citet{2024AJ....168..104S}, rather than relying on any single grid retrieval result. This conservative decision is motivated by the fact that individual models tend to have very tight constraints that do not account for uncertainties such as cloud average and temperature variation. As shown in Figure~\ref{fig:emission}, the chemically-derived P-T region only overlaps with the upper atmosphere between 10$^{-4}$ and 10$^{-1}$ bar, which is expected since all three retrievals assumed chemical equilibrium. This consistency demonstrates that our method agrees with existing equilibrium chemistry models. However, because the data only constrain the observable region, we conclude that we cannot constrain the T$\rm_{int}$ for WASP-69 b. Future work should aim to also retrieve the chemical abundances of this planet's atmosphere using the emission spectra without assuming chemical equilibrium. Upcoming transmission spectroscopy data from JWST programs GO 3712 and 5924 may provide additional constraints on T$\rm_{int}$, given that the planet has a high inferred C/O ratio \citep{2022AA...665A.104G}.

WASP-80 b is one of the few exoplanets with both emission and transmission spectra measured over a wide JWST wavelength range \citep{2023Natur.623..709B, 2025PNAS..12216193W}. Unlike WASP-69 b, it shows no evidence for limb asymmetry. Thus, here we generate its P-T profiles assuming uniform redistribution of heat. The emission data of WASP-80 b were analyzed using free retrievals, and we find that the free retrieval data broadly lead to results consistent with those from the transmission retrieval data: all T$\rm_{int}$ values are possible to explain the observed abundances of C-H-O species. However, the chemically-derived P-T region based on the emission data extends to lower pressures compared to the transmission data (see Figure~\ref{fig:yes_targets}). This is because the emission spectrum covers longer wavelength data past 4~$\mu$m, where CO$_2$ and CO are detected, while the transmission spectrum only spans 2.4-4.0~$\mu$m \citet{2023Natur.623..709B}. The emission retrieval thus yields a substantially higher CO abundance (log(VMR)~=~-2.49 versus -7.05 in transmission), which pushes the max-P curve to lower pressures than in the transmission case (see Equation~\ref{eq:K1_P}). 

\begin{figure}
    \centering
    \includegraphics[width=0.8\textwidth]{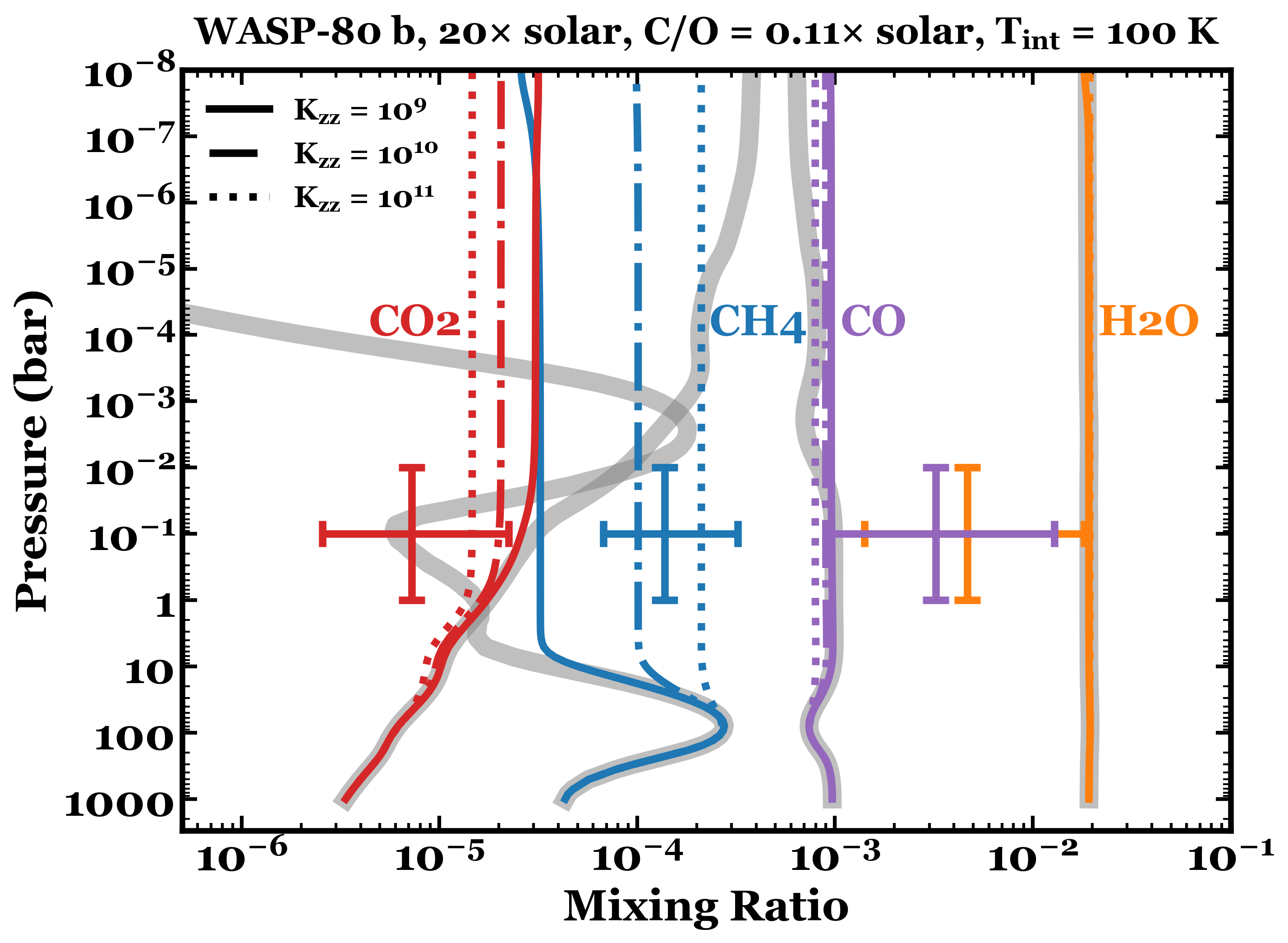}
    \caption{Forward modeling results for WASP-80 b using JWST-observed abundances (with 1$\sigma$ error bars) from \citet{2025PNAS..12216193W}. The notations are the same as in Figure~\ref{fig:wasp107b_forward}. Solid, dashed-dotted, and dotted lines correspond to K$\rm_{deep}$ values of 10$^{9}$, 10$^{10}$, and 10$^{11}$~cm$^2$~s$^{-1}$, respectively.}
    \label{fig:wasp80_forward}
\end{figure}

Although \citet{2025PNAS..12216193W} reported a tightly constrained T$\rm_{int}$ of 381$\pm$38~K from grid retrievals, we caution that this T$\rm_{int}$ range is likely too narrow. Our analysis shows that both lower and higher T$\rm_{int}$ can explain the 1$\sigma$ retrieved abundances of C-H-O species with some values of vertical mixing and C/O ratio. For example, using a P-T profile with T$\rm_{int} = 100$~K, K$\rm_{zz}=10^{10}$~cm$^2$~s$^{-1}$, and C/O ratio of 0.11$\times$ solar, a forward model can also reproduce the observed abundances of key C-H-O species, see Figure~\ref{fig:wasp80_forward}. Much lower T$\rm_{int}$ values remain viable, and the T$\rm_{int}$ range inferred in \citet{2025PNAS..12216193W} may be too narrow due to the limited range of vertical mixing strength and compositional parameters explored from grid retrievals.

\section{Discussion} \label{sec:discussion}

\subsection{Examination of Population-Level Trend for T$\rm_{int}$}

\begin{deluxetable*}{lccccc}
\tablewidth{0pt}
\tablecaption{Comparing Derived T$_{\mathrm{int}}$ in This Work and from the Literature. \label{table:comparison}}
\tablehead{
\colhead{Target\tablenotemark{a}} & 
\colhead{Min T$_{\mathrm{int}}$ [K], this work} & 
\colhead{Previous T$_{\mathrm{int}}$ [K]} & 
\colhead{Ref.} &
\colhead{Eccentricity} &
\colhead{Ref.} 
}
\startdata
HD 189733 b & 900 & $>$500 & \citetalias{2024Natur.632..752F} & 0.027$^{+0.021}_{-0.018}$ & \citetalias{2021ApJS..255....8R} \\
HIP 67522 b & 750 & not provided & \citetalias{2024AJ....168..297T} & 0.0640$^{+0.1870}_{-0.0490}$ & \citetalias{2024AJ....168..297T} \\
HAT-P-12 b & 650 & $>$800 & \citetalias{2025AA...703A.264C} & 0.026$^{+0.026}_{-0.018}$ & \citetalias{2014ApJ...785..126K} \\
HAT-P-18 b & 300 & not provided & \citetalias{2022ApJ...940L..35F} & 0.106$^{+0.150}_{-0.084}$; $<0.048$\tablenotemark{b} & \citetalias{2014ApJ...785..126K}; \citetalias{2025arXiv251107746Y} \\
\multirow{2}{*}{WASP-80 b} & 0 & not provided & \citetalias{2023Natur.623..709B} & \multirow{2}{*}{0.0020$^{+0.010}_{-0.002}$} & \multirow{2}{*}{\citetalias{2015MNRAS.450.2279T}} \\
 & 0 & 381$\pm$38 & \citetalias{2025PNAS..12216193W} & & \\
\multirow{2}{*}{WASP-107 b} & 450 & $460\pm40$ & \citetalias{2024Natur.630..831S} & 0.06$\pm$0.04; $<0.081$\tablenotemark{b}; 0.09$\pm$0.02 & \citetalias{2021AJ....161...70P}; \citetalias{2025arXiv251107746Y}; \citetalias{2026ApJ...996L..28W}  \\
 & 170 or 270 & $>345$ & \citetalias{2024Natur.630..836W} & 0.06$\pm$0.04 & \citetalias{2021AJ....161...70P} \\
V1298 Tau b & 350 & $\sim$500 & \citetalias{Barat2025} & 0.134$\pm$0.075  & \citetalias{2022NatAs...6..232S} \\
GJ 3470 b & 0 & $250\pm200$ & \citetalias{2024ApJ...970L..10B} & 0.114$\pm$0.051  & \citetalias{2019AJ....157...97K} \\
TOI-270 d & 0 & not provided & \citetalias{2024arXiv240303325B} & 0.032$\pm$0.023 & \citetalias{2021MNRAS.507.2154V} \\
\enddata
\tablenotetext{a}{Note that we cannot constrain the T$\rm_{int}$ values for HD 209458 b, WASP-166 b, and WASP-69 b using our framework based on existing JWST data (see Section~\ref{sec:crowd}).}
\tablenotetext{b}{3$\sigma$ upper limit value.}
\end{deluxetable*}

Among the 12 targets that we investigated, we are able to provide constraints on the minimum T$\rm_{int}$ for 9 targets. The results are summarized in Table~\ref{table:comparison}. Note that these constraints only apply if observed species abundances reflect dynamically quenched chemical disequilibria from the deeper atmosphere. This is the core assumption of our approach. For planets with existing T$\rm_{int}$ constraints from grid modeling (HAT-P-12 b, WASP-107 b, V1298 Tau b, and GJ 3470 b), our results are generally consistent with previous findings, although our minimum T$\rm_{int}$ values are typically lower, and therefore more conservative, than those previously reported in the literature. Our methodology casts a wide net since it does not explicitly impose a K$\rm_{zz}$ constraint; as long as chemical equilibrium for the four key C-H-O species can be achieved at some depth, there exists some corresponding K$\rm_{zz}$ values, whether physical or not, that would allow quenching to be achieved from that depth. 

\begin{figure}
  \centering
  \begin{subfigure}{.5\textwidth}
    \centering
    \includegraphics[width=\textwidth]{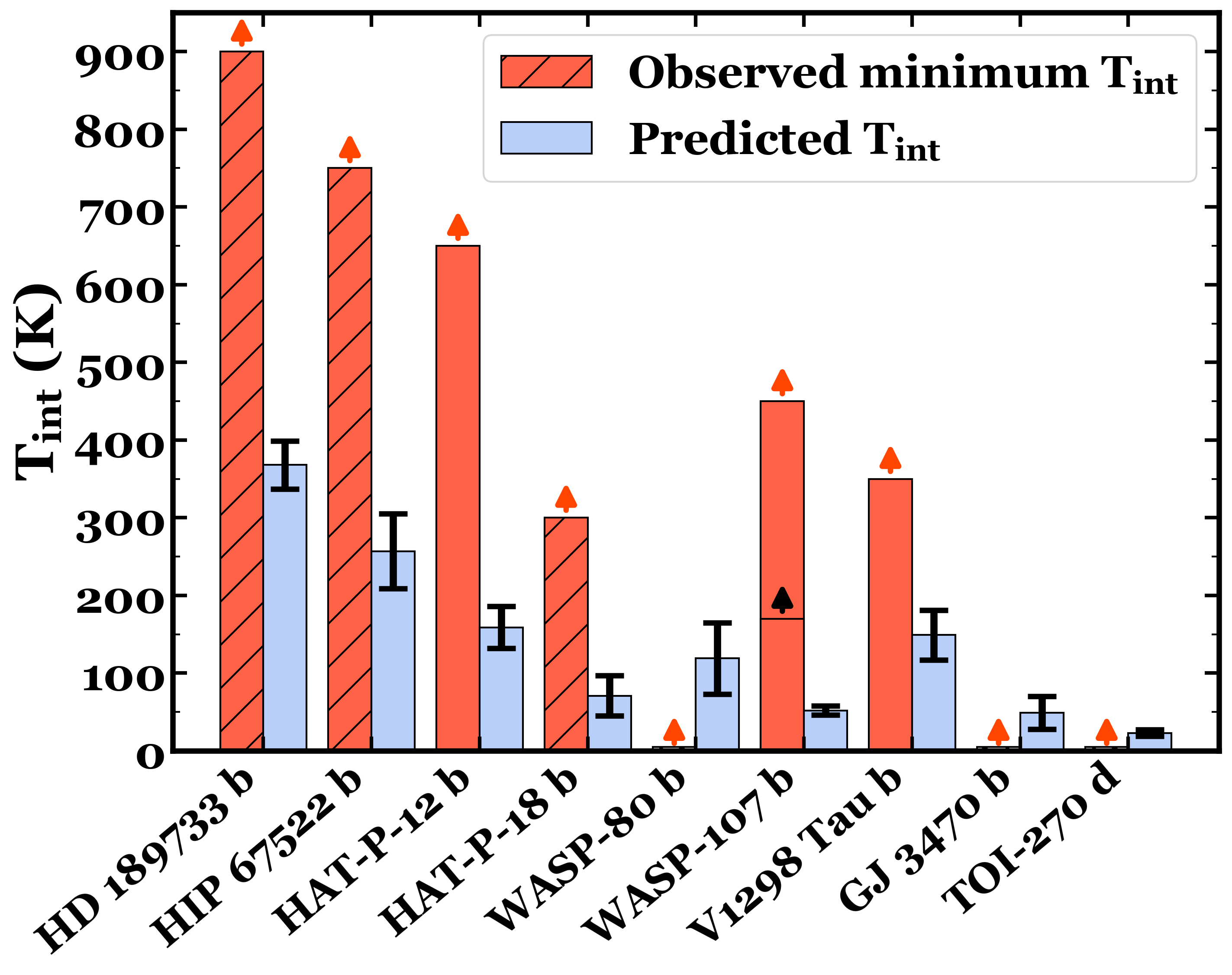}
  \end{subfigure}
  \begin{subfigure}{.5\textwidth}
    \centering
   \includegraphics[width=\textwidth]{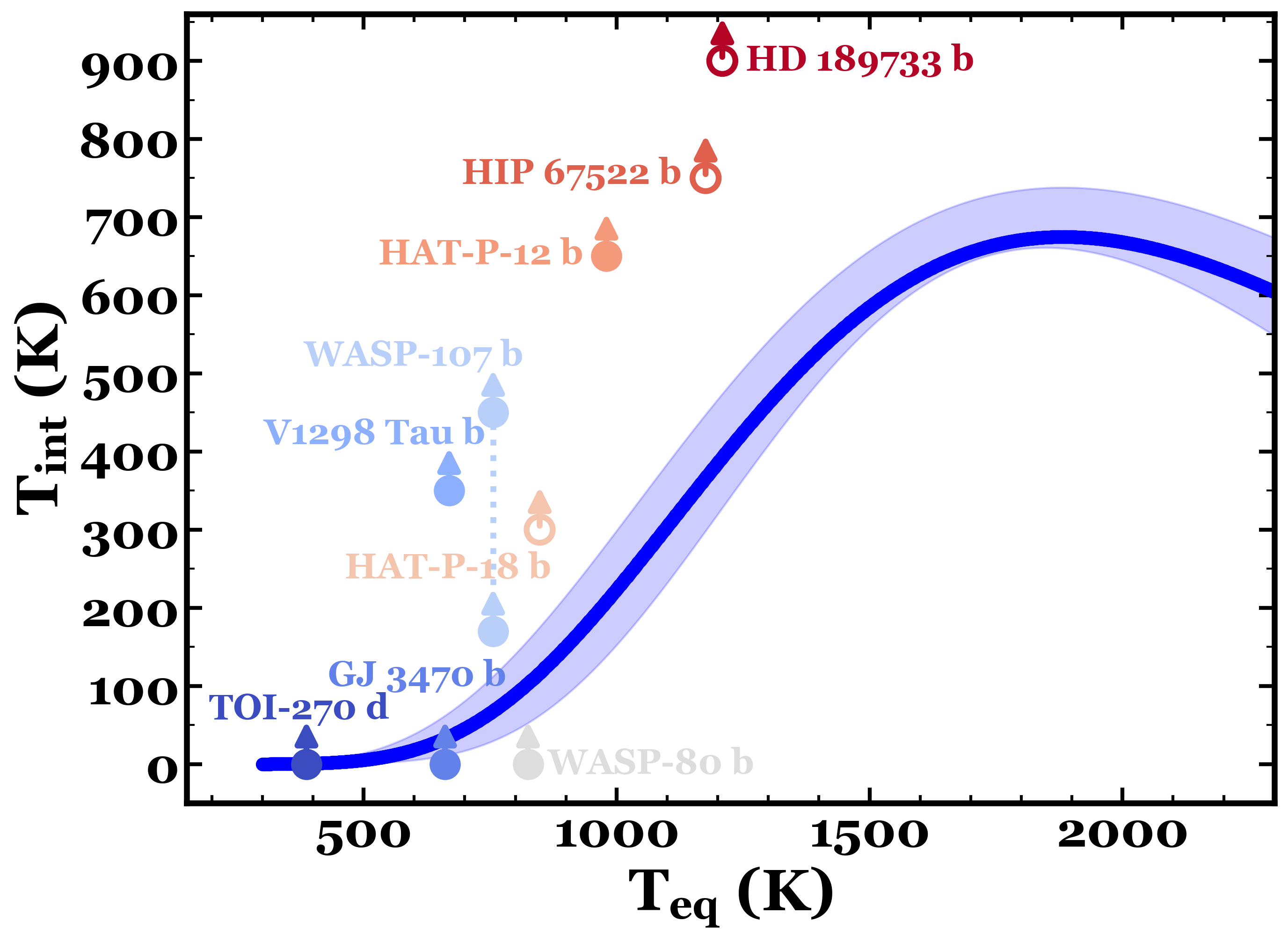}
  \end{subfigure}
    \begin{subfigure}{.5\textwidth}
    \centering
    \includegraphics[width=\textwidth]{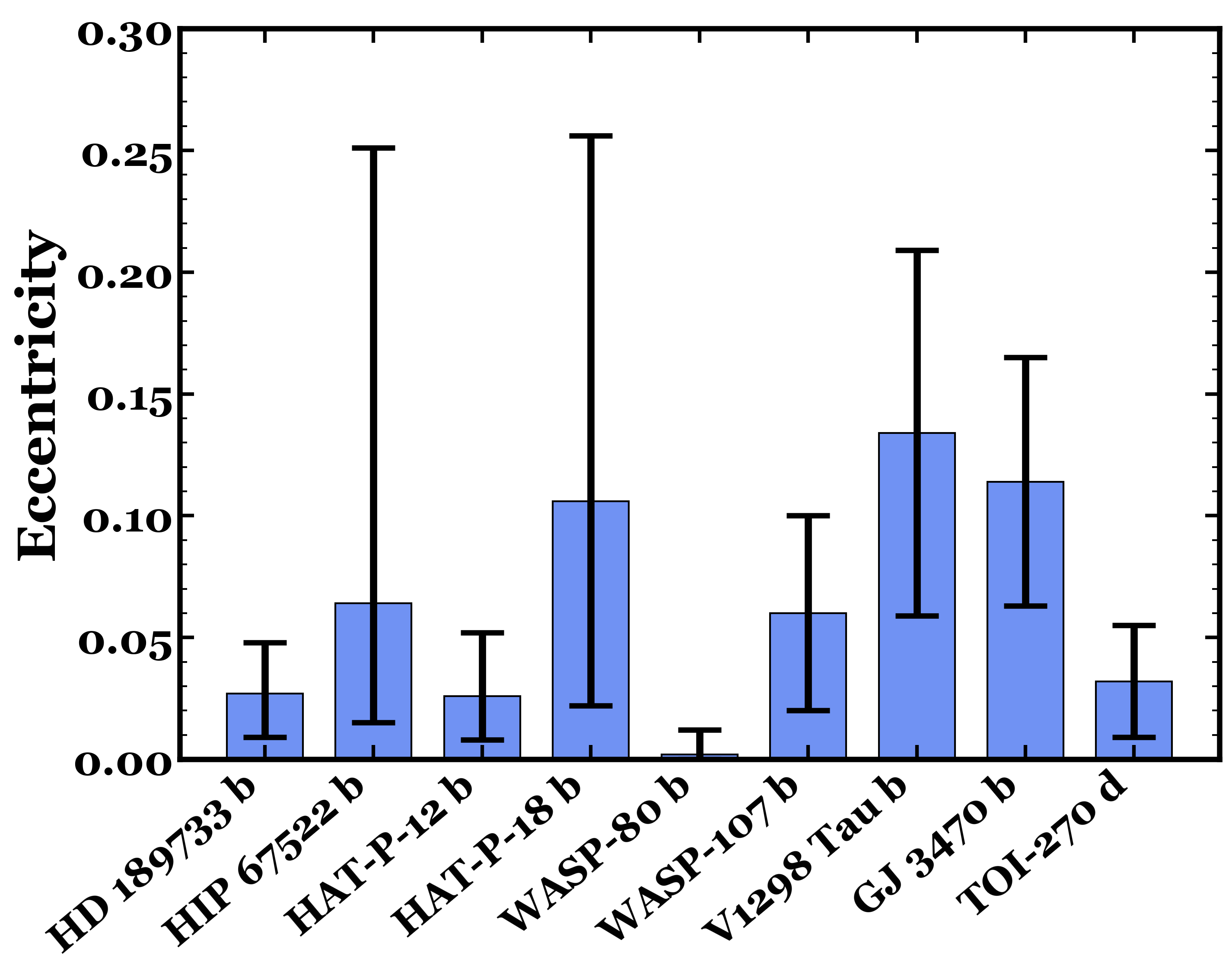}
   \end{subfigure}
    \caption{(Top) Comparison between the minimum T$\rm_{int}$ values derived using the method in this work and the T$\rm_{int}$ values predicted by thermal evolution models. Red bars denote the minimum T$\rm_{int}$ values, with the upward arrows indicating that the true T$\rm_{int}$ may be higher because we determined lower limit values. Blue bars denote the evolution model predictions. Hatched red bars identify planets for which alternative solutions (e.g., photochemistry or in-situ thermochemical equilibrium) can also reproduce the observed atmospheric composition, indicating weaker constraints on T$\rm_{int}$. (Middle) Inferred minimum T$\rm_{int}$ values for the nine exoplanets compared with the empirical T$\rm_{eq}$-T$\rm_{int}$ relationship from \citet{2018AJ....155..214T, 2019ApJ...884L...6T}. Note that the points correspond to the same minimum T$\rm_{int}$ values shown in the top panel, with open symbols denoting planets with degenerate solutions. The top panel shows planet-by-planet individual model comparisons, while the middle panel places them in the context of a population-level trend. (Bottom) Orbital eccentricities of the exoplanets with T$\rm_{int}$ constrained in this work, along with their 1$\sigma$ uncertainties, using data from Table~\ref{table:comparison}. Downward triangles indicate 3$\sigma$ upper limit values from \citet{2025arXiv251107746Y}.}
  \label{fig:evolution}
\end{figure}

For example, our forward modeling shows that matching the 1$\sigma$ retrieval results from WASP-107 b's AURORA and CHIMERA retrievals requires extraordinarily high K$\rm_{zz}$ values based on our derived minimum T$\rm_{int}$ (Figure~\ref{fig:WASP107b_aurora}). The discrepancy between the grid-model results and our minimum T$\rm_{int}$ tends to be smaller when the quench points are achieved at lower pressures (e.g., WASP-107 b's ATMO retrieval, Figure~\ref{fig:wasp107b_atmo}), because the K$\rm_{zz}$ ranges used in grid retrievals are typically large enough to encompass the values needed to achieve quenching at those pressures. For HD 189733 b, HIP 67522 b, and HAT-P-18 b, while our high T$\rm_{int}$ solutions could explain the atmospheric observations, there are degeneracies with alternative explanations such as thermochemical equilibrium (HD 189733 b and HIP 67522 b) or photochemistry (HAT-P-18 b) in the upper atmosphere. We therefore regard the T$\rm_{int}$ constraints for these targets as weaker and distinguish the confidence levels in the subsequent trend analysis (Figure~\ref{fig:evolution}).


With these caveats in mind, we seek a population-level understanding of the interior heating mechanisms for these exoplanets by examining the trend of their constrained minimum T$\rm_{int}$ values. What is the bigger picture? Guided by the thermal evolution model of \citet{2007ApJ...659.1661F, 2016ApJ...831...64T, 2019ApJ...874L..31T}, we calculated the expected T$\rm_{int}$ for each exoplanet at its current age. The results are shown in the top panel of Figure~\ref{fig:evolution}. It is evident that several exoplanets show apparent T$\rm_{int}$ values exceeding those predicted by standard evolution models. We use the term apparent T$\rm_{int}$ to emphasize that these values are derived from current atmospheric constraints and may evolve as more data and improved retrieval models become available. Planets with elevated apparent T$\rm_{int}$ include HD 189733 b, HIP 67522 b, HAT-P-12 b, HAT-P-18 b, WASP-107 b (when using the minimum T$\rm_{int}$ derived using the \citet{2024Natur.630..831S} dataset and the ATMO free retrieval), and V1298 Tau b. For WASP-107 b, this elevated apparent T$\rm_{int}$ is a known result \citep{2024Natur.630..831S} indicating that the planet may be tidally heated \citep[e.g.][]{2020ApJ...897....7M}; HAT-P-18 b may be in a similar situation \citep{2024ApJ...972..159Y, 2025arXiv251107746Y}; however, broader wavelength coverage data is required to resolve degeneracy with the photochemical interpretation \citep{2022ApJ...940L..35F}. The warm Saturn HAT-P-12 b shows surprisingly high apparent T$\rm_{int}$, consistent with forward modeling results \citep{2025AA...703A.264C}, but without any obvious reason for why that might be. HIP 67522 b and V1298 Tau b are reported as quite young \citep{2024ApJ...973L..30B, Barat2025}, so the difference between theory and the observed minimum may reflect an underestimated initial entropy or some unmodelled physics that causes decreased early cooling rates. On the other hand, the hot Jupiter HD 189733 b is a puzzling case, showing a very high apparent T$\rm_{int}$ despite the planet's unremarkable density. Taken together, these elevated apparent T$\rm_{int}$ suggest that additional interior heating sources beyond standard evolution are likely required for at least a subset of the population. 

Several heating mechanisms have been proposed to explain the elevated interior heating and inflated radii of irradiated warm Neptunes and hot Jupiters, including tidal heating \citep{2001ApJ...548..466B}, Ohmic dissipation \citep{2010ApJ...714L.238B}, downward transport of kinetic energy \citep{2002A&A...385..166S}, enhanced atmospheric opacities \citep{2007ApJ...661..502B}, layered convection \citep{2007ApJ...661L..81C}, and the advection of potential temperature \citep{2017ApJ...841...30T, 2019A&A...632A.114S}. The optimal strategy would be to combine the effects of all of these heating mechanisms and then compare them directly with our chemically-derived apparent T$\rm_{int}$ values. However, in practice, we have only limited exoplanet properties to perform this task, and it is challenging to quantitatively evaluate some of these proposed heating mechanisms. 

\citet{2018AJ....155..214T} used Bayesian analysis to extract the T$\rm_{int}$ from the mass-radius observations and concluded that Ohmic dissipation is likely the dominant heating mechanism in hot-Jupiters. Thus, in this work, we decided to compare our derived minimum T$\rm_{int}$ values with the empirical T$\rm_{int}$-T$\rm_{eq}$ relationship from \citet{2018AJ....155..214T,2019ApJ...884L...6T} as a baseline. The results are shown in the middle panel of Figure~\ref{fig:evolution}. Even though most of our targets follow the general positive trend of this T$\rm_{int}$-T$\rm_{eq}$ relation, several exoplanets show significantly higher T$\rm_{int}$ that deviates from the predicted T$\rm_{int}$. These gaps suggest that even though Ohmic dissipation (or some other processes that mimic its trend) may dominate the internal heating for the general population of hot-Jupiters, additional heating mechanisms appear to be needed to explain the excess T$\rm_{int}$ for some individual targets. As an example, HIP 67522 b has much higher internal heat that is well beyond the predicted T$\rm_{int}$ for Ohmic dissipation. And it is important to note that the system does have a second c planet that is in a 2:1 resonance with planet b \citep{2024ApJ...973L..30B}, which can maintain tidal heating. This type of heating seems inescapable as both planets have non-zero eccentricities. 

Previous works have shown that eccentricity-driven tidal heating may be important on some exoplanets with missing methane (and often also inflated radii) \citep[e.g.,][]{2020ApJ...897....7M}. Indeed, compiling eccentricities for our sample (Table~\ref{table:comparison} and Figure~\ref{fig:evolution} bottom panel) reveals that many of our targets with unusually hot interiors also have non-zero eccentricities, including HD 189733 b, HIP 67522 b, HAT-P-12 b, HAT-P-18 b, WASP-107 b, and V1298 Tau b, which is consistent with a tidal heating contribution. Note also that our methodology can only examine the regime of the \citet{2018AJ....155..214T} T$\rm_{int}$-T$\rm_{eq}$ relationship where T$\rm_{int}$ increases with T$\rm_{eq}$. As exoplanets become increasingly hot, the difference between the atmospheric compositions under chemical equilibrium and disequilibrium at the observed pressure levels gradually diminishes. Thus, more precise constraints on atmospheric composition are needed to distinguish local equilibrium chemistry from disequilibrium chemistry due to quenching, and identifying the internal heat of these hotter exoplanets becomes progressively more difficult the hotter they are \citep{2023MNRAS.522.2525A}.

\subsection{Caveats} 

\subsubsection{The Effect of Photochemistry}\label{sec:photochem}
It is established that thermochemical equilibrium will be reached in the deep, hot part of the atmospheres of giant planets. But at altitudes where vertical transport is faster than the chemical kinetics reaction timescale that restores thermochemical equilibrium, the mixing ratio of a species could be ``quenched." However, photochemical processes can subsequently alter the mixing ratios of species in the upper atmosphere above the quench point \citep[e.g.,][]{2014RSPTA.37230073M}. The exact region where photochemistry affects species abundances is determined by the balance of incident stellar flux and the strength of vertical transport \citep[e.g.,][]{2021ApJ...921...27H}. 

Weaker vertical transport and more incident stellar short-wavelength photons or other high-energy radiation would extend the photochemically active layer deeper into the atmosphere, typically leading to the destruction of species with weak bonds (among the four species that matter here, CH$_4$ and H$_2$O) and the formation of new photochemical products (among the four species, CO and CO$_2$). Thus, the photochemically-active region would lead to enhanced mixing ratios of CO and CO$_2$ and decreased mixing ratios of CH$_4$ and H$_2$O relative to their quenched mixing ratios. By examining Equation~\ref{eq:K1_P}, we can conclude that, to first order, at the same temperature, photochemistry would lead to a decreased pressure if the observed region is photochemically active. The chemically-derived P-T region would shift down, and thus it may require lower T$\rm_{int}$ to allow crossing to happen. When subject to strong stellar UV flux and weak vertical transport strength, ``minimum" T$\rm_{int}$ estimates from our approach may not be low enough for some exoplanet targets. 

Photochemistry tends to ``fry" an atmosphere, producing oxidized species that can look like the products of deeper cooking. Our model is an endmember of quenched thermochemistry, which is the most common chemical process across all giant planets (and brown dwarfs). Nevertheless, in cases where a strong photochemical influence may be suspected, such as unusually high abundances of sulfur dioxide (SO$_2$) \citep{2009ApJ...701L..20Z, 2021ApJ...923..264T, 2021MNRAS.506.3186H, 2023Natur.617..483T, 2023A&A...670A.161P}, hydrogen cyanide (HCN) and higher-order hydrocarbons such as acetylene (C$_2$H$_2$) \citep{2004ApJ...605L..61L, 2010ApJ...717..496L, 2011ApJ...737...15M,2011ApJ...727...65S, 2011ApJ...738...32L, 2012A&A...546A..43V, 2012ApJ...745....3M, 2012ApJ...745...77K, 2012ApJ...758...36M, 2013ApJ...763...25M, 2014A&A...562A..51V, 2014RSPTA.37230073M, 2015A&A...577A..33V, 2016ApJ...829...66M, 2018ApJ...853....7K, 2019MNRAS.487.2242H, 2021ApJ...921...27H, 2023ApJ...956..125O, 2024A&A...682A..52V}, or a lack of NH$_3$ on planets around G/K type stars \citep{2021ApJ...921...27H} (however, we should proceed cautiously where there is absence of evidence), grid modeling that includes disequilibrium processes of both transport and photochemistry would be necessary to pin down the exact T$\rm_{int}$ using the observations. It would be profitable to pursue the low-hanging fruit first that minimize this complication.

Note that our method can provide hints on whether photochemistry is important. If the crossing point is deep in the atmosphere, that means strong K$\rm_{zz}$ is needed, thus the impact of photochemistry is likely small and the minimum T$\rm_{int}$ estimate should be representative, which is the case for most of our targets. On the other hand, caution is warranted when the crossing occurs at shallower pressures, which indicates weak vertical transport and photochemistry can more easily affect the carbon speciation of the observable atmosphere. For planets where no intersection is found between the chemically-derived P-T region and the atmospheric P-T profiles, photochemistry should be explored before concluding that the retrieval results are unphysical.

\subsubsection{The Effect of Aerosols}\label{sec:cloud}

In this work, the choice of parameters in the climate model can significantly affect the resulting T$\rm_{int}$ values. Because our goal was to find the minimum T$\rm_{int}$ that conforms to observations, we selected the highest metallicity among the range that agrees with observations, so the P-T profiles are the hottest they can be. Even though the C/O ratio minimally affects the P-T profile, we selected the lowest C/O ratio that is allowed by our choice of high metallicity in our climate modeling to make P-T profiles as hot as possible. Employing similar logic, we assumed these planets have clear atmospheres and used zero albedo for all of our climate models with the goal of making the hottest possible P-T profile. 

However, clouds and hazes are almost ubiquitously found in exoplanet atmospheres \citep{2016Natur.529...59S, 2017AJ....154..261C, 2017ApJ...847L..22F, 2022ApJ...937...90D, 2022ApJ...941L...5E, 2024ApJ...961L..23B}. Clouds form as a result of either condensation of vapor species or thermochemistry in the deep atmosphere and typically have simple compositions \citep{2021JGRE..12606655G}. Photochemical hazes, on the other hand, have much more complex chemical compositions which are strongly affected by the gas composition of the background atmosphere \citep{2020PSJ.....1...17M}. Introducing clouds and hazes would increase the albedo of the planet. However, the effect of clouds on the planet's P-T profile is not simply decreasing the temperature of the atmosphere at every pressure but is found to be non-linear and also dependent on various planetary parameters \citep{2005ApJ...627L..69F, 2013ApJ...775...33M, 2015ApJ...813L...1C, 2019ApJ...872....1R, 2020ApJ...899...53M, 2021ApJ...908..101R}. For example, \citet{2020ApJ...899...53M} showed that a cloudy exoplanet with T$\rm_{int}$ between 800-1200~K can have an increased temperature in the deeper atmosphere with P$\gtrsim0.1$~bar and the upper atmosphere (P$\lesssim10^{-3}$~bar) but decreased temperature in between, compared to the cloudless case. Thus, a cloudy exoplanet may call for lower T$\rm_{int}$ values to satisfy the observational constraints, as cloud opacity shifts the adiabatic to lower pressures \citep[as suggested by][]{2014ApJ...797...41Z}. Indeed, if clouds can produce sufficient warming where quenching occurs, then internal heating would be less important. 

Future work on using climate models with the capacity of incorporating the effects of clouds is needed to understand how the minimum T$\rm_{int}$ responds to clouds. The effects of photochemical hazes are expected to be even more complex \citep{1991Sci...253.1118M, 2008P&SS...56..648T, 2015ApJ...815..110M, 2015NatCo...610231Z, 2016AsBio..16..873A, 2017Natur.551..352Z, 2021MNRAS.502.5643L} and likely to require laboratory experiments to determine their optical properties and assess their effects on the thermal structure and the subsequent impact on the estimate of T$\rm_{int}$. Note that both our methodology and disequilibrium grid retrievals would encounter this type of uncertainty, as grid retrievals typically do not consider the effects of clouds and hazes on the thermal profile, as well.


\subsubsection{The Effect of Horizontal Transport}
Exoplanets are three-dimensional in their dynamics, which means quenching of atmospheric species can occur both vertically and horizontally \citep{2006ApJ...649.1048C}. In this work, we adopted 1D P-T profiles assuming full heat redistribution, thereby neglecting any day-night asymmetries. Transmission spectra primarily probe the limbs of the atmosphere, which is an average of the dayside and nightside. The key question, then, is whether the observed atmospheric composition reflects local limb abundances or those influenced by faster horizontal transport from the dayside?

Previous work found that horizontal quenching is important only for planets with equilibrium temperatures from stellar heating above $\sim$1400~K \citep{2021MNRAS.505.5603B}. For cooler planets, the atmospheric composition tends to be horizontally homogeneous and reflect vertically quenched chemical compositions. This situation develops because, the horizontal transport timescale is only smaller than the vertical transport timescale below the quench point, which is where chemical equilibrium dominates (since the chemical timescale is shorter than both transport timescales there). Above the quench point at lower pressures, horizontal transport becomes slower than vertical mixing, and vertical quenching thus dominates in the observable limb region.

Among our sample, nearly all planets have equilibrium temperatures below 1400~K and are thus expected to exhibit horizontally homogeneous compositions, shaped primarily by vertical quenching. An exception is HD 209458 b (T$_\mathrm{eq}=1449$~K), where previous 3D modeling has shown that horizontal advection can significantly influence the limb composition \citep{2013A&A...558A..91P, 2018ApJ...855L..31D, 2018ApJ...866....2Z}. For the subset of targets from which we extract robust lower limits on T$_\mathrm{int}$ (T$_\mathrm{eq} \lesssim 1200$~K), horizontal transport is not expected to play a major role.

\section{Conclusion}\label{sec:conclusion}

In this paper, we present a simple theoretical tool that allows one to quickly calculate P-T regions that are compatible with atmospheric compositional observations of H$_2$O, CH$_4$, CO, and CO$_2$ in H$_2$-rich exoplanet atmospheres. Assuming the abundances of species from the retrieved pressure levels are minimally affected by photochemistry, we demonstrate that this new tool allows one to:
\begin{itemize}
    \item Quickly infer the minimum T$\rm_{int}$ that is required to explain the retrieved atmospheric abundances of H$_2$O, CH$_4$, CO, and CO$_2$ for those exoplanets that need elevated internal heat to explain the upper-atmosphere observations, mostly exoplanets with ``missing methane."
    \item Our simple model does not invoke the usage of parameterized vertical transport (K$\rm_{zz}$), and it only relies on the abundances of C-H-O species observed in the upper atmosphere to find the temperature and pressure where equilibrium is locked in at the quench point. This capability can enable future work to quickly constrain the P-T profiles that agree with observations and estimate K$\rm_{zz}$. It is not meant to replace grid retrieval; instead, our model can complement the latter method by providing a separate and quick check either prior to or after grid retrievals. Our approach is not biased by the choice of the range of K$\rm_{zz}$ that is used in grid retrievals. Because our method does not explicitly include the use of K$\rm_{zz}$, this parameter can only be inferred after the crossing region is determined for a certain P-T profile. Thus, it can be anticipated that future grid retrievals that include our methodology will find a larger T$\rm_{int}$ range, if they initially prescribe a smaller range for K$\rm_{zz}$, as demonstrated by WASP-107 b, HAT-P-12 b, and V1298 Tau b in our dataset.
\end{itemize}

Our method allows us to apply this framework to 12 warm-to-hot exoplanet targets that have key carbon- and oxygen-bearing species abundances constrained by JWST observations (specifically, H$_2$O, CO$_2$, CO, and CH$_4$). For each planet, we constrain the minimum T$\rm_{int}$ required to drive the observed depletion of methane. We find that these chemistry-constrained minimum T$\rm_{int}$ values often exceed those predicted by standard planet evolution models, indicating that additional heating mechanisms may be necessary to raise the intrinsic temperatures of these exoplanets. The size of our sample also allows us to examine the population-level trend of derived minimum T$\rm_{int}$ as a function of planetary equilibrium temperature. When comparing it to the empirical T$\rm_{eq}$-T$\rm_{int}$ trend derived from the Hot-Jupiter mass-radius population by \citet{2018AJ....155..214T}, which attributes Hot Jupiter internal heating primarily due to Ohmic dissipation, several of our targets lie significantly above the expected relation. This suggests that while Ohmic dissipation may dominate the internal heating for the general population, additional heating processes, such as tidal heating, play important roles for some individual systems. These findings underscore the diversity of thermal histories among warm-to-hot exoplanets and highlight the diagnostic power of atmospheric composition as a complementary constraint on exoplanet interiors. 

Future studies can build upon this framework by using the inferred quench points (i.e., the crossing points between the chemically-derived P-T relationships and the atmospheric P-T profiles) to deduce the range of vertical mixing strengths (K$\rm_{zz}$) that are consistent with observations. Degeneracies involving photochemistry and clouds should also be explored in greater detail, as both can mimic the effects of a hot interior. Photochemistry can destroy CH$_4$ to make CO$_2$ and CO, while clouds may locally warm the atmosphere near the quench pressures. Forward models that incorporate both photochemistry and cloud feedbacks, guided by the minimum T$\rm_{int}$ values derived using our framework, will help disentangle these processes. For the planets with elevated interior heating (i.e., higher apparent T$\rm_{int}$ than evolution model predictions), future work should test whether the chemically-derived T$\rm_{int}$ are consistent with the observed mass-radius of the planet, i.e., does interior heating drive the radius inflation of the planet? Finally, investigating the physical origin of interior heating, such as tidal heating, Ohmic dissipation, or delayed cooling, will be essential for linking atmospheric chemistry to planetary interior evolution.

In closing, we should be mindful that as the field evolves with more transmission spectroscopy observations of warm-to-hot exoplanets, our interpretations might change. The present paper makes progress on this topic but is undoubtedly not the final word. Thus, we make a publicly available website archiving all the data presented in this study: https://exoplanethaziness.shinyapps.io/thermalweb/. This website allows us to add new observations and to keep track of the updated T$\rm_{int}$ trend.

\section{Acknowledgments} \label{sec:Acknowledgements}

X. Yu is supported by NASA Habitable Worlds Program Grant 80NSSC24K0075, NASA Planetary Science Early Career Award 80NSSC23K1108, and the Heising-Simons Foundation grant 2023-3936. C. Glein wishes to acknowledge funding from the Heising-Simons Foundation grant 2023-4657 and the SwRI IR\&D grant 15-R6576. D. Thorngren thanks Johns Hopkins University for support via the Davis Fellowship.  We thank colleagues for their very helpful discussions, including Jonathan Fortney, Peter Gao, Yaqin Wu, Kevin Stevenson, Maryta Bryan (Maryta, thank you for inviting me to the University of Toronto seminar so I could talk to you and your wonderful colleagues), and Gongjie Li. Special thanks to Peter Gao, who helps me learn secrets about the climate model. This work also benefited from the 2025 Exoplanet Summer Programs in the Other Worlds Laboratory (OWL) at the University of California, Santa Cruz, where I had wonderful discussions with the brilliant Kazumasa Ohno, Yayatti Chachan, Bryanna Lacy, and Ben Lew. I would also like to thank Shang-Min Tsai for hosting me with delicious Taiwanese food and great discussions that benefit me by thinking more deeply about this methodology. We thank Qiao Xue, Pa Chia Thao, Yoav Rotman, Everett Schlawin, Saugata Barat, Sagnick Mukherjee, Eva-Maria Ahrer, and Luis Welbanks for providing JWST-retrieved species abundances that are used in this work.

\newpage
\appendix
\restartappendixnumbering

\section{Testing the one-reaction approach}\label{sec:one_eq}

The simplest approach would be to assume only the CO-CH$_4$ equilibrium without considering the CO-CO$_2$ equilibrium. This has been the typical method that is assumed in the literature \citep[e.g.,][]{2020AJ....160..288F}. In this case, we ignore CO$_2$ in the hydrogen mole fraction calculations and revise Equation~\ref{eq:XH2} to be:
\begin{equation}
X_{\rm H_2} = [1 - (X_{\rm H_2O} + X_{\rm CH_4} +  X_{\rm CO}]/1.2.
\label{eq:XH2_v2}
\end{equation}

\begin{figure}[!htb]
  \centering
    \begin{subfigure}{.5\textwidth}
    \centering
    \includegraphics[width=\textwidth]{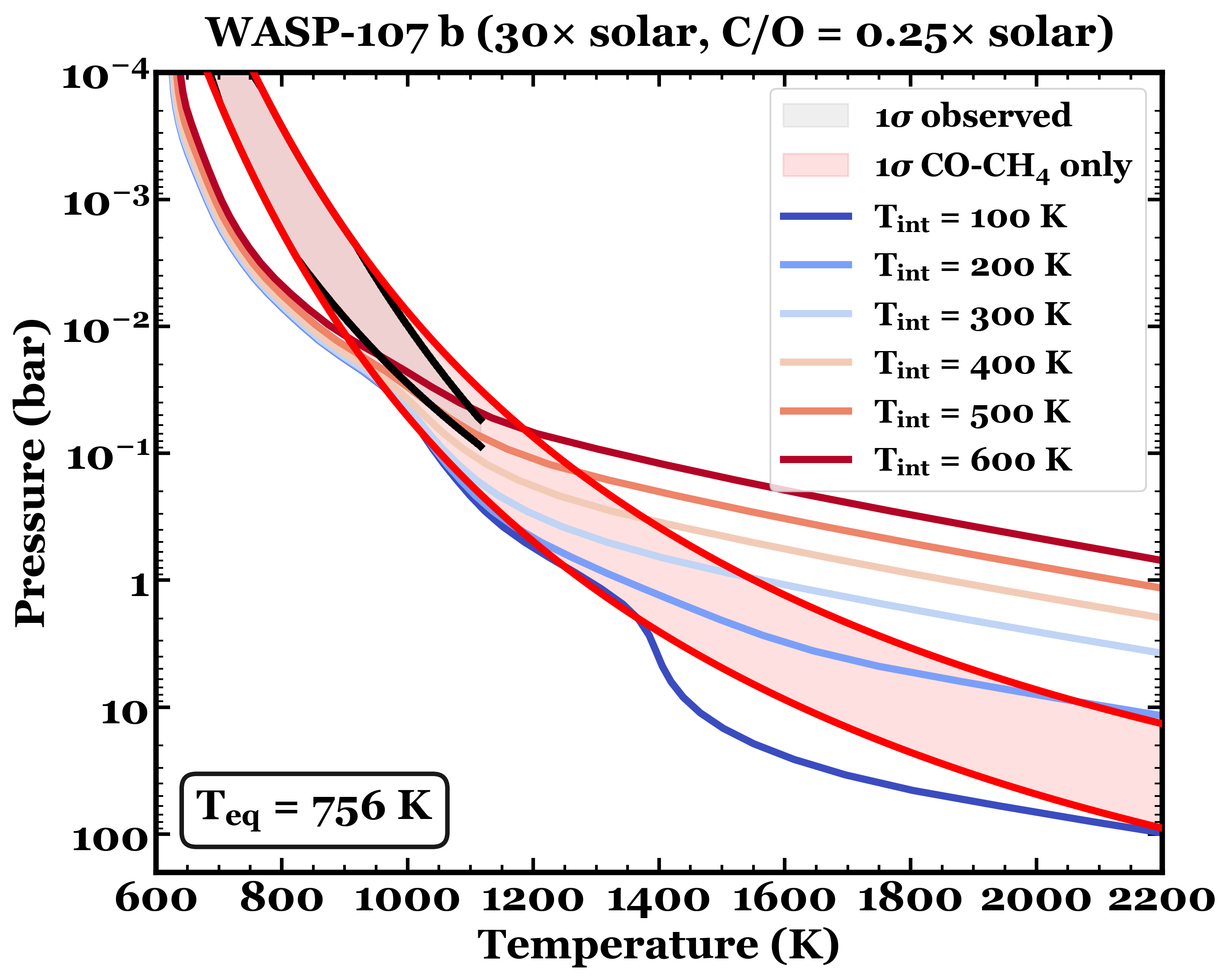}
  \end{subfigure}
  \hfill
  \begin{subfigure}{.5\textwidth}
    \centering
    \includegraphics[width=\textwidth]{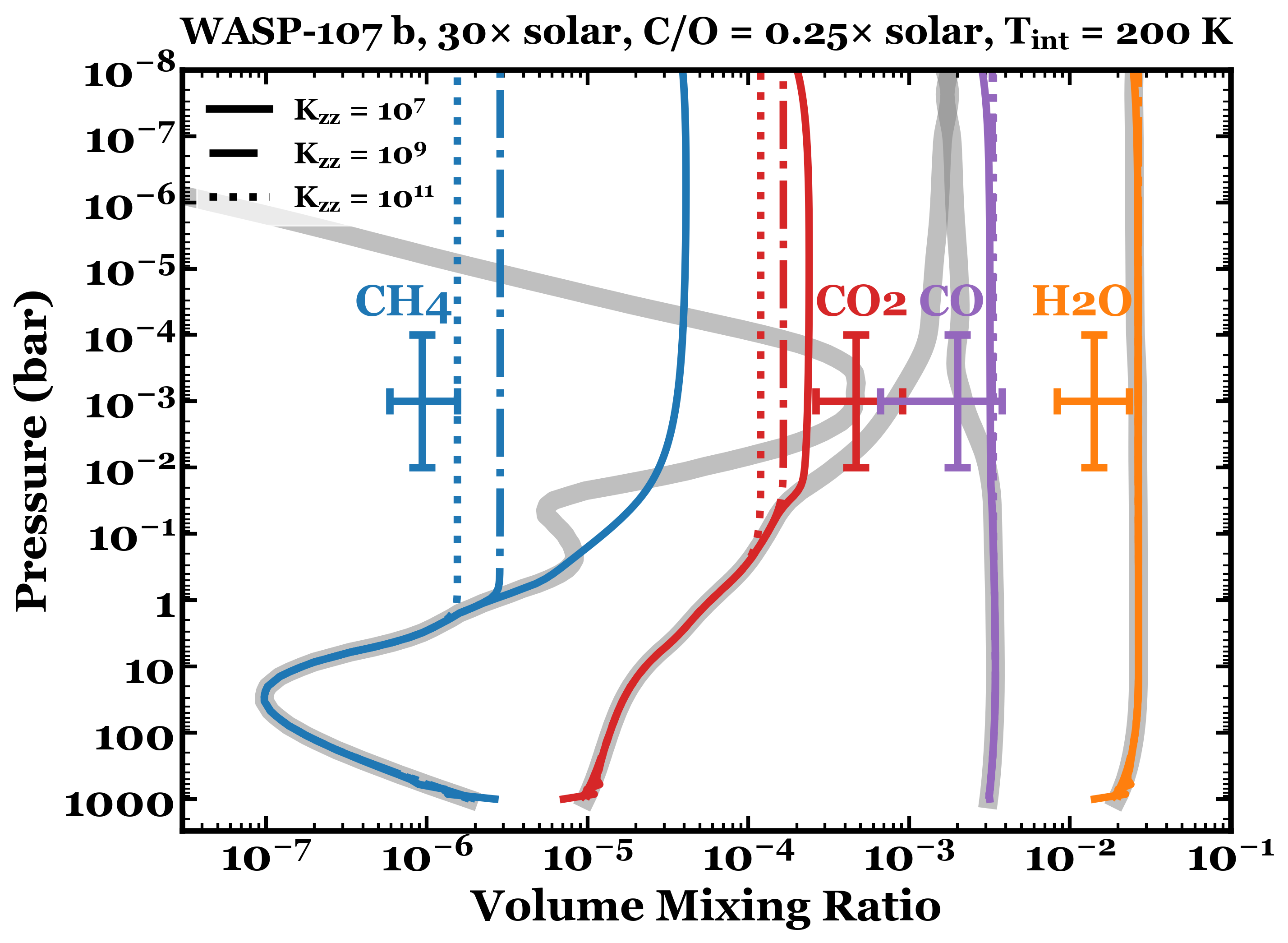}
  \end{subfigure}
  \caption{(Top) Inferring T$\rm_{int}$ of WASP-107 b using an alternative approach, assuming only CO-CH$_4$ quenching at depth. (Bottom) Forward modeling results for WASP-107 b using JWST-observed abundances (with 1$\sigma$ error bars) from \citet{2024Natur.630..831S}. The notations are the same as in Figures~\ref{fig:wasp107b_atmo} and \ref{fig:wasp107b_forward}, but the P-T profile we used in the bottom plot has T$\rm_{int}=200$~K instead of 450~K. Solid, dashed-dotted, and dotted lines correspond to K$\rm_{deep}$ values of 10$^{7}$, 10$^{9}$, and 10$^{11}$~cm$^2$~s$^{-1}$, respectively.}
  \label{fig:alternative-COCH4}
\end{figure}

Then, we can calculate the chemical equilibrium pressure using Equation~\ref{eq:K1_P} with a grid of H$_2$O-CO-CH$_4$ combinations that are allowable from JWST observations. The results using this simple method, with WASP-107 b being an example, are shown in Figure~\ref{fig:alternative-COCH4}(Top). This simplest approach leads to a much more extended observationally allowed P-T region, compared to the two chemical equilibria approach in the main text since an observational constraint has been removed. For WASP-107 b's ATMO retrieval from \citet{2024Natur.630..831S}, the chemically-derived apparent T$\rm_{int}$ drops to 100~K. We demonstrate that these low T$\rm_{int}$ solutions are not compatible with the observed abundances by performing forward modeling for the T$\rm_{int}=200$~K case, with results shown in Figure~\ref{fig:alternative-COCH4}(Bottom). When the K$\rm_{zz}$ is high enough to match the observed CH$_4$ abundances, it underpredicts CO$_2$. Although for low K$\rm_{zz}$ that matches the observed CO$_2$ abundance, we overpredict CH$_4$. 

The two-equilibria approach that we adopted in the main text is more consistent because combinations of H$_2$O-CO-CH$_4$ are discarded if they do not lead to CO$_2$ abundances that are within the observed range. This example shows the importance of considering both CO-CH$_4$ and CO-CO$_2$ equilibria in determining the observationally allowed P-T region that is physically plausible. The CO-CO$_2$ reaction has also been shown to be important in previous works such as \citet{2018ApJ...862...31T}. Indeed, previous works that invoked the analytical quench timescale using only CO-CH$_4$ equilibrium \citep{2002Icar..159...95B, 2011ApJ...737...15M, 2011ApJ...738...72V} already described some caveats of this single reaction approach. If data are sufficient, it makes sense to consider both equilibria.

\section{Sensitivity test of the main model}\label{sec:sensitivity}

\begin{figure}[!htb]
  \centering
  \begin{subfigure}{.45\textwidth}
    \centering
\includegraphics[width=\textwidth]{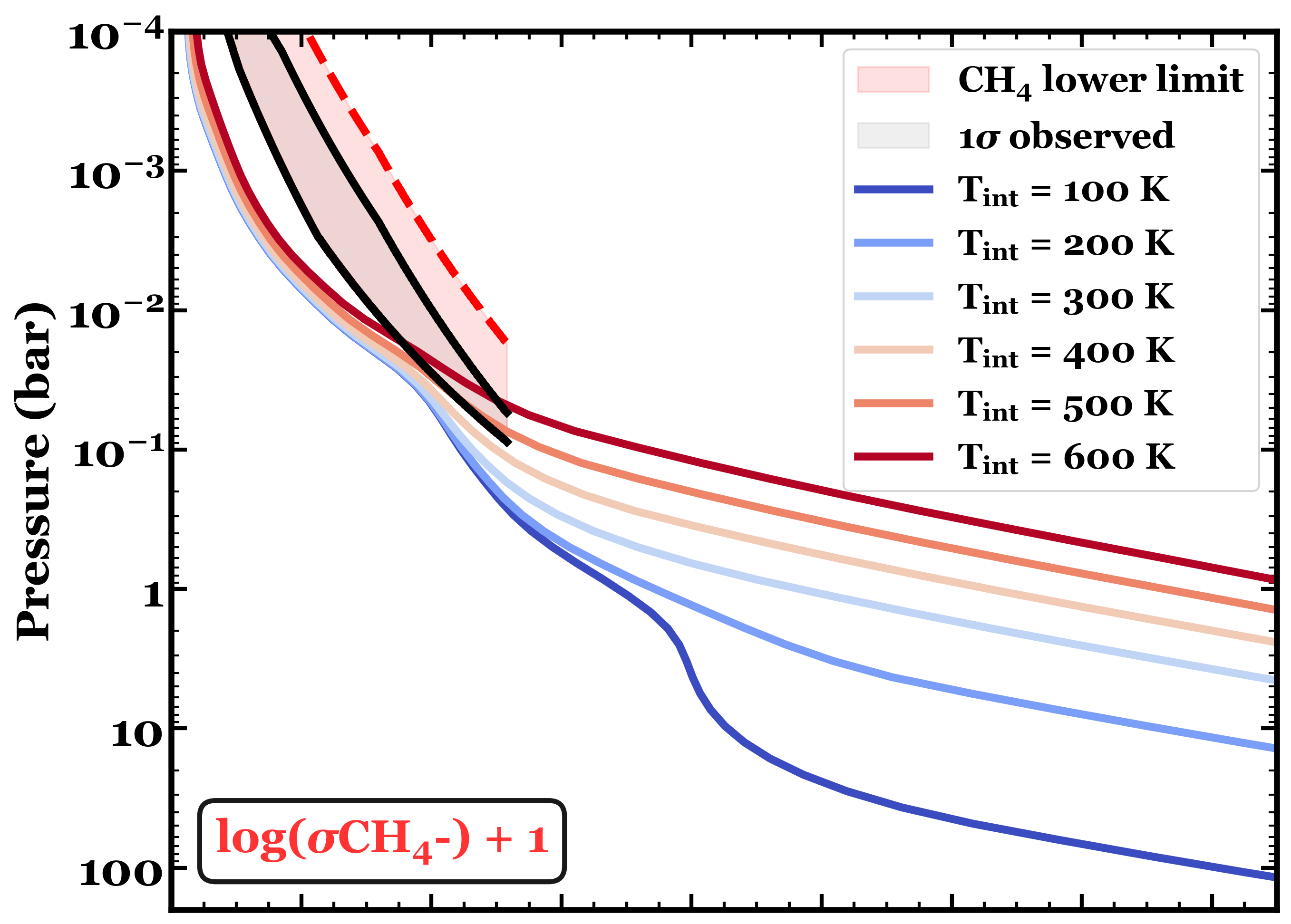}
  \end{subfigure}
  \hfill
  \begin{subfigure}{.45\textwidth}
    \centering
    \includegraphics[width=\textwidth]{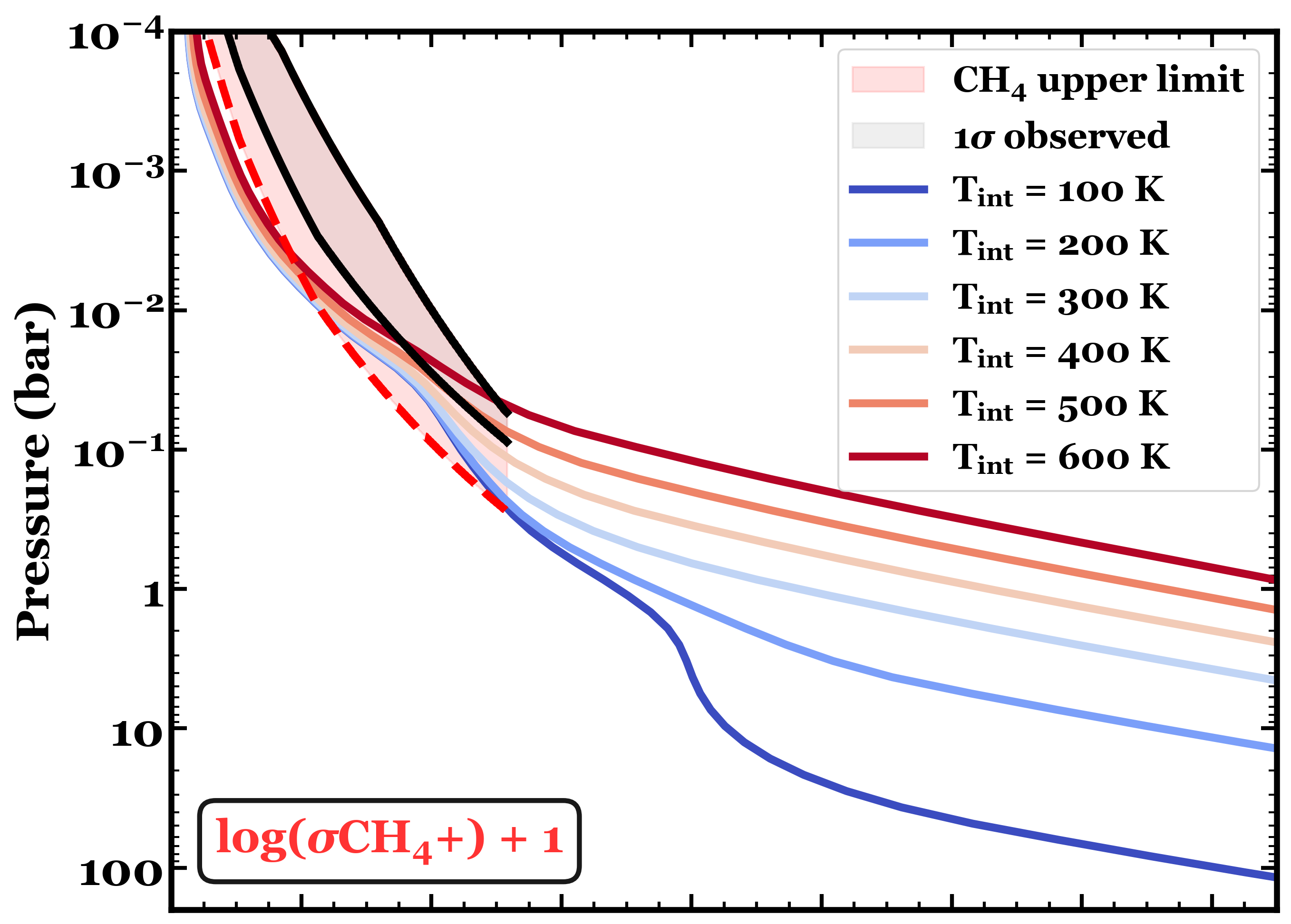}
  \end{subfigure}


  \begin{subfigure}{.45\textwidth}
    \centering
    \includegraphics[width=\textwidth]{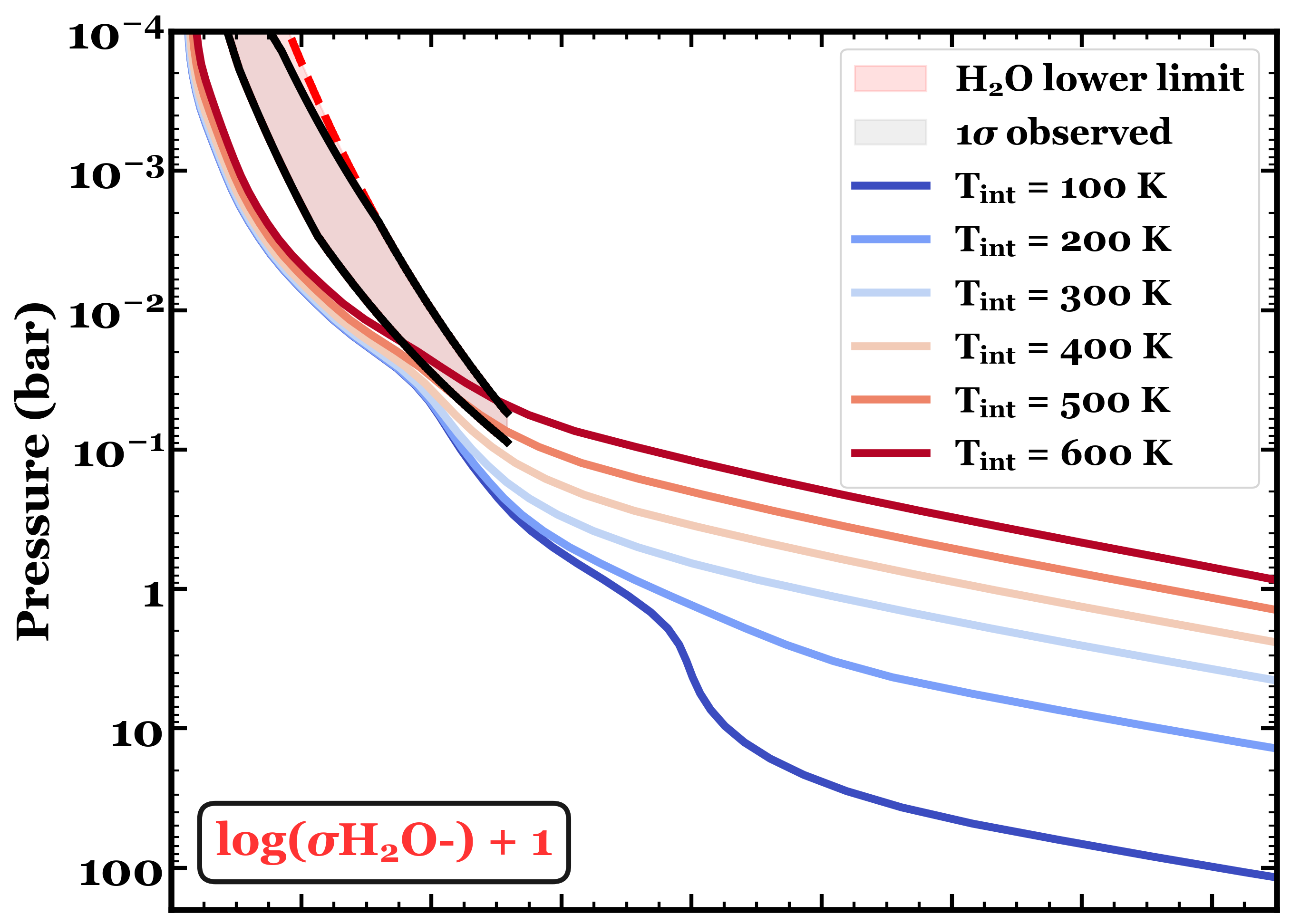}
  \end{subfigure}
  \hfill
  \begin{subfigure}{.45\textwidth}
    \centering
    \includegraphics[width=\textwidth]{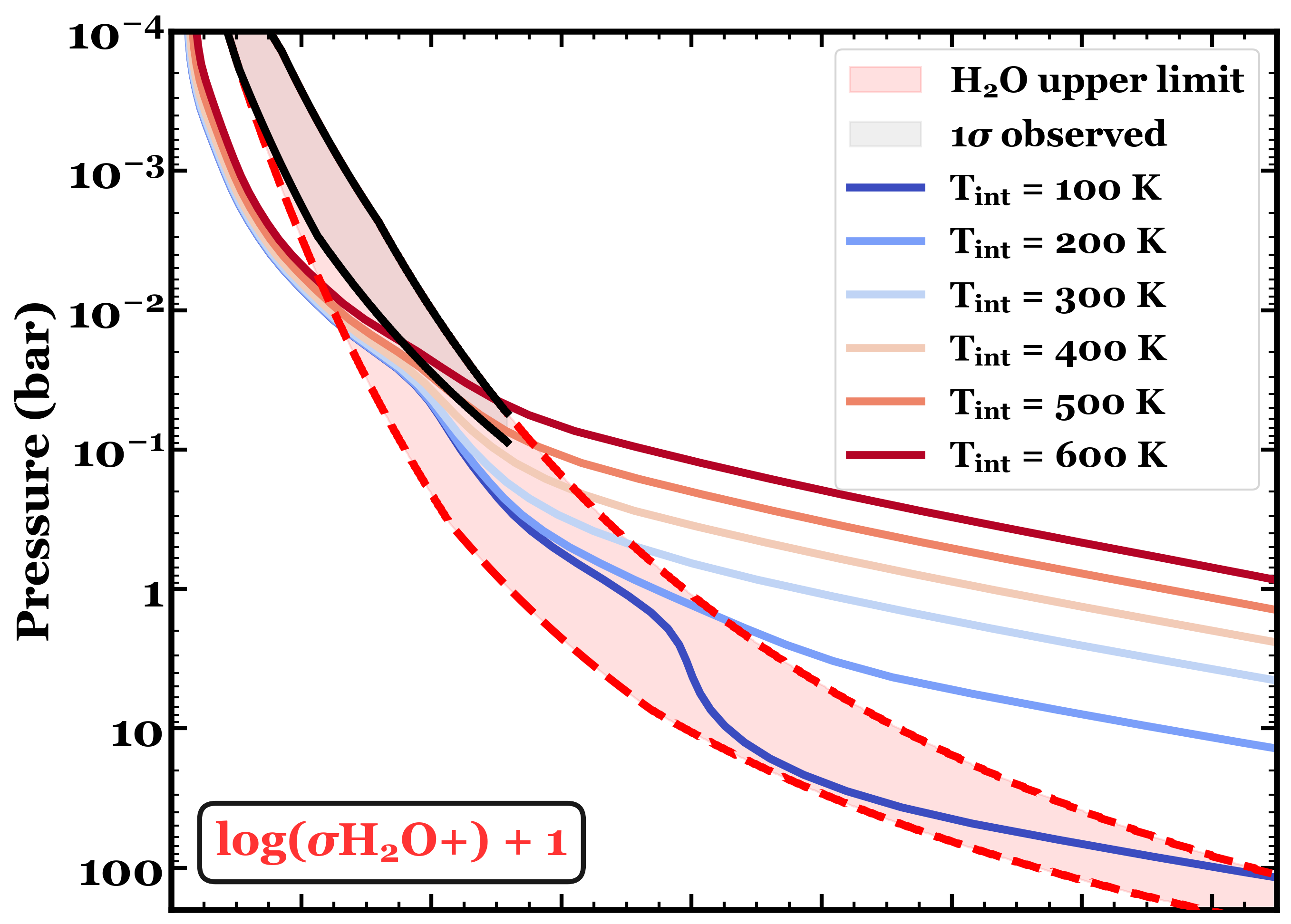}
  \end{subfigure}

  \begin{subfigure}{.45\textwidth}
    \centering
    \includegraphics[width=\textwidth]{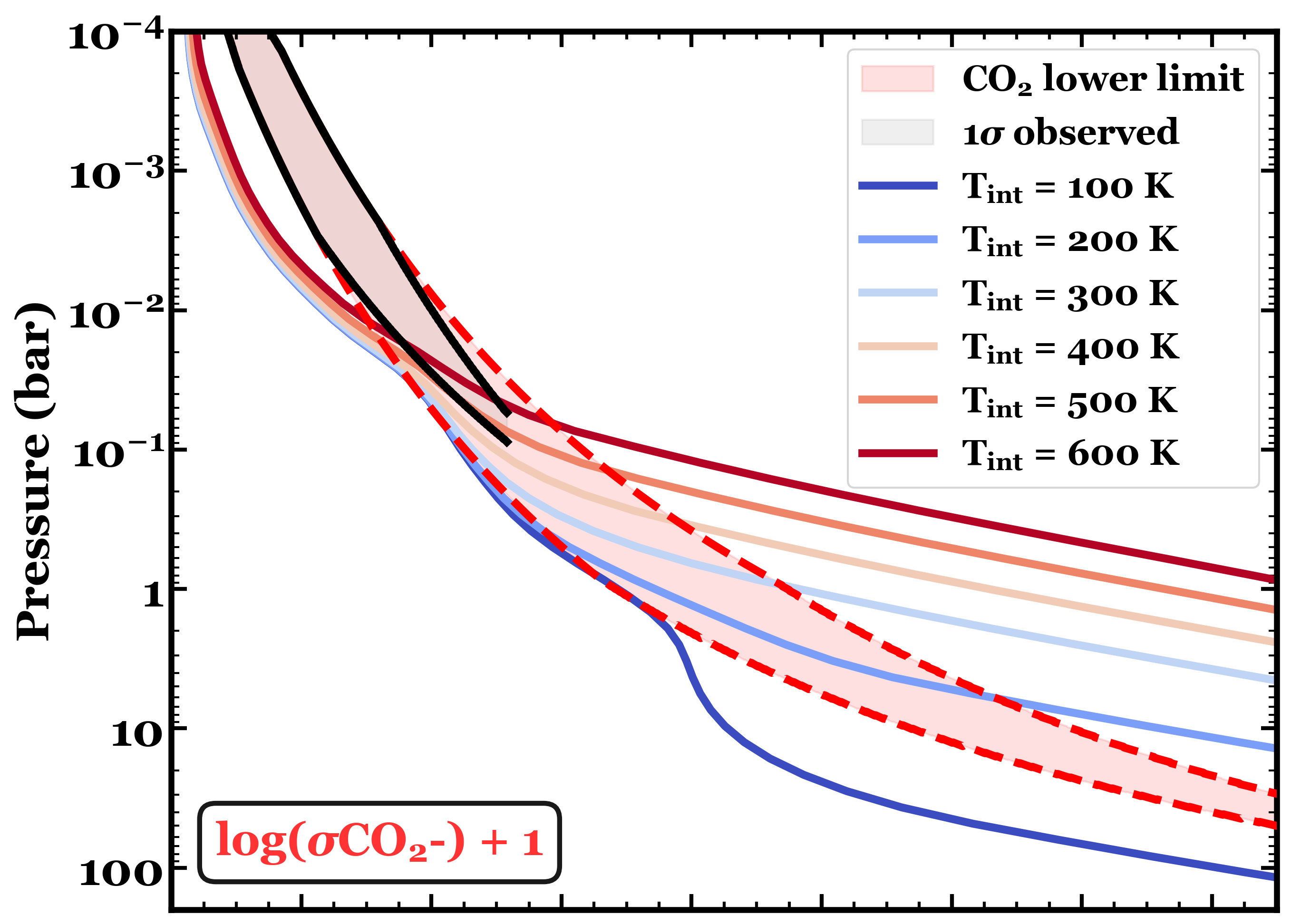}
  \end{subfigure}
  \hfill
  \begin{subfigure}{.45\textwidth}
    \centering
    \includegraphics[width=\textwidth]{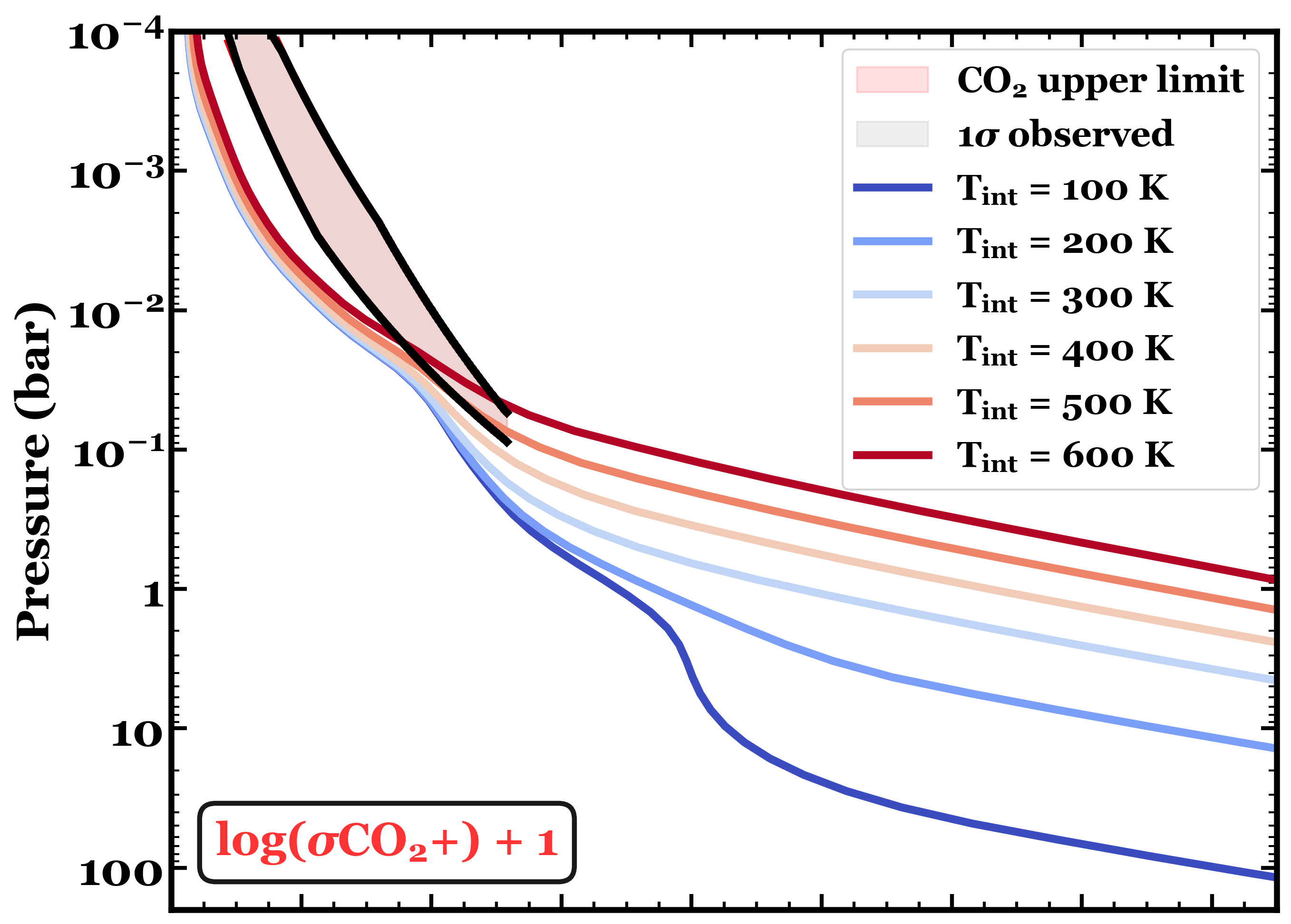}
  \end{subfigure}

    \begin{subfigure}{.45\textwidth}
    \centering
    \includegraphics[width=\textwidth]{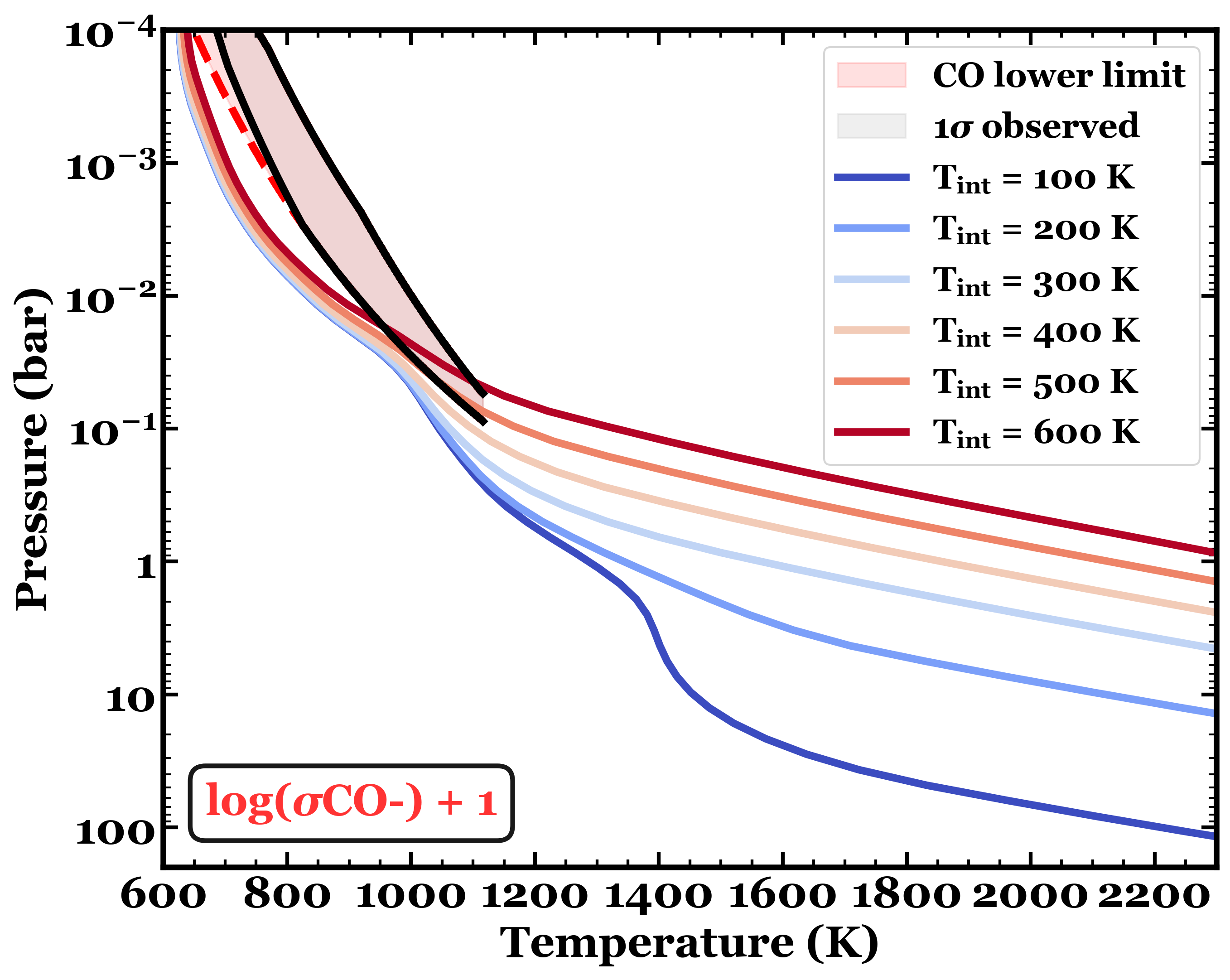}
  \end{subfigure}
  \hfill
  \begin{subfigure}{.45\textwidth}
    \centering
    \includegraphics[width=\textwidth]{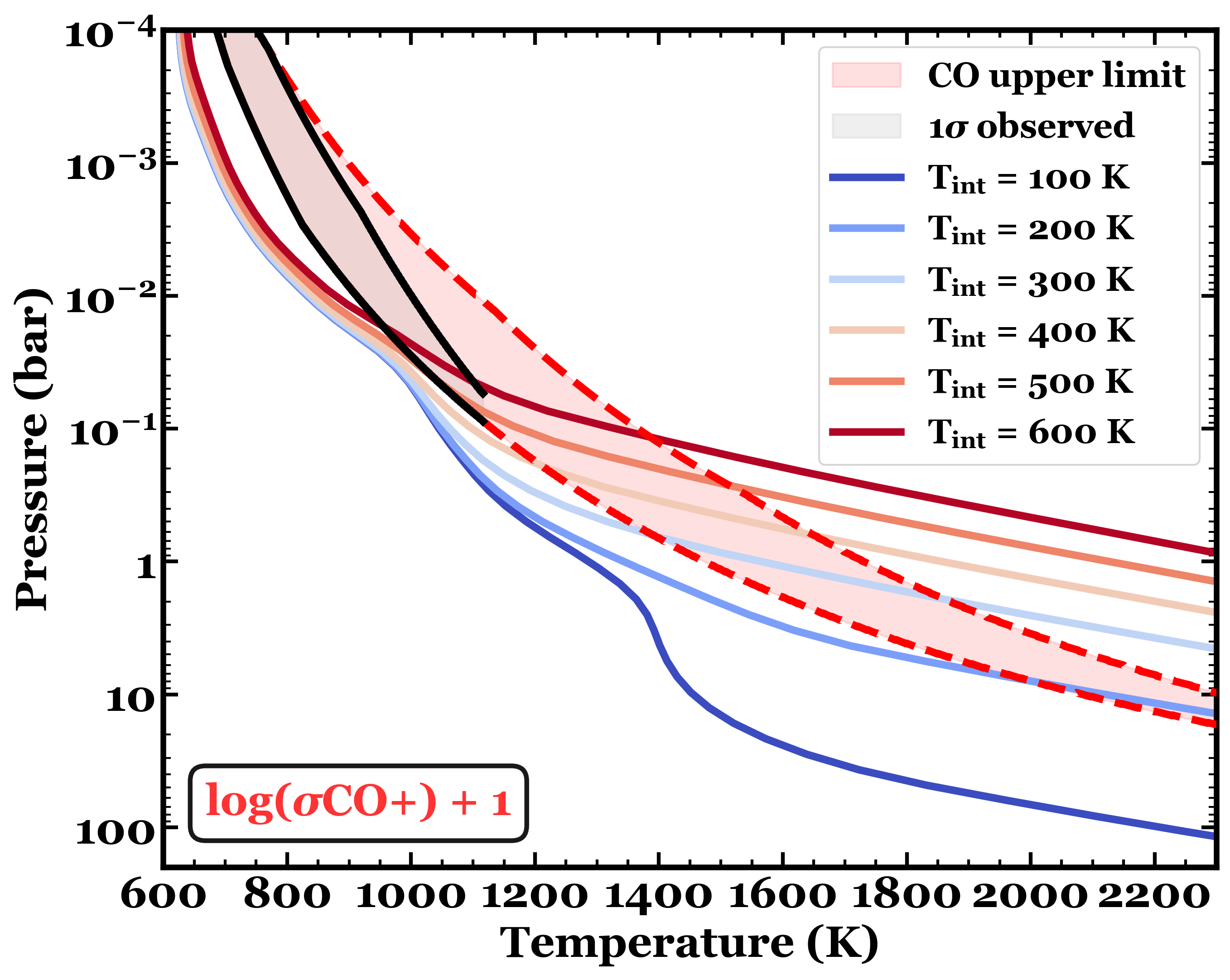}
  \end{subfigure}
  \caption{Sensitivity tests on how increasing the upper limit and decreasing the lower limit by 1 order of magnitude (the dashed lines and the pink region enclosed) for each species affects the minimum T$\rm_{int}$ determination compared to the nominal 1$\sigma$ for WASP-107 b. The rest of this figure is the same as Figure~\ref{fig:wasp107b_atmo}.}
  \label{fig:sensitivity}
\end{figure}

\section{Additional forward models}\label{sec:add_forward}

\begin{figure}[!htb]
  \centering
    \begin{subfigure}{.49\textwidth}
    \centering
    \includegraphics[width=\textwidth]{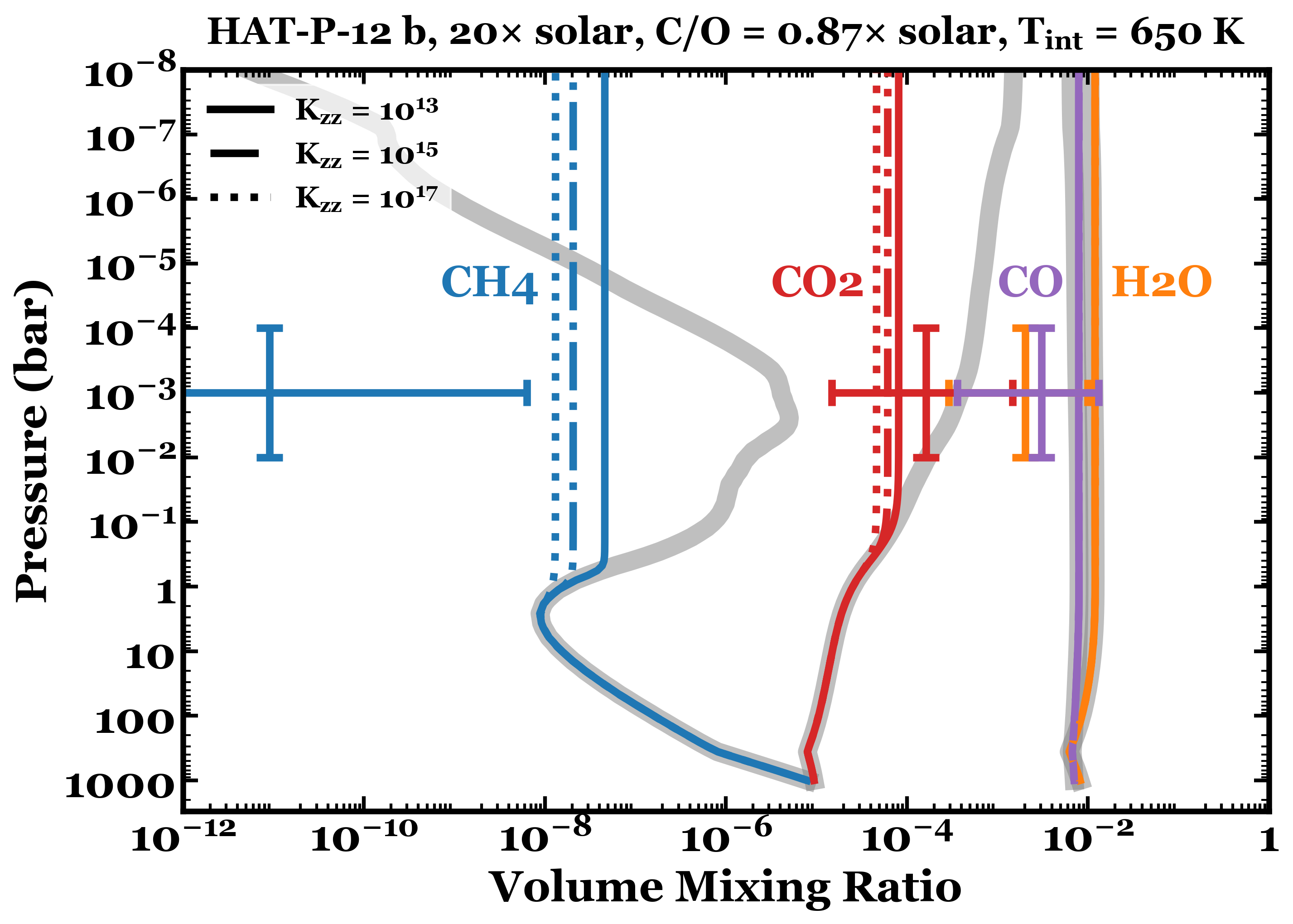}
  \end{subfigure}
  \hfill
  \begin{subfigure}{.49\textwidth}
    \centering
    \includegraphics[width=\textwidth]{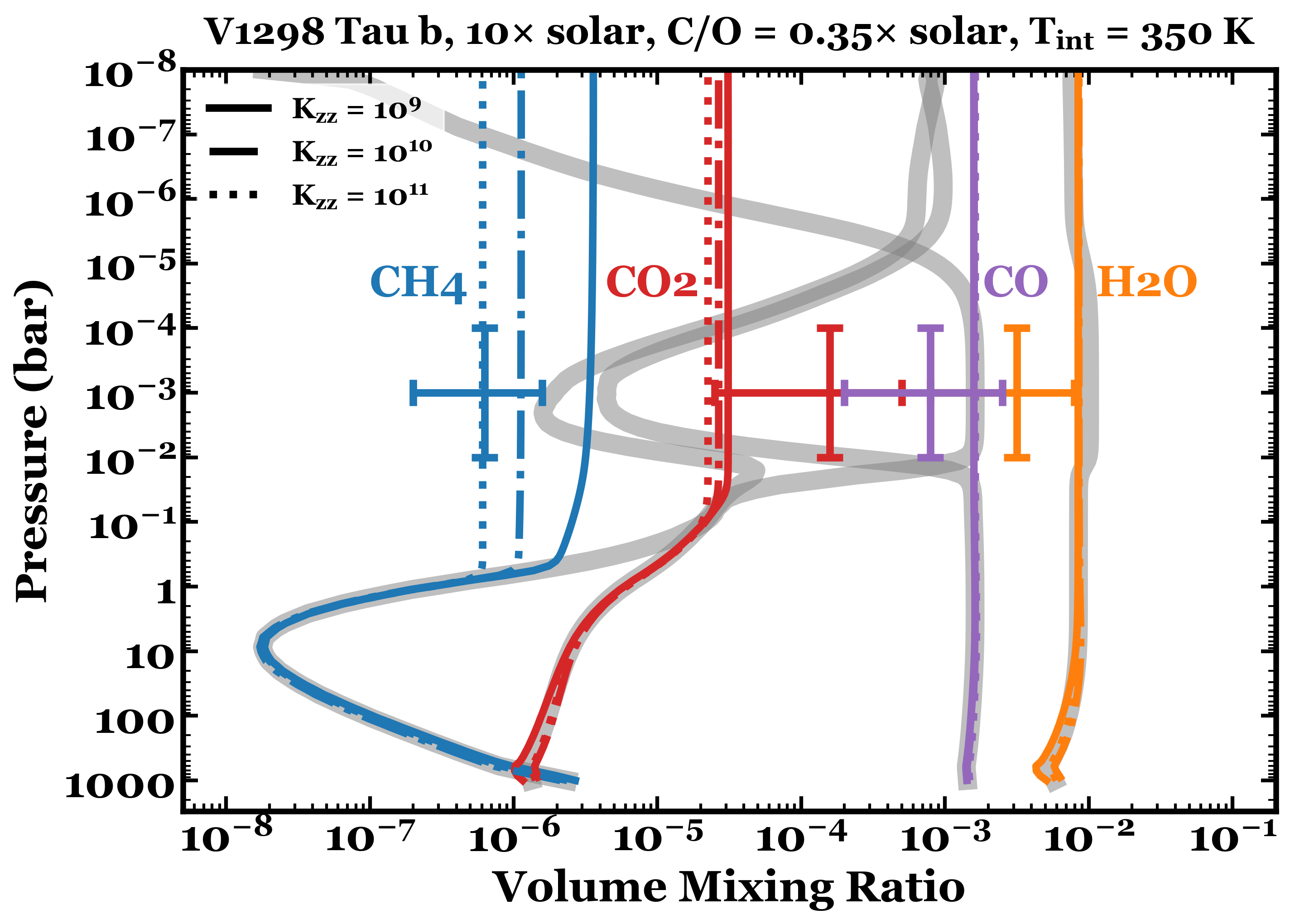}
  \end{subfigure}
  \caption{Forward modeling results for HAT-P-12 b using JWST-observed abundances (with 1$\sigma$ error bars) from \citet{2025AA...703A.264C} (Left) and for V1298 Tau b using JWST-observed abundances (with 1$\sigma$ error bars) from \citet{Barat2025} (Right). The notations are the same as in Figure~\ref{fig:wasp107b_forward}. For HAT-P-12 b, solid, dashed-dotted, and dotted lines correspond to K$\rm_{deep}$ values of 10$^{13}$, 10$^{15}$, and 10$^{17}$~cm$^2$~s$^{-1}$, respectively. For V1298 Tau b, solid, dashed-dotted, and dotted lines correspond to K$\rm_{deep}$ values of 10$^{9}$, 10$^{10}$, and 10$^{11}$~cm$^2$~s$^{-1}$, respectively.}
  \label{fig:HATP12b}
\end{figure}

\section{Additional models for HAT-P-18 b}\label{sec:add_HATP18b}
\begin{figure}[!htb]
  \centering
    \begin{subfigure}{.49\textwidth}
    \centering
    \includegraphics[width=\textwidth]{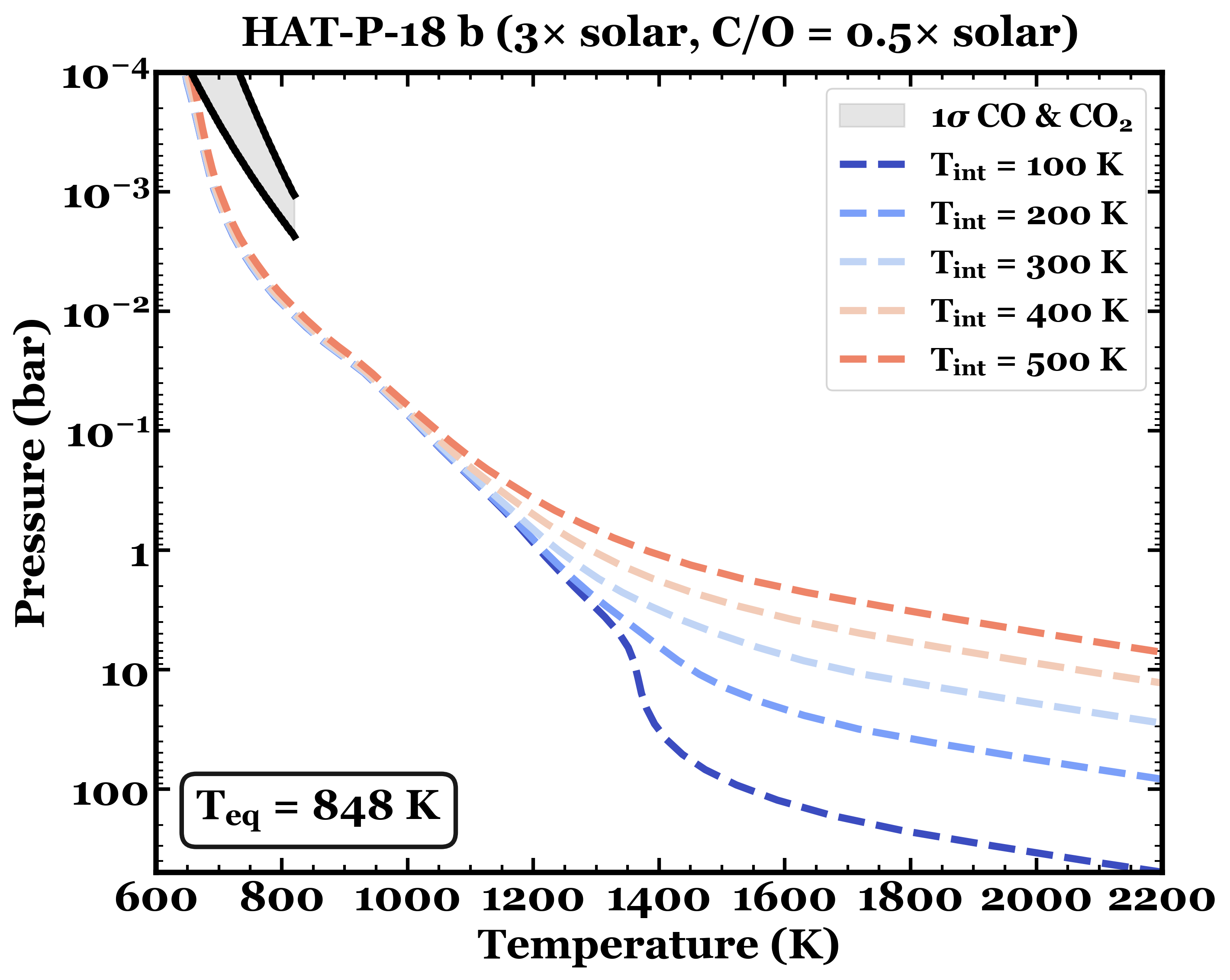}
  \end{subfigure}
  \hfill
  \begin{subfigure}{.49\textwidth}
    \centering
    \includegraphics[width=\textwidth]{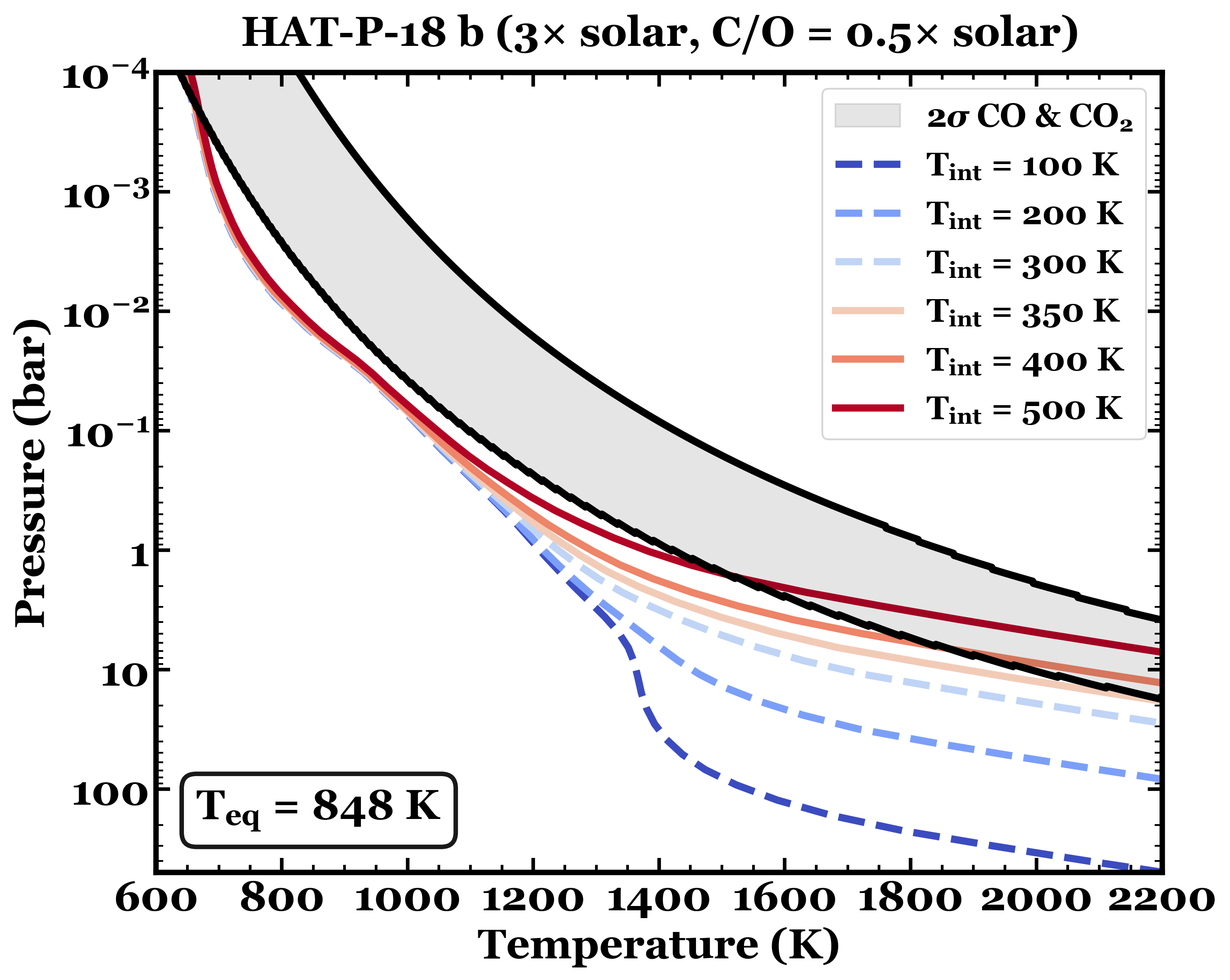}
  \end{subfigure}
  \caption{Additional results for HAT-P-18 b with chemically-derived P-T regions calculated using 1$\sigma$ retrieved C-H-O species abundances (left) and 2$\sigma$ CO and CO$_2$ abundances (right) from \citet{2022ApJ...940L..35F}. The subplots follow the same notations as in Figure~\ref{fig:wasp107b_atmo}.}
  \label{fig:HATP18b}
\end{figure}

\newpage
\section{Additional tables}\label{sec:app_tables}

\begin{deluxetable*}{lccccccc}
\tablewidth{0pt}
\tablecaption{Summary of Physical and Orbital Properties of the 12 Exoplanet Targets Listed by Equilibrium Temperature. \label{table:target_property}}
\tablehead{
\colhead{Target} & 
\colhead{$T_{\mathrm{eq}}$ [K]} & 
\colhead{Radius [$R_\oplus$]} & 
\colhead{Mass [$M_\oplus$]} & 
\colhead{Semi-major axis [$a$] (AU)} & 
\colhead{Ref.\tablenotemark{a}} & 
\colhead{Age (Gyr)} &
\colhead{Ref.\tablenotemark{b}}
}
\startdata
HD 209458 b & $1449\pm16$ & $15.23^{+0.18}_{-0.21}$ & $217^{+4}_{-5}$ & $0.04707^{+0.00045}_{-0.00047}$ & \citetalias{Bonomo2017} & $4\pm0.8$ & \citetalias{Fossati2023} \\
WASP-166 b & $1273\pm35$ & $7.1\pm0.3$ & $32.1\pm1.6$ & $0.0641^{+0.0011}_{-0.0011}$ & \citetalias{2019MNRAS.488.3067H} & $2.1\pm0.9$ & \citetalias{2019MNRAS.488.3067H} \\
HD 189733 b & $1209\pm19$ & $12.54\pm0.43$ & $370.6^{+16.5}_{-15.6}$ & $0.03106^{+0.00051}_{-0.00049}$ & \citetalias{2019PASP..131k5003A} & $7.4\pm2.7$ & \citetalias{Fossati2023} \\
HIP 67522 b & $1176\pm40$ & $9.763^{+0.493}_{-0.504}$ & $13.8\pm1.0$ & $0.0747^{+0.0034}_{-0.0038}$ & \citetalias{2024ApJ...973L..30B} & $0.017\pm0.002$ & \citetalias{2024ApJ...973L..30B} \\
HAT-P-12 b & $980\pm18$ & $10.749^{+0.325}_{-0.235}$ & $67.059\pm3.814$ & $0.0384^{+0.0003}_{-0.0003}$ & \citetalias{2009ApJ...706..785H} & $2.5\pm2.0$ & \citetalias{2009ApJ...706..785H} \\
WASP-69 b & $961\pm21$ & $11.85\pm0.19$ & $82.64\pm5.88$ & $0.04525^{+0.00075}_{-0.00075}$ & \citetalias{Casasayas2017} & $7.6\pm4.0$ & \citetalias{Fossati2023} \\
HAT-P-18 b & $848\pm26$ & $11.153\pm0.583$ & $62.61\pm4.132$ & $0.0559^{+0.0007}_{-0.0007}$ & \citetalias{2011ApJ...726...52H} & $5.2\pm2.3$ & \citetalias{Fossati2023}  \\
WASP-80 b & $825\pm25$ & $11.2^{+0.34}_{-0.35}$ & $171\pm11$ & $0.0344^{+0.0010}_{-0.0011}$ & \citetalias{2015MNRAS.450.2279T} & $1.6\pm2.3$ & \citetalias{Fossati2023} \\
WASP-107 b & $756\pm22$ & $10.8\pm0.3$ & $30.5\pm1.7$ & $0.0553^{+0.0013}_{-0.0013}$ & \citetalias{2021AJ....161...70P} & $3.4\pm0.7$ & \citetalias{2021AJ....161...70P} \\
V1298 Tau b & $669\pm21$ & $10.27^{+0.58}_{-0.53}$ & $12\pm1$ & $0.1688^{+0.0026}_{-0.0026}$ & \citetalias{Barat2025, 2019ApJ...885L..12D} & $0.02\pm0.01$ & \citetalias{Barat2025} \\
GJ 3470 b & $661\pm21$  & $4.22\pm0.09$ & $11.1\pm0.9$ & $0.0355^{+0.0019}_{-0.0019}$ & \citetalias{2019AJ....157...97K} & $1.1\pm1.6$ & \citetalias{Fossati2023} \\
TOI-270 d & $387\pm10$  & $2.133\pm0.058$ & $4.78\pm0.43$ & $0.0721^{+0.0005}_{-0.0005}$ & \citetalias{2021MNRAS.507.2154V} & $>1$ & \citetalias{2019NatAs...3.1099G} \\
\enddata
\tablenotetext{a}{Mass, radius, and semi-major axis references.}
\tablenotetext{b}{Age references.}
\tablecomments{Planets are ordered by decreasing equilibrium temperature ($T_{\mathrm{eq}}$), assuming zero albedo. Radius and mass are in Earth units.}
\end{deluxetable*}

\begin{deluxetable*}{lcccccc}
\tablecaption{Adopted log$_{10}$ Volume Mixing Ratios (VMRs) and 1$\sigma$ Uncertainties for the JWST Targets. \label{table:vmr}}
\tablehead{
\colhead{Target} &
\colhead{Retrieval Method} &
\colhead{$\log(X_{\mathrm{H_2O}})$} &
\colhead{$\log(X_{\mathrm{CH_4}})$} &
\colhead{$\log(X_{\mathrm{CO_2}})$} &
\colhead{$\log(X_{\mathrm{CO}})$} &
\colhead{Reference}
}
\startdata
HD 209458 b & PLATON free retrieval & $-2.75^{+0.38}_{-0.29}$ & $-8.14^{+1.16}_{-1.30}$ & $-6.13^{+0.42}_{-0.33}$ & $-3.59^{+0.27}_{-0.28}$ & \citet{2024ApJ...963L...5X} \\
WASP-166 b & POSEIDON free retrieval & $-1.30^{+0.21}_{-0.23}$ & $-7.93^{+1.34}_{-1.37}$ & $-2.23^{+0.31}_{-0.37}$ & $-6.36^{+2.35}_{-2.40}$ & \citet{2025AJ....170...50M} \\
HD 189733 b & CHIMERA free retrieval & $-3.48^{+0.93}_{-0.63}$ & $-10.13^{+1.35}_{-1.15}$ & $-6.09^{+0.83}_{-0.55}$ & $-4.02^{+1.08}_{-0.70}$ & \citet{2024Natur.632..752F} \\
HIP 67522 b & CHIMERA grid retrieval & $-3.56^{+0.81}_{-0.96}$ & $-10.22^{+1.18}_{-1.07}$ & $-5.56^{+0.81}_{-0.95}$ & $-3.91^{+1.01}_{-1.09}$ & \citet{2024AJ....168..297T} \\
HAT-P-12 b & ARCiS free retrieval & $-2.69^{+0.69}_{-0.85}$ & $-11.04^{+2.84}_{-2.69}$ & $-3.79^{+0.96}_{-1.04}$ & $-2.51^{+0.63}_{-0.93}$ & \citet{2025AA...703A.264C} \\
WASP-69 b & PICASO equilibrium grid retrieval & $-2.68^{+0.23}_{-0.23}$ &  $-7.19^{+0.92}_{-0.92}$ & $-4.51^{+0.15}_{-0.15}$ & $-2.46^{+0.30}_{-0.30}$ & \citet{2024AJ....168..104S} \\
HAT-P-18 b & ATMO free retrieval & $-3.03^{+0.31}_{-0.25}$ & $-5.08^{+0.35}_{-0.34}$ & $-4.19^{+0.40}_{-0.40}$ & $-4.76^{+1.65}_{-1.94}$ & \citet{2022ApJ...940L..35F} \\
\multirow{2}{*}{WASP-80 b} & AURORA free retrieval & $-1.80^{+0.55}_{-0.94}$ & $-4.22^{+0.47}_{-0.74}$ & $-8.00^{+2.52}_{-2.48}$ & $-7.05^{+3.13}_{-2.96}$ & \citet{2023Natur.623..709B} \\
 & CHIMERA free retrieval & $-2.33^{+0.59}_{-0.52}$ & $-3.86^{+0.37}_{-0.31}$ & $-5.14^{+0.49}_{-0.45}$ & $-2.49^{+0.60}_{-0.55}$ & \citet{2025PNAS..12216193W} \\
\multirow{4}{*}{WASP-107 b} & ATMO free retrieval & $-1.85^{+0.22}_{-0.23}$ & $-6.03^{+0.22}_{-0.20}$ & $-3.33^{+0.29}_{-0.25}$ & $-2.70^{+0.28}_{-0.48}$ & \citet{2024Natur.630..831S} \\
 & NEMESIS free retrieval & $-1.72^{+0.26}_{-0.25}$ & $-6.14^{+0.42}_{-1.70}$ & $-2.62^{+0.36}_{-0.34}$ & $-3.85^{+0.62}_{-0.87}$ & \citet{2024Natur.630..831S} \\
 & AURORA free retrieval & $-2.60^{+0.70}_{-0.60}$ & $-6.10^{+0.50}_{-0.40}$ & $-4.40^{+0.80}_{-0.70}$ & $-3.00^{+0.70}_{-0.70}$ & \citet{2024Natur.630..836W} \\
 & CHIMERA free retrieval & $-2.10^{+0.20}_{-0.30}$ & $-5.80^{+0.20}_{-0.20}$ & $-3.90^{+0.30}_{-0.30}$ & $-1.90^{+0.20}_{-0.20}$ & \citet{2024Natur.630..836W} \\
V1298 Tau b & PICASO free retrieval & $-2.50^{+0.40}_{-0.60}$ & $-6.20^{+0.40}_{-0.50}$ & $-3.80^{+0.50}_{-0.80}$ & $-3.10^{+0.50}_{-0.60}$ & \citet{Barat2025} \\
GJ 3470 b & AURORA free retrieval & $-1.08^{+0.43}_{-0.52}$ & $-4.05^{+0.25}_{-0.27}$ & $-2.47^{+0.61}_{-0.43}$ & $-0.96^{+0.18}_{-0.70}$ & \citet{2024ApJ...970L..10B} \\
TOI-270 d & SCARLET free retrieval & $-1.10^{+0.31}_{-0.92}$ & $-1.64^{+0.38}_{-0.36}$ & $-1.67^{+0.40}_{-0.60}$ & $-1.46^{+0.00}_{-3.00}$ & \citet{2024arXiv240303325B} \\
\enddata
\tablecomments{All values are log$_{10}$ volume mixing ratios with their 1$\sigma$ uncertainties derived from the JWST retrievals listed in Table~\ref{table:target}.}
\end{deluxetable*}

\begin{deluxetable*}{lccccccccc}
\tablecaption{Elemental Ratio Constraints for the Exoplanet Sample. \label{table:elementalratio}}
\tablehead{
\colhead{Target} & 
\colhead{[O/H]} & 
\colhead{[C/H]} & 
\colhead{[C/O]} & 
\colhead{Adopted [M/H]} & 
\colhead{[C/O] for adopted} & 
\colhead{Adopted [C/O]} & 
\colhead{Reference} & 
\colhead{Retrieval Method} \\
\colhead{} & 
\colhead{($\times$ solar)} & 
\colhead{($\times$ solar)} & 
\colhead{($\times$ solar)} & 
\colhead{($\times$ solar)} & 
\colhead{[M/H] ($\times$ solar)} & 
\colhead{($\times$ solar)} & 
\colhead{} & 
\colhead{}
}
\startdata
HD 209458 b & 1.04-4.70 & 0.29-1.04 & 0.07-0.75 & 5 & 0.18-0.22 & 0.25 & \citet{2024ApJ...963L...5X} & PLATON free retrieval for CH$_4$ \\
WASP-166 b & 34.07-103.92 & 5.36-26.36 & 0.06-0.49 & 100 & 0.22-0.74 & 0.25 & \citet{2025AJ....170...50M} & POSEIDON free retrieval \\
HD 189733 b & 0.10-3.94 & 0.04-2.50 & 0.01-2.05 & 5 & 0.58-0.65 & 0.5 & \citet{2024Natur.632..752F} & CHIMERA free retrieval \\ 
HIP 67522 b & 0.04-3.04 & 0.02-2.76 & 0.01-2.13 & 3 & 0.85-0.94 & 1 & \citet{2024AJ....168..297T} & CHIMERA grid retrieval \\ 
HAT-P-12 b & 0.67-26.21 & 0.82-32.16 & 0.08-2.13 & 20 & 0.87-1.68 & 1 & \citet{2025AA...703A.264C} & ARCiS free retrieval \\ 
WASP-69 b & 2.99-10.51 & 3.81-15.15 & 0.72-1.85 & 10 & 1.41-1.47 & 1.5 & \citet{2024AJ....168..104S} & PICASO equilibrium grid retrieval \\ 
HAT-P-18 b & 0.57-2.98 & 0.06-2.07 & 0.03-1.32 & 3 & 0.66-0.70 & 0.5 & \citet{2022ApJ...940L..35F} & ATMO free retrieval \\
\multirow{2}{*}{WASP-80 b} & 1.80-55.17 & 0.02-0.65 & 0.0004-0.34 & 50 & 0.0004-0.012 & 0.25 & \citet{2023Natur.623..709B} & AURORA free retrieval \\
 & 2.31-31.09 & 2.12-28.94 & 0.11-2.02 & 20 & 0.11-1.53 & 0.25 & \citet{2025PNAS..12216193W} & CHIMERA free retrieval \\
\multirow{4}{*}{WASP-107 b} & 9.40-28.77 & 1.99-10.22 & 0.08-0.74 & 30 & 0.31-0.35 & 0.25 & \citet{2024Natur.630..831S} & ATMO free retrieval \\
 & 12.78-45.74 & 2.40-13.21 & 0.07-0.60 & 30 & 0.08-0.45 & 0.25 & \citet{2024Natur.630..831S} & NEMESIS free retrieval \\
 & 0.84-17.97 & 0.45-11.44 & 0.04-1.94 & 10 & 0.04-1.14 & 0.25 & \citet{2024Natur.630..836W} & AURORA free retrieval \\ 
 & 12.01-33.30 & 17.40-44.54 & 0.85-1.82 & 30 & 1.21-1.53 & 1.5 & \citet{2024Natur.630..836W} & CHIMERA free retrieval \\ 
V1298 Tau b & 1.03-11.36 & 0.49-6.53 & 0.06-1.65 & 10 & 0.34-0.64 & 0.25 & \citet{Barat2025} & PICASO free retrieval \\
GJ 3470 b & 49.90-477.61 & 49.05-471.09 & 0.20-1.89 & 100 & 0.51-1.58 & 0.5 & \citet{2024ApJ...970L..10B} & AURORA free retrieval \\ 
TOI-270 d & 18.74-314.33 & 31.98-312.77 & 0.19-6.48 & 100 & 0.32-2.60 & 0.25 & \citet{2024arXiv240303325B} & SCARLET free retrieval \\
\enddata
\tablecomments{Ranges of elemental ratios (O/H, C/H, C/O) relative to protosolar are derived from JWST retrievals in Table~\ref{table:vmr}. ``Adopted" values are those used to generate the self-consistent P-T profiles in this study.}
\end{deluxetable*}

\bibliography{refererence}{}
\bibliographystyle{aasjournalv7}

\end{CJK*}
\end{document}